\documentclass[11pt,a4paper]{article}
\makeatletter
\IfFileExists{jheppub.sty}{\usepackage{jheppub}}{%
\usepackage{amsmath,amssymb,graphicx}
\usepackage[numbers,sort&compress]{natbib}
\usepackage[a4paper,top=26mm,bottom=27mm,left=29mm,right=29mm]{geometry}
\usepackage[T1]{fontenc}
\usepackage{lmodern}
\usepackage{microtype}
\usepackage{fancyhdr}
\usepackage{xcolor}
\usepackage[colorlinks=true,citecolor=blue,linkcolor=blue,urlcolor=blue]{hyperref}
\setlength{\parindent}{1.25em}\setlength{\parskip}{0pt}
\setcounter{secnumdepth}{3}
\pagestyle{fancy}\fancyhf{}\fancyfoot[C]{-- \thepage\ --}\fancypagestyle{plain}{\fancyhf{}\fancyfoot[C]{-- \thepage\ --}}
\renewcommand{\abstract}[1]{\gdef\jhep@abstract{##1}}
\newcommand{\keywords}[1]{\gdef\jhep@keywords{##1}}
\newcommand{\emailAdd}[1]{\gdef\jhep@email{##1}}
\newcommand{\affiliation}[2][]{\gdef\jhep@affiliation{##2}}
\newcommand{\arxivnumber}[1]{}

\def\jhep@abstract{}\def\jhep@keywords{}\def\jhep@email{}\def\jhep@affiliation{}
\renewcommand{\maketitle}{\thispagestyle{empty}\vspace*{1.7cm}\begin{center}{\LARGE\bfseries\@title\par}\vspace{1.4cm}{\normalsize\@author\par}\vspace{0.3cm}\ifx\jhep@affiliation\@empty\else{\small\jhep@affiliation\par}\fi\ifx\jhep@email\@empty\else{\small E-mail: \href{mailto:\jhep@email}{\jhep@email}\par}\fi\end{center}\vspace{1.25cm}\noindent{\bfseries Abstract:}\quad\jhep@abstract\par\vspace{0.6cm}\noindent{\small\bfseries Keywords:}\quad{\small\jhep@keywords}\par\vspace{0.55cm}\tableofcontents\clearpage}
}
\makeatother
\usepackage[utf8]{inputenc}
\usepackage{mathtools,bm}
\usepackage{booktabs,longtable,tabularx,array,multirow}
\usepackage{tikz}
\usetikzlibrary{arrows.meta,positioning,calc,fit,shapes.geometric}
\usepackage{xcolor}
\usepackage{enumitem}
\usepackage{physics}
\usepackage{siunitx}
\usepackage{csquotes}
\usepackage[nameinlink,capitalise]{cleveref}
\usepackage{caption,subcaption,float}
\usepackage{url}
\usepackage{graphicx}
\definecolor{accent}{RGB}{18,73,120}
\definecolor{soft}{RGB}{238,244,249}
\setlist{itemsep=0.2em,topsep=0.25em}
\allowdisplaybreaks[2]
\numberwithin{equation}{section}

\newcommand{\YB}{Y_B}
\newcommand{\helicity}{\mathcal H}

\title{Electroweak-Scale Cosmic Magnetism: Magnetogenesis, Baryogenesis, Phase Transitions and Gravitational Waves}
\author{Fayez Abu-Ajamieh}
\emailAdd{fayezabuajamieh@gmail.com}
\affiliation{(Foremly)L The Indian Institute for Science}
\abstract{Magnetic fields, the baryon asymmetry of the Universe, and stochastic gravitational waves provide complementary probes of physics in the early Universe. Near the electroweak scale, these phenomena are not necessarily independent. The electroweak anomaly allows hypermagnetic helicity to exchange charge with chiral fermions, sphalerons can change baryon and lepton number, and first-order phase transitions in extensions of the Standard Model can drive plasma motion, magnetogenesis, and gravitational waves. In this review, we explain these connections and examine the literature needed to make quantitative predictions. We begin with the properties of primordial magnetic fields and the electroweak plasma, then discuss the mechanisms that produce magnetic fields, the conversion between helicity and baryon number, magnetic backreaction, and the evolution of the field to the present day. We also review the observational constraints and recent developments, including gauge-invariant descriptions of magnetic flux through the crossover and reconnection-controlled field evolution. Particular attention is given to the assumptions behind commonly used formulas and to the distinction between established results and model-dependent estimates. We argue that the most useful way forward is to predict the baryon abundance, present-day magnetic-field spectrum, and gravitational-wave signal from the same microscopic model, with consistent treatment of the intervening plasma dynamics.}

\begin{document}
\maketitle

\section{Scope, motivation and the literature landscape}
\label{sec:scope}

The electroweak epoch offers a particularly interesting connection between magnetic fields, the excess of matter over antimatter, and gravitational radiation. These phenomena are often studied independently, but the Standard Model anomaly and the dynamics of the electroweak plasma allow them to influence one another. We first explain the connection and then review the theoretical and observational ingredients needed to calculate it.

\subsection{Three relics that should often be calculated together}

One measured relic of early-Universe physics is the excess of matter over antimatter. We characterize it by the baryon-to-entropy ratio
\begin{equation}
Y_B\equiv \frac{n_B-n_{\bar B}}{s}\simeq 8.6\times 10^{-11},
\end{equation}
or, after electron--positron annihilation, by the baryon-to-photon ratio $\eta_B\simeq 6.1\times10^{-10}$. A second possible relic is a primordial magnetic field, which may contribute to the fields observed in galaxies, clusters, and cosmic voids. A third is a stochastic gravitational-wave background generated by phase transitions, fluid motion, turbulence, or magnetic stresses.

The three observables are commonly studied one at a time. This separation is useful in many applications, but it is not always justified near the electroweak scale. The electroweak theory contains anomalous global currents; the plasma supports hypercharge gauge fields above electroweak symmetry breaking; nonperturbative sphaleron transitions violate $B+L$; chiral chemical potentials source parity-odd currents; and any strongly first-order electroweak or nearby beyond-the-Standard-Model (BSM) transition can generate bulk flows, turbulence, magnetic fields and gravitational waves. Thus a primordial field can both carry information from the early Universe and actively influence its later evolution.

A compact representation of the subject is
\begin{equation}
\boxed{\begin{gathered}
\text{EW dynamics}\longleftrightarrow\text{magnetogenesis}
\longleftrightarrow\text{helicity/chiral anomaly}\\
\updownarrow\hspace{5.5cm}\updownarrow\\[-0.8ex]
\text{GW/IGMF observables}\longleftrightarrow\text{MHD evolution}
\longleftrightarrow\text{BAU}
\end{gathered}}
\end{equation}
The arrows are genuinely two-way in several places. A chiral asymmetry can generate a helical magnetic field through the chiral plasma instability, whereas decaying hypermagnetic helicity can source fermion number through the Abelian anomaly. Strong hypermagnetic fields can modify the electroweak free energy and the sphaleron barrier, while the electroweak transition itself can generate magnetic structure. Helical turbulence can source circularly polarized gravitational radiation, so CP-odd microphysics can in principle leave a macroscopic parity-odd gravitational-wave signature.

\begin{figure}[H]
\centering
\begin{tikzpicture}[node distance=1.65cm and 1.7cm, every node/.style={align=center},box/.style={draw=accent,rounded corners,fill=soft,minimum width=2.9cm,minimum height=0.9cm}, arr/.style={-{Latex[length=2mm]},thick,accent}]
\node[box] (micro) {BSM / EW\\microphysics};
\node[box,right=of micro] (pt) {EW crossover or\\first-order transition};
\node[box,right=of pt] (mag) {Hypermagnetic /\\magnetic fields};
\node[box,below=of mag] (mhd) {MHD evolution\\and helicity};
\node[box,left=of mhd] (bau) {Anomalies, sphalerons\\and baryogenesis};
\node[box,left=of bau] (gw) {Gravitational-wave\\sources};
\draw[arr] (micro)--(pt);
\draw[arr] (pt)--(mag);
\draw[arr] (mag)--(mhd);
\draw[arr] (mhd)--(bau);
\draw[arr] (bau)--(pt);
\draw[arr] (pt)--(gw);
\draw[arr] (mhd)--(gw);
\draw[arr,bend left=12] (bau) to (mag);
\draw[arr,bend left=12] (mag) to (bau);
\end{tikzpicture}
\caption{The coupled system reviewed in this article. The important conceptual point is that magnetic fields are not merely a late-time observable: around the electroweak epoch they can participate directly in charge transport and nonperturbative gauge dynamics.}
\label{fig:network}
\end{figure}
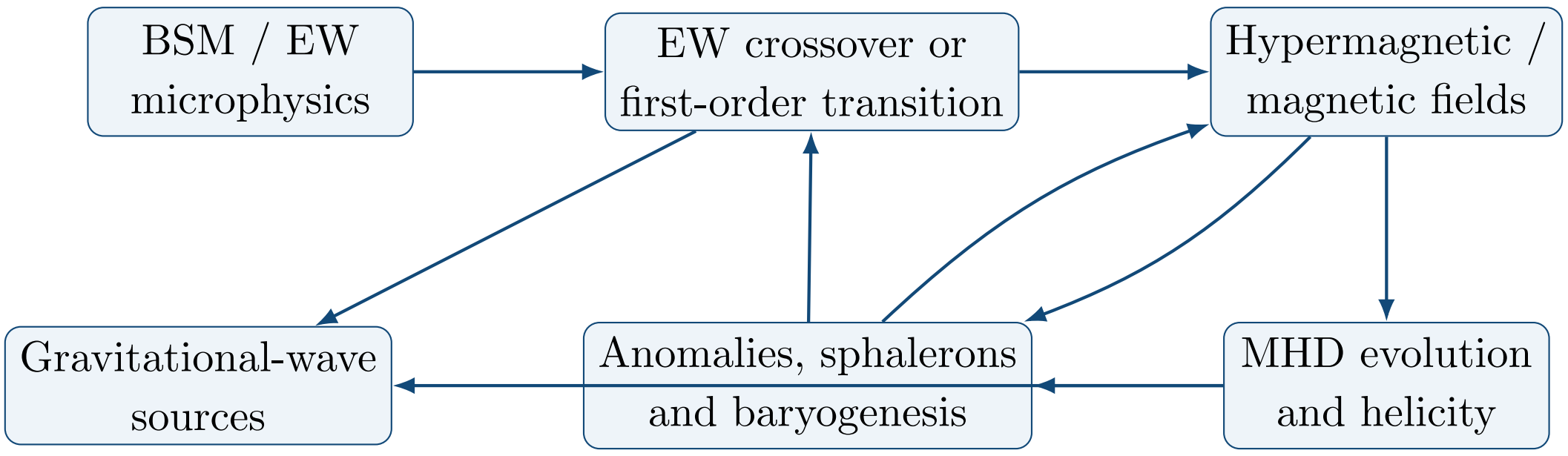

\subsection{Why another review is justified}

The basic connections between primordial fields, anomalous currents, and baryogenesis have been known for decades. Grasso and Rubinstein discussed many of these ideas in their early review, while later reviews developed the cosmological evolution and observation of magnetic fields in considerable detail. The relevant starting points include Refs.~\cite{GrassoRubinstein2001,KandusKunzeTsagas2011,DurrerNeronov2013,Subramanian2016,Vachaspati2021Review,KamadaYamamotoYang2023}. Our purpose is to bring together the developments that are now needed for a quantitative electroweak calculation \cite{VandeVis2025}.

The motivation for a new review is the progress in quantitative methods, not the claim that the basic connection is new. Results from thermal field theory, lattice studies, plasma physics, gravitational-wave phenomenology, and gamma-ray astronomy are now needed together.

Several developments are especially important:
\begin{itemize}
    \item Lattice simulations have established the Standard Model electroweak crossover and provided a quantitative sphaleron rate and freeze-out temperature \cite{DOnofrio2014,DOnofrioRummukainen2016}.
    \item The 2010s produced quantitative kinetic treatments of baryogenesis from evolving hypermagnetic helicity, including evolution through the crossover \cite{FujitaKamada2016,KamadaLong2016a,KamadaLong2016b}.
    \item Chiral magnetohydrodynamics has developed from a qualitative instability argument into a mature dynamical framework with analytic and numerical results \cite{JoyceShaposhnikov1997,Boyarsky2012,Brandenburg2017,Schober2018,KamadaYamamotoYang2023}.
    \item Baryon-isocurvature and Big-Bang-nucleosynthesis constraints have shown that producing the mean BAU is not sufficient; spatial fluctuations can be fatal to otherwise attractive primordial-field scenarios \cite{GiovanniniShaposhnikov1998BBN,KamadaUchidaYokoyama2021}.
    \item The long-time decay of primordial magnetic fields is being reexamined in terms of reconnection-controlled dynamics, changing the mapping from early-Universe field parameters to present-day $(B_0,\lambda_0)$ \cite{HoskingSchekochihin2023,UchidaKamadaTashiro2025}.
    \item First-principles calculations have revisited the response of electroweak sphalerons to background magnetic fields \cite{AnnalaRummukainen2023}, while real-time studies are beginning to include magnetic fields directly in first-order electroweak transition dynamics \cite{DiBianCai2026}.
    \item The 2025--2026 literature contains several mechanisms in which magnetogenesis, baryogenesis and gravitational waves are intentionally predicted together rather than appended as separate calculations \cite{DomckeEmaPatel2025,ArteagaGhoshalStrumia2025,LiuQinBian2026,BorahDevGhoshal2026}.
    \item An updated 2026 analysis finds that the older baryon-isocurvature exclusions need not rule out maximally helical primordial fields, although a nonhelical loophole appears to demand extreme tuning \cite{HamadaMukaidaUchida2026Constraints}.
    \item Most importantly for helicity-driven baryogenesis, Hamada, Mukaida and Uchida have recently reformulated magnetic-flux evolution through the hot Standard Model in terms of generalized symmetries and a gauge-invariant mixing angle. Their analysis finds order-one modifications and an additional nonperturbative process, and explicitly cautions that older estimates may carry sizable uncertainties \cite{HamadaMukaidaUchida2026}.
\end{itemize}

The result is a field in which many papers use the same words---``helical field'', ``inverse cascade'', ``electroweak transition'', ``sphaleron washout''---while making materially different approximations. One purpose of this review is to make those approximations visible.

\subsection{What this review includes and excludes}

We focus on magnetic fields that either originate near electroweak symmetry breaking or influence the plasma during this period. This includes fields generated much earlier if their helicity affects electroweak baryogenesis, as well as nearby BSM transitions whose magnetic and gravitational-wave signals survive until today.

The following topics are treated in detail:
\begin{enumerate}
    \item the statistical description of stochastic primordial magnetic fields and helicity;
    \item the electroweak thermal history, sphalerons and hypercharge/electromagnetic mixing;
    \item electroweak magnetogenesis from Higgs gradients, bubble collisions, defects, turbulence and CP-odd dynamics;
    \item the chiral magnetic effect (CME), chiral vortical effect (CVE), chiral plasma instability (CPI), and anomalous MHD;
    \item baryogenesis from hypermagnetic helicity and the reverse conversion of chiral charge into helicity;
    \item magnetic feedback on the electroweak transition and on the sphaleron barrier;
    \item inflationary and high-scale hypermagnetogenesis insofar as it feeds electroweak baryogenesis;
    \item turbulent and reconnection-controlled field evolution;
    \item gravitational-wave production by phase transitions and magnetic/turbulent stresses, including circular polarization;
    \item CMB, BBN, spectral-distortion and gamma-ray constraints;
    \item a common BAU--PMF--GW phenomenological pipeline and the major unsolved theoretical problems.
\end{enumerate}

We do not attempt to reproduce all of cosmic magnetism. Astrophysical dynamos, detailed cluster magnetism, stellar magnetic fields and late-time battery mechanisms are referenced only to clarify whether a primordial origin is required. Likewise, this is not a complete review of electroweak baryogenesis, inflationary magnetogenesis or gravitational waves from first-order phase transitions. For those subjects we emphasize the pieces needed to understand their magnetic interfaces and refer to dedicated reviews \cite{DurrerNeronov2013,Subramanian2016,KamadaYamamotoYang2023,CapriniFigueroa2018,Caprini2020,Athron2024,VandeVis2025}.

\subsection{A hierarchy of confidence}

Not all results in this subject have the same degree of theoretical control. The anomaly equations, the existence of sphalerons, and the basic helicity constraints are firmly established. Predicting the final baryon asymmetry or today's magnetic field requires additional approximations. We will therefore distinguish exact structural results from numerical results tied to specific assumptions and from questions that remain unresolved.

The most firmly established ingredients are the Standard Model anomaly equations, electroweak sphalerons, the helicity realizability bound, the chiral plasma instability in its appropriate dynamical regime, and the sourcing of tensor perturbations by anisotropic stress. These results provide the basic physical foundation of the subject. Other results are quantitatively well developed but depend on specific assumptions. Examples include lattice determinations of the Standard Model sphaleron rate, MHD decay laws within identified turbulence regimes, and gamma-ray constraints for specified source populations and cascade models.

Several important questions remain less settled. These include the precise efficiency with which hypermagnetic helicity is converted into the final baryon asymmetry during the electroweak crossover, the late-time decay of primordial magnetic spectra across different dynamical regimes, the helicity initially produced by realistic first-order phase transitions, and the interpretation of gamma-ray cascade non-observations when beam--plasma instabilities may be active.

This distinction is not semantic. A common phenomenological formula may be very useful and yet fail to be a controlled precision result when extrapolated outside the assumptions under which it was obtained.

\subsection{Notation and thermal milestones}

We use natural units, $\hbar=c=k_B=1$, throughout. The cosmological scale factor is $a(t)$, $H=\dot a/a$ is the Hubble rate, and $d\tau=dt/a$ defines conformal time. We denote physical electric and magnetic fields by $\bm E$ and $\bm B$, and use $\bm B_Y$ for the hypermagnetic field before electroweak symmetry breaking.

Two Standard Model temperatures recur throughout the discussion. For the observed Higgs mass, lattice simulations find a sharp crossover centered around
\begin{equation}
T_{\rm cross}\simeq 159.5\pm1.5\ \mathrm{GeV},
\end{equation}
while electroweak sphaleron freeze-out occurs later,
\begin{equation}
T_{\rm sph}\simeq 131.7\pm2.3\ \mathrm{GeV},
\end{equation}
with the precise numbers depending on the thermal history and Standard Model parameters \cite{DOnofrioRummukainenPRD2016,DOnofrio2014}. The separation between these temperatures is central to helicity-driven baryogenesis: hypermagnetic/electromagnetic conversion and sphaleron shutoff overlap but are not identical events.

\subsection{The central phenomenological question}

The main question is whether a particle-physics model can predict the baryon asymmetry, primordial magnetic fields, and gravitational waves with a common set of assumptions. Ideally one would start from the microscopic parameters and follow the sequence
\begin{equation}
\begin{aligned}
\bm\theta_{\rm BSM}
&\longrightarrow\left\{T_*,\alpha,\frac{\beta}{H_*},v_w,R_*H_*\right\}
\longrightarrow\left\{B_*,\lambda_*,f_{h,*}\right\}\\
&\longrightarrow Y_B
\longrightarrow\left\{B_0,\lambda_0,f_{h,0}\right\}\\
&\longrightarrow\left\{\Omega_{\rm GW}(f),\Pi_{\rm GW}(f)\right\}.
\end{aligned}
\label{eq:masterpipeline}
\end{equation}
where $f_h$ is a helicity fraction and $\Pi_{\rm GW}$ characterizes circular polarization. In the present literature, almost every arrow in \cref{eq:masterpipeline} is studied somewhere, but very few calculations control the entire chain in one model.

That gap is scientifically useful. It gives a concrete definition of what ``comprehensive'' future work should mean: not adding more mechanisms to a list, but reducing the matching uncertainty between the mechanisms already known to exist.

\section{Magnetic-field primer for high-energy theorists}
\label{sec:primer}

Before discussing individual mechanisms, we establish the magnetic quantities used throughout the review. The main distinction from ordinary particle phenomenology is that a primordial field is a stochastic, expanding, plasma-coupled system. Its strength alone is not enough to describe it.

A particle-physics reader can follow most of the literature once a small set of magnetic quantities and conventions are fixed. Unfortunately, conventions differ significantly between cosmology and plasma-physics papers, so we make them explicit here.

\subsection{Physical and comoving fields}

The expansion of the Universe makes it important to distinguish a physical field from a comoving one. We work with the spatially flat Friedmann--Robertson--Walker metric
\begin{equation}
 ds^2=a^2(\tau)(-d\tau^2+d\bm x^2),
\end{equation}
Maxwell theory is conformally invariant in vacuum. It is therefore convenient to define comoving electric and magnetic fields
\begin{equation}
\bm B_c=a^2\bm B_{\rm phys},\qquad \bm E_c=a^2\bm E_{\rm phys}.
\end{equation}
In ideal MHD without turbulent processing, flux freezing implies $\bm B_{\rm phys}\propto a^{-2}$, so the comoving field is constant. A physical correlation length scales as $\lambda_{\rm phys}\propto a$ under pure expansion, while the comoving length $\lambda_c=\lambda_{\rm phys}/a$ remains fixed. Any departure from these scalings signals diffusion, turbulence, inverse transfer, viscous processing, or a source term.

Radiation domination is especially convenient because after the usual conformal rescalings the relativistic MHD equations closely resemble flat-space MHD. This observation underlies much of the standard numerical treatment of primordial fields \cite{DurrerNeronov2013,Subramanian2016}.

\subsection{Energy density and field amplitude}

The first useful measure of a primordial magnetic field is its energy density. In rationalized natural units, it is
\begin{equation}
 \rho_B=\frac{B^2}{2},
\end{equation}
whereas Gaussian-unit expressions conventionally use $B^2/(8\pi)$. Cosmology papers often quote the magnetic energy fraction
\begin{equation}
 \epsilon_B\equiv \frac{\rho_B}{\rho_{\rm rad}},\qquad
 \rho_{\rm rad}=\frac{\pi^2}{30}g_*(T)T^4.
\end{equation}
At generation, this is frequently a more portable quantity than a field strength in Gauss because it factors out the enormous redshift between the early Universe and today.

The present-day magnetic field can be quoted either by its smoothed amplitude $B_\lambda$ at a chosen comoving scale, often $1\,\mathrm{Mpc}$, or by $B_0$ at the integral scale $\lambda_0$. These quantities need not be close, especially for causal spectra with most of their power at short wavelengths. For a steeply blue spectrum, $B_{1\,\mathrm{Mpc}}$ can be tiny even when the total magnetic energy near the integral scale is substantial.

\subsection{Stochastic fields and power spectra}

A primordial magnetic field is not usually a single uniform vector. It is better described statistically, through its correlation functions and its distribution of power over wavelengths. For an approximately homogeneous and isotropic divergence-free field, the Fourier-space two-point function takes the form
\begin{align}
\left\langle B_i(\bm k)B_j^*(\bm q)\right\rangle
= (2\pi)^3\delta^{(3)}(\bm k-\bm q)
\left[\frac{P_B(k)}{2}P_{ij}(\hat{\bm k})
+\frac{iP_H(k)}{2}\epsilon_{ijl}\hat k_l\right],
\label{eq:twopoint}
\end{align}
where $P_{ij}=\delta_{ij}-\hat k_i\hat k_j$. The parity-even spectrum $P_B$ determines the magnetic energy, while the parity-odd spectrum $P_H$ determines magnetic helicity. Different papers absorb factors of $2$, $4\pi$, or $k^2$ into the spectral definitions; it is safer to compare integrated quantities than raw spectral normalizations.

One useful energy-spectrum convention is
\begin{equation}
 \rho_B=\int_0^\infty dk\,E_B(k),
\end{equation}
and similarly
\begin{equation}
 h_B\equiv\frac{1}{V}\int_V d^3x\,\bm A\cdot\bm B
 =\int_0^\infty dk\,H_B(k).
\end{equation}
The integral scale can be defined by
\begin{equation}
 \xi_B\equiv\frac{\int dk\,k^{-1}E_B(k)}{\int dk\,E_B(k)}.
\label{eq:integralscale}
\end{equation}
Because several definitions of ``coherence length'' appear in the literature, a quantitative comparison should always state the weighting convention.

\subsection{Magnetic helicity}

The field strength does not tell us whether magnetic lines are twisted or linked. That information is contained in magnetic helicity,
\begin{equation}
 \helicity_B=\int d^3x\,\bm A\cdot\bm B,
\end{equation}
measures the linkage and handedness of magnetic flux. In a periodic box or for sufficiently well-behaved fields it is gauge invariant. In open domains a relative-helicity construction is required. Cosmological calculations normally assume periodic/statistically homogeneous conditions, for which the spectral helicity is unambiguous.

Energy and helicity obey a realizability condition,
\begin{equation}
 |H_B(k)|\le \frac{2E_B(k)}{k}.
\label{eq:realizability}
\end{equation}
A maximally helical field saturates this inequality at the scales carrying most of the energy. It is convenient to introduce a fractional helicity $f_h$ through
\begin{equation}
 h_B=2f_h\,\xi_B\rho_B,\qquad |f_h|\le1,
\end{equation}
with the understanding that a single integrated $f_h$ suppresses scale dependence.

Helicity matters for two independent reasons in this review. First, in high-conductivity MHD it is approximately conserved much better than magnetic energy, strongly influencing the decay law and promoting growth of the coherence scale. Second, above electroweak symmetry breaking the Abelian hypermagnetic helicity is related by the Standard Model anomaly to fermionic quantum numbers. Thus $h_Y$ is simultaneously an MHD quantity and a charge reservoir.

\subsection{Causal spectra}

If a magnetic field is produced by a causal process after inflation, it cannot have correlations extending beyond the causal horizon at its formation. Combined with the condition $\nabla\cdot\bm B=0$, this constrains the large-scale power spectrum to be blue. In the standard convention, one obtains
\begin{equation}
 E_B(k)\propto k^4
\end{equation}
for a divergence-free causal field on scales much larger than the stirring scale; equivalent statements are often written as $P_B(k)\propto k^2$ depending on the spectral convention \cite{DurrerCaprini2003,DurrerNeronov2013}. This is sometimes called the Batchelor spectrum.

The consequence is important. A causal electroweak field may have substantial energy on a microscopic integral scale while contributing very little power at $1\,\mathrm{Mpc}$. Comparing such a model directly with a scale-invariant CMB limit quoted as $B_{1\rm Mpc}<\mathrm{few\ nG}$ can therefore be misleading.

\subsection{Conductivity and magnetic Reynolds number}

An early-Universe plasma conducts electricity extremely well. Its large conductivity is the reason magnetic fields can survive for cosmological times. Neglecting anomalous currents for the moment, Ohm's law can be written as
\begin{equation}
 \bm J=\sigma(\bm E+\bm v\times\bm B)+\bm J_{\rm anom},
\label{eq:ohm}
\end{equation}
where $\bm J_{\rm anom}$ contains parity-odd contributions such as the CME. Combining \cref{eq:ohm} with Maxwell's equations gives the induction equation,
\begin{equation}
 \frac{\partial\bm B}{\partial t}
 =\nabla\times(\bm v\times\bm B)
 +\frac{1}{\sigma}\nabla^2\bm B
 +\text{anomalous terms}.
\label{eq:induction}
\end{equation}
The magnetic Reynolds number at scale $L$ is
\begin{equation}
 \mathrm{Re}_M\equiv vL\sigma.
\end{equation}
For \(\mathrm{Re}_M\gg 1\), magnetic advection dominates over Ohmic diffusion. This follows from the induction equation, where the ratio of the advection term to the diffusion term is approximately \(\mathrm{Re}_M=vL\sigma\). Physically, this means that the plasma transports and deforms the magnetic field much faster than resistive effects can dissipate it. Indeed, Ohm's law for a conducting plasma gives
\[
\mathbf{E}+\mathbf{v}\times\mathbf{B}=\frac{\mathbf{J}}{\sigma}.
\]
Since \(\mathbf{J}\sim \mathbf{B}/L\), the resistive term is suppressed relative to \(\mathbf{v}\times\mathbf{B}\) by a factor of order \(1/\mathrm{Re}_M\). Consequently, in the limit of large magnetic Reynolds number, Ohm's law reduces to \(\mathbf{E}+\mathbf{v}\times\mathbf{B}\simeq 0\). Combining this relation with Faraday's law shows that the magnetic flux through any surface moving with the plasma is approximately conserved, \(d\Phi_B/dt\simeq 0\). This is known as *magnetic flux freezing*: the magnetic field is effectively carried along with the plasma rather than diffusing through it. The field strength can nevertheless change as the plasma expands, compresses, or stretches the magnetic-field configuration. Flux freezing is approximate, and resistive effects or magnetic reconnection can still modify the field topology.
The kinetic Reynolds number is
\begin{equation}
 \mathrm{Re}\equiv \frac{vL}{\nu},
\end{equation}
with kinematic viscosity $\nu$. The magnetic Prandtl number
\begin{equation}
 \mathrm{Pr}_M\equiv\frac{\nu}{\eta}=\nu\sigma,
\end{equation}
where $\eta=1/\sigma$ is magnetic diffusivity, controls the relative kinetic and magnetic dissipation scales. In the early Universe $\mathrm{Pr}_M$ can be very large, although its temperature dependence and the relevant transport coefficients vary across particle thresholds \cite{BaymHeiselberg1997,ArnoldMooreYaffe2000,DurrerNeronov2013}.

\subsection{Why the coherence length is usually harder than the field strength}

Producing a large field is not enough. A field generated near the electroweak scale initially occupies a very small physical region compared with a megaparsec today. During radiation domination, the expansion rate is
\begin{equation}
 H_*\simeq1.66\sqrt{g_*}\frac{T_*^2}{M_{\rm Pl}}.
\end{equation}
A causal magnetogenesis mechanism satisfies
\begin{equation}
 \lambda_*\lesssim H_*^{-1},
\end{equation}
and a first-order transition usually produces a still smaller characteristic scale set by the bubble separation,
\begin{equation}
 R_*\sim \frac{v_w}{\beta}
\end{equation}
up to order-one convention-dependent factors. Pure redshift is generally not enough to produce a Mpc-scale field today. The viability of electroweak magnetogenesis therefore depends sensitively on nonlinear MHD evolution.

Consequently, a prediction for the present-day field amplitude $B_0$ and coherence length $\lambda_0$ is meaningful only if the calculation specifies the initial spectrum, the helicity fraction, and the assumed turbulent and viscous history. It must also identify whether the subsequent evolution follows conventional eddy-turnover scaling or a reconnection-controlled prescription. These choices can substantially change the predicted position in the $B$--$\lambda$ plane.

\subsection{Magnetic field, hypermagnetic field and the electroweak basis}

Before electroweak symmetry breaking, it is the Abelian hypercharge field $Y_\mu$, rather than the ordinary photon field, that provides the relevant long-lived gauge field. The photon emerges as the unbroken combination of $SU(2)_L$ and $U(1)_Y$ fields below the electroweak transition:
\begin{equation}
 A_\mu=\sin\theta_W\,W^3_\mu+\cos\theta_W\,Y_\mu,
\end{equation}
while
\begin{equation}
 Z_\mu=\cos\theta_W\,W^3_\mu-\sin\theta_W\,Y_\mu.
\end{equation}
In vacuum $\theta_W$ is the usual weak mixing angle. In the hot plasma, especially across the crossover, the physically relevant interpolation between unconfined magnetic fluxes is subtler than simply inserting a temperature-dependent Higgs expectation value. This issue was treated phenomenologically in earlier baryogenesis calculations and has been revisited in a gauge-invariant framework by Hamada, Mukaida and Uchida \cite{KamadaLong2016b,HamadaMukaidaUchida2026}. We return to it in \cref{sec:baryogenesis}.

\subsection{A translation table}

\begin{table}[H]
\centering
\caption{Common quantities appearing in the literature. Precise normalization conventions differ between authors.}
\begin{tabularx}{\textwidth}{@{}lX@{}}
\toprule
Quantity & Meaning \\
\midrule
$B_*$ & physical or comoving field amplitude at generation\\
$B_0$ & present-day characteristic field amplitude, often at the integral scale \\
$B_{1\rm Mpc}$ & present comoving field smoothed on $1\,\mathrm{Mpc}$ \\
$\lambda_*,\xi_*$ & initial correlation or integral length \\
$\lambda_0$ & present-day correlation/integral length \\
$E_B(k)$ & magnetic energy per unit wavenumber \\
$H_B(k)$ & magnetic helicity per unit wavenumber \\
$f_h$ & integrated helicity fraction, $|f_h|\le1$ \\
$\epsilon_B$ & magnetic energy fraction $\rho_B/\rho_{\rm rad}$ \\
$\mathrm{Re}_M$ & magnetic Reynolds number \\
$\mathrm{Pr}_M$ & magnetic Prandtl number \\
$B_Y,h_Y$ & hypermagnetic field and helicity above the EW crossover \\
\bottomrule
\end{tabularx}
\end{table}

With this dictionary in hand, the rest of the subject reduces to understanding three coupled dynamical layers: the electroweak plasma, anomalous charge transport, and nonlinear MHD.

\section{Electroweak plasma: crossover, anomalies and sphalerons}
\label{sec:electroweak}

The electroweak plasma determines how primordial hypermagnetic fields can exchange helicity with fermion number. Three ingredients are central: the crossover between the electroweak phases, anomalous current conservation laws, and sphaleron processes.

\subsection{The Standard Model transition is a crossover}

An essential starting point is that the Standard Model does not undergo a first-order electroweak phase transition for the observed Higgs mass. Nonperturbative calculations show that the first-order region ends at much smaller Higgs masses, so at $m_h\simeq125\,\mathrm{GeV}$ symmetry breaking occurs through a smooth crossover \cite{Kajantie1996,Kajantie1997}. This matters because bubble-driven magnetogenesis then requires physics beyond the Standard Model \cite{DOnofrio2014}.

This basic fact divides electroweak magnetic cosmology into two broad classes. Mechanisms relying on ordinary bubble nucleation at the electroweak transition require BSM physics. By contrast, helicity conversion, anomalous transport and sphaleron dynamics can operate through the Standard Model crossover itself.

\subsection{Anomalous global currents}

The link between magnetic fields and matter--antimatter asymmetry begins with the chiral anomaly. Classically, baryon and lepton currents appear conserved in the Standard Model, but quantum effects change their divergences. Summing over the fermion generations gives, schematically,
\begin{equation}
\partial_\mu j_{B+L}^{\mu}
= N_g\left[\frac{g^2}{16\pi^2}W^a_{\mu\nu}\widetilde W^{a\mu\nu}
-\frac{g'^2}{16\pi^2}Y_{\mu\nu}\widetilde Y^{\mu\nu}\right],
\label{eq:BplusLanomaly}
\end{equation}
up to convention-dependent normalization factors, whereas $B-L$ is anomaly free within the Standard Model. Here
\begin{equation}
\widetilde F^{\mu\nu}=\frac12\epsilon^{\mu\nu\rho\sigma}F_{\rho\sigma}.
\end{equation}
For the Abelian field,
\begin{equation}
Y_{\mu\nu}\widetilde Y^{\mu\nu}\propto -4\bm E_Y\cdot\bm B_Y
\end{equation}
Thus, the evolution of hypermagnetic helicity is directly related to the generation of fermionic chiral asymmetries through the Abelian anomaly. In particular, the anomaly implies that the fermionic charge associated with a chiral current is not conserved in the presence of hypercharge gauge fields, but instead evolves according to
\[
\partial_\mu j_f^\mu=-C_f\frac{g'^2}{4\pi^2}\mathbf E_Y\cdot\mathbf B_Y,
\]
where \(C_f\) depends on the fermion's chirality and hypercharge. On the other hand, the hypermagnetic helicity, defined as \(H_Y=\int d^3x\,\mathbf A_Y\cdot\mathbf B_Y\), evolves according to
\[
\frac{dH_Y}{dt}=-2\int d^3x\,\mathbf E_Y\cdot\mathbf B_Y,
\]
where boundary contributions have been neglected. Since both equations depend on the same quantity, \(\mathbf E_Y\cdot\mathbf B_Y\), a change in hypermagnetic helicity necessarily induces an anomalous change in the corresponding fermionic charge. In other words, magnetic helicity can act as a reservoir of fermionic chiral asymmetry, allowing an exchange between the gauge-field configuration and the fermionic sector. For example, when a helical hypermagnetic field loses helicity through resistive effects, the anomaly can generate an imbalance between fermionic species and their antiparticles. These asymmetries can subsequently be redistributed by Standard Model interactions, including electroweak sphalerons, and may contribute to the observed baryon asymmetry. Importantly, in the ideal-MHD limit, \(\mathbf E_Y\cdot\mathbf B_Y=0\), so helicity is conserved and this anomalous source vanishes. Therefore, it is the **change in hypermagnetic helicity**, rather than simply the presence of a helical field, that drives this mechanism of baryogenesis.
The non-Abelian term has a different character. Vacuum configurations of $SU(2)_L$ are labeled by a Chern--Simons number, and transitions between topologically distinct sectors change $B+L$. At zero temperature the transition is exponentially suppressed by tunneling. At high temperature it proceeds through thermal activation over the sphaleron barrier \cite{KuzminRubakovShaposhnikov1985,Manton1983,KlinkhamerManton1984}. The standard electroweak-baryogenesis and anomalous-$B+L$ framework was developed and reviewed extensively in the foundational literature \cite{CohenKaplanNelson1993,RubakovShaposhnikov1996,RiottoTrodden1999}.

\subsection{Sphaleron rates}

At high temperature, nonperturbative $SU(2)_L$ transitions can change baryon and lepton number rapidly. In the electroweak-symmetric plasma, the sphaleron rate per unit volume scales as
\begin{equation}
\Gamma_{\rm sph}\sim \kappa\,\alpha_w^5T^4,
\end{equation}
where the coefficient must be obtained nonperturbatively. Lattice simulations find \cite{DOnofrio2014}
\begin{equation}
\frac{\Gamma_{\rm sph}}{T^4}\simeq(18\pm3)\alpha_w^5
\qquad (T>T_{\rm cross}).
\end{equation}
Below the crossover the rate becomes exponentially suppressed as the Higgs expectation value grows. In the temperature interval most relevant for freeze-out, the same lattice study provides the numerical fit
\begin{equation}
\log\left(\frac{\Gamma_{\rm sph}}{T^4}\right)
\simeq 0.83\frac{T}{\mathrm{GeV}}-147.7,
\end{equation}
with quoted uncertainties in the coefficients \cite{DOnofrio2014}.

The practical lesson is that ``electroweak symmetry breaking'' and ``sphaleron freeze-out'' are not a single event. There is a temperature interval over which the composition of the long-range gauge field and the efficiency of $B+L$ violation are both changing. Helicity-driven baryogenesis is highly sensitive to their relative timing.

\subsection{Chemical equilibrium and spectator effects}

An anomaly source does not act on baryon number in isolation. At electroweak temperatures, gauge interactions, Yukawa reactions, and sphalerons redistribute asymmetries among particle species. The resulting chemical-potential constraints determine how much of a source ultimately appears as baryon number. The corresponding chemical-potential bookkeeping is worked out in \cref{sec:chemicalnetwork}.

The exact reduced network depends on temperature. At very high temperatures the electron Yukawa interaction, for example, is too slow to equilibrate $e_R$; at lower temperatures it becomes important. Modern spectator-rate estimates place its equilibration scale in a standard radiation-dominated history at order $10^5\,\mathrm{GeV}$ (with a transition region rather than a sharp switch), whereas some older magnetic-baryogenesis literature used values closer to $10\,\mathrm{TeV}$ \cite{BodekerSchroder2019,GarbrechtSchwaller2014,DomckeKamadaMukaida2023}. This is why a right-handed-electron asymmetry can act as a long-lived chiral reservoir in several high-scale magnetogenesis scenarios \cite{Campbell1992,JoyceShaposhnikov1997,Kamada2018}. The coupled evolution of the fermionic asymmetries can be expressed in compact matrix notation as
\[
\frac{d\mathbf Y}{d\ln T}
=\mathbf S_Y\mathcal E_Y+\boldsymbol{\Gamma}(T)\mathbf Y,
\]
where \(\mathbf Y\) is a vector containing the charge-to-entropy ratios of the different fermionic species. The first term describes the generation of fermionic asymmetries through the Abelian anomaly, with \(\mathcal E_Y\propto\langle\mathbf E_Y\cdot\mathbf B_Y\rangle\) and \(\mathbf S_Y\) containing the corresponding anomaly coefficients. The second term accounts for Standard Model interactions that redistribute or wash out these asymmetries, including Yukawa interactions, strong sphalerons, and weak sphalerons. The matrix \(\boldsymbol{\Gamma}(T)\) contains their temperature-dependent rates, normalized to the Hubble expansion rate. Since different interactions enter thermal equilibrium at different temperatures, the resulting equations depend on which processes are assumed to be sufficiently fast. This explains why different treatments in the literature can lead to different predictions for the final baryon asymmetry.

\subsection{The electroweak mixing problem in a thermal medium}

It is tempting to describe the conversion of a hypermagnetic field into an electromagnetic field by simply making the vacuum weak mixing angle temperature dependent. This is useful phenomenologically, but the definition of such an angle in a non-Abelian thermal medium requires care. In particular, screening and nonperturbative magnetic dynamics mean that the naive interpolation need not be gauge invariant \cite{KamadaLong2016b}.

Hamada, Mukaida and Uchida revisited the issue using the dimensionally reduced three-dimensional effective theory and the language of generalized symmetries \cite{HamadaMukaidaUchida2026}. They identify the magnetic symmetries of the hot theory, define a gauge-invariant interpolation for unconfined magnetic flux, and argue that approximate conservation of that flux alters the conventional conversion picture. Their analysis finds two order-one corrections to the standard helicity-decay estimate and points to an additional nonperturbative process in a background magnetic flux.

The conservative interpretation is not that earlier calculations are invalid in all regimes, but that the conversion coefficient relating a given primordial hypermagnetic configuration to the final $Y_B$ is currently not known at precision level. This issue deserves to be separated from uncertainties in magnetogenesis and MHD.

\subsection{A thermal timeline}

\begin{figure}[htbp]
\centering
\includegraphics[width=0.98\linewidth]{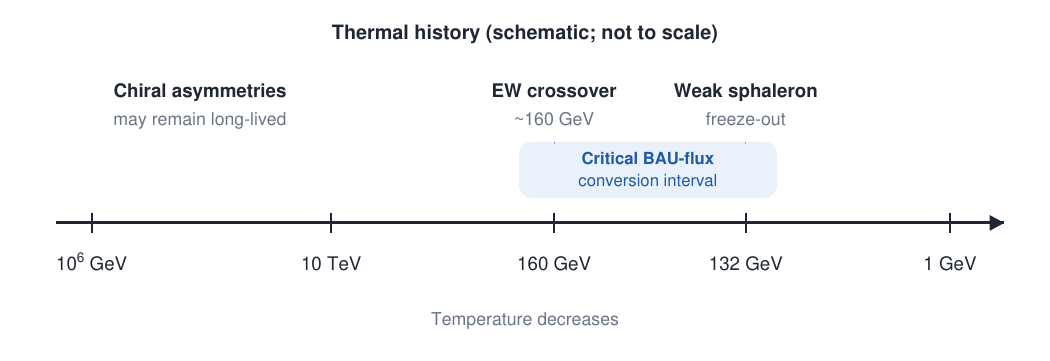}
\caption{Schematic thermal timeline. Temperature decreases from left to right; horizontal positions are not to scale. The electroweak crossover and sphaleron freeze-out temperatures are Standard Model values; BSM first-order transitions can occur at different temperatures.}
\label{fig:timeline}
\end{figure}

\subsection{Why the Sakharov conditions look different in magnetic baryogenesis}

Magnetic baryogenesis satisfies the same basic Sakharov requirements as conventional baryogenesis, but the physical sources are different. The anomaly and sphalerons violate baryon number; a helical field can carry information about parity and CP breaking; and time-dependent helicity, magnetic diffusion, or chiral instability can drive the system away from equilibrium.

This is why statements such as ``$B-L=0$ is always washed out'' require qualification. In equilibrium and without a persistent source, sphalerons erase $B+L$. But if a hypermagnetic source continuously drives the anomalous charge while sphalerons remain active, a nonzero quasi-steady asymmetry can survive until freeze-out \cite{KamadaLong2016a,KamadaLong2016b}.

For magnetic baryogenesis, the final baryon abundance depends on the complete thermal history rather than on the asymmetry generated at a single temperature. One must follow the source $\bm E_Y\cdot\bm B_Y$, the spectator interactions, the conversion of hypermagnetic into electromagnetic flux, and the sphaleron rate $\Gamma_{\rm sph}(T)$ throughout the crossover.

\section{A practical first-order phase-transition primer}
\label{sec:foptprimer}

In the Standard Model the electroweak transition is a crossover. Many beyond-the-Standard-Model theories instead predict a first-order transition, which can supply out-of-equilibrium plasma motion and seed magnetic fields. We summarize the quantities needed to connect a transition calculation with magnetogenesis and gravitational waves.

A substantial part of electroweak cosmic magnetism concerns beyond-the-Standard-Model dynamics that turns electroweak symmetry breaking, or a nearby hidden-sector transition, into a first-order phase transition (FOPT). Readers entering from particle phenomenology often know the effective-potential language but not which quantities are actually passed to magnetohydrodynamic and gravitational-wave calculations. This section supplies that bridge. Detailed modern reviews of thermal phase transitions and electroweak baryogenesis should be consulted for precision calculations \cite{Caprini2020,Athron2024,VandeVis2025}; here the emphasis is on the variables that control magnetic initial conditions.

\subsection{From a thermal potential to bubble nucleation}

A first-order phase transition occurs when the free energy develops two competing minima separated by a barrier. In particle-physics models, this is studied using the finite-temperature effective potential for a background scalar field $\phi$. A commonly used one-loop expression is
\begin{equation}
V_{\rm eff}(\phi,T)=V_0(\phi)+V_{\rm CW}(\phi)+V_T(\phi,T)+V_{\rm resum}(\phi,T),
\label{eq:fopt_veff}
\end{equation}
where first is the tree-level potential, the second term is the Coleman-Weinberg potential that account for zero-temperature quantum fluctuations, the third is is the thermal fluctuations, and the final encodes the resummation of infrared-sensitive diagrams. We explain below how each term is obtained; a more extensive discussion can be found in Ref.~\cite{Athron2024}.

The \emph{tree-level potential} $V_0$ follows directly from the scalar interactions in the Lagrangian. For example, along the neutral Higgs direction in the Standard Model, $H=(0,\phi/\sqrt{2})^T$, one has
\[
V_0(\phi)=-\frac{\mu^2}{2}\phi^2+\frac{\lambda}{4}\phi^4.
\]
In an extension with additional scalars, the corresponding mass terms, portal couplings, and scalar self-interactions must be included. The starting point for the loop calculation is then the set of \emph{field-dependent masses} $m_i(\phi)$. For scalars, the squared masses are the eigenvalues of the Hessian of $V_0$ with respect to the scalar fields, evaluated on the chosen background; for gauge bosons and fermions, they follow from their gauge and Yukawa couplings to that background. One must also specify the multiplicity $n_i$ of each degree of freedom.

The \emph{Coleman--Weinberg potential} $V_{\rm CW}$ is the one-loop correction at zero temperature. In the $\overline{\rm MS}$ scheme, and for the usual Landau-gauge expression, it can be written as
\[
V_{\rm CW}(\phi)=\sum_i\frac{(-1)^{F_i}n_i}{64\pi^2}m_i^4(\phi)
\left[\ln\!\left(\frac{m_i^2(\phi)}{Q^2}\right)-C_i\right]+V_{\rm ct}(\phi),
\]
where $F_i=0$ ($1$) for bosons (fermions), $Q$ is the renormalization scale, and $C_i=3/2$ for scalars and fermions and $5/6$ for gauge bosons in this convention. The optional finite counterterm contribution $V_{\rm ct}$ depends on the renormalization prescription; it is often chosen to preserve specified zero-temperature masses and vacuum expectation values. In practice, one computes the mass spectrum as a function of $\phi$, inserts these masses into the sum, and fixes the renormalized parameters using the desired vacuum conditions. If the scalar curvature masses become negative away from a minimum, the resulting imaginary part of the perturbative potential needs a specified treatment rather than being interpreted as a physical free-energy contribution.

The \emph{thermal correction} $V_T$ comes from the same fluctuations in a heat bath at temperature $T$. Evaluating the one-loop thermal determinant, or equivalently summing the bosonic and fermionic Matsubara frequencies, gives
\[
V_T(\phi,T)=\frac{T^4}{2\pi^2}
\left[\sum_{i\in B}n_i J_B\!\left(\frac{m_i^2(\phi)}{T^2}\right)
-\sum_{i\in F}n_i J_F\!\left(\frac{m_i^2(\phi)}{T^2}\right)\right],
\]
with
\[
\begin{aligned}
J_B(y)&=\int_0^\infty\!dx\,x^2\ln\!\left(1-e^{-\sqrt{x^2+y}}\right),\\
J_F(y)&=\int_0^\infty\!dx\,x^2\ln\!\left(1+e^{-\sqrt{x^2+y}}\right).
\end{aligned}
\]
The functions $J_B$ and $J_F$ can be evaluated numerically without assuming that $m_i/T$ is small. At high temperature, their expansions generate terms proportional to $T^2m_i^2(\phi)$, which tend to restore symmetry, while the bosonic functions also contain a term $-Tm_i^3(\phi)/(12\pi)$. When a bosonic mass is approximately proportional to $\phi$, this cubic term can help generate a barrier between two phases. Fermions do not generate an analogous cubic term at one loop. However, a barrier can also arise already at tree level in extended scalar sectors, and the high-temperature expansion is unreliable when important masses are comparable to or larger than $T$.

Finally, the \emph{daisy (ring) contribution} accounts for the infrared sensitivity of bosonic modes at finite temperature. In particular, bosonic zero Matsubara modes can become light, so that successive insertions of their thermal self-energies are not adequately described by an unresummed one-loop expansion. These self-energies are calculated from finite-temperature two-point diagrams and produce thermal mass corrections $\Pi_i(T)$, typically of order $g^2T^2$, $\lambda T^2$, or $y^2T^2$. For illustration, the leading Standard Model Higgs thermal mass, neglecting the smaller Yukawa couplings, is
\[
\Pi_H(T)\simeq T^2\left(\frac{3g^2+g'^2}{16}+\frac{\lambda}{2}+\frac{y_t^2}{4}\right).
\]
The resummed squared masses are therefore $M_i^2(\phi,T)=m_i^2(\phi)+\Pi_i(T)$, or the eigenvalues of the corresponding corrected mass matrix when fields mix.

In the commonly used Arnold--Espinosa prescription, only the infrared-sensitive bosonic zero modes are resummed. The last term in Eq.~\eqref{eq:fopt_veff} then becomes
\[
\begin{aligned}
V_{\rm resum}(\phi,T)&=V_{\rm daisy}(\phi,T)\\
&=-\frac{T}{12\pi}\sum_{i\in B_0}n_i
\left\{\big[m_i^2(\phi)+\Pi_i(T)\big]^{3/2}
-\big[m_i^2(\phi)\big]^{3/2}\right\},
\end{aligned}
\]
where $B_0$ includes the scalar zero modes and the longitudinal gauge-boson modes, with their appropriate multiplicities. This expression replaces the unresummed zero-mode contribution without double counting it. Another frequently used prescription, due to Parwani, replaces the masses inside the one-loop thermal functions by thermally corrected masses; it is an alternative resummation scheme and should not be combined indiscriminately with the Arnold--Espinosa ring term. The two prescriptions can lead to different numerical predictions at a fixed perturbative order. More systematic treatments use dimensional reduction or higher-order thermal calculations, especially when the transition is strongly coupled or highly supercooled \cite{Athron2024}.

Once $V_{\rm eff}(\phi,T)$ is known, its local minima are followed as the Universe cools. The critical temperature $T_c$ is the temperature at which the competing phases have equal free energy, $V_{\rm eff}(\phi_{\rm false},T_c)=V_{\rm eff}(\phi_{\rm true},T_c)$. This does not imply that the transition occurs at $T_c$: the system can remain trapped in a metastable minimum until thermal fluctuations nucleate bubbles of the stable phase. For a single canonically normalized scalar, the leading thermal tunneling configuration is an $O(3)$-symmetric critical bubble with action
\begin{equation}
S_3(T)=4\pi\int_0^\infty dr\,r^2\left[\frac12\left(\frac{d\phi}{dr}\right)^2+V_{\rm eff}(\phi,T)-V_{\rm eff}(\phi_{\rm false},T)\right],
\end{equation}
subject to
\begin{equation}
\phi'(0)=0,\qquad \phi(r\to\infty)=\phi_{\rm false}.
\end{equation}
The corresponding bounce equation is $\phi''+(2/r)\phi'=\partial V_{\rm eff}/\partial\phi$, with $\phi'(0)=0$ and the field approaching the false vacuum at large $r$. For several scalar fields, the same calculation is performed along a multi-field bounce trajectory with the appropriate kinetic terms. The thermal nucleation rate per unit volume has the semiclassical form
\begin{equation}
\Gamma(T)\simeq A(T)e^{-S_3(T)/T},
\end{equation}
with a prefactor $A(T)$ of order $T^4$ times fluctuation determinants. Determining when nucleation becomes appreciable requires comparing this rate with the cosmological expansion history rather than imposing a universal fixed value of $S_3/T$.

The temperatures used to describe a first-order transition refer to different physical events. The \emph{critical temperature} $T_c$ is where the equilibrium free energies of the phases are equal. It is therefore determined by solving
\begin{equation}
V_{\rm eff}(\phi_{\rm false},T_c)
=V_{\rm eff}(\phi_{\rm true},T_c),
\end{equation}
where $\phi_{\rm false}$ and $\phi_{\rm true}$ are the local minima of the effective potential at $T_c$.

The \emph{nucleation temperature} $T_n$ characterizes the onset of appreciable bubble nucleation, with the precise criterion depending on the cosmological history. It is commonly defined by requiring approximately one bubble to nucleate per Hubble volume,
\begin{equation}
\int_{t_c}^{t_n}dt\,\frac{\Gamma(T)}{H^3(T)}
\simeq 1,
\end{equation}
where $\Gamma(T)\simeq A(T)e^{-S_3(T)/T}$ is the nucleation rate per unit volume. Assuming adiabatic cooling during radiation domination, with approximately constant relativistic degrees of freedom, this becomes
\begin{equation}
\int_{T_n}^{T_c}\frac{dT}{T}
\frac{\Gamma(T)}{H^4(T)}\simeq 1.
\end{equation}
The frequently used approximation $\Gamma(T_n)\sim H^4(T_n)$ is a simpler estimate, but the integral criterion is more appropriate for transitions with nontrivial nucleation histories.

The \emph{percolation temperature} $T_p$ is associated with the growth and overlap of bubbles, and need not coincide with $T_n$. It is commonly determined from the fraction of space remaining in the false vacuum,
\begin{equation}
P_{\rm false}(t)=e^{-I(t)},
\end{equation}
where
\begin{equation}
I(t)=\frac{4\pi}{3}\int_{t_c}^{t}dt'\,
\Gamma(t')a^3(t')
\left[\int_{t'}^t dt''\frac{v_w(t'')}{a(t'')}\right]^3.
\end{equation}
Here $a(t)$ is the scale factor and $v_w$ is the bubble-wall velocity. A commonly adopted percolation criterion is
\begin{equation}
I(T_p)\simeq 0.34,
\qquad P_{\rm false}(T_p)\simeq 0.7,
\end{equation}
corresponding to approximately $30\%$ of space having converted to the true vacuum. For strongly supercooled transitions, one must additionally verify that the physical false-vacuum volume decreases so that the transition can complete.

For strongly supercooled transitions, the reheating temperature $T_{\rm reh}$ can differ appreciably from all of them. If the released energy is rapidly thermalized, energy conservation gives
\begin{equation}
\rho_{\rm rad}(T_{\rm reh})
\simeq \rho_{\rm rad}(T_p)+\Delta\rho(T_p),
\end{equation}
where $\Delta\rho$ is the energy-density difference between the false and true phases. For approximately constant relativistic degrees of freedom, this reduces to
\begin{equation}
T_{\rm reh}\simeq T_p(1+\alpha_p)^{1/4},
\qquad
\alpha_p\equiv\frac{\Delta\rho(T_p)}{\rho_{\rm rad}(T_p)}.
\end{equation}
This approximation assumes rapid conversion and thermalization of the released energy; otherwise, reheating must be determined from the coupled energy-transfer and expansion equations.

A magnetogenesis calculation should specify which temperature is used for the subsequent redshifting of field amplitudes and correlation lengths. Finally, since the perturbative effective potential is generally gauge- and renormalization-scheme dependent at finite order, the predicted transition temperatures and bounce action should be quoted together with an assessment of these uncertainties \cite{Athron2024,MorrisseyRamseyMusolf2012}.

\subsection{The parameters $\alpha$ and $\beta/H$}

Two quantities appear in almost every phenomenological study of a first-order transition. The first is its strength, usually characterized by the released energy relative to the radiation energy density:
\begin{equation}
\alpha\equiv \frac{\Delta\rho}{\rho_{\rm rad}},
\end{equation}
where $\Delta\rho$ denotes an appropriate released vacuum/thermal energy density evaluated near the transition. Several definitions are used in the literature, including latent heat, trace-anomaly and pseudotrace prescriptions. They coincide parametrically for weak transitions but can differ for strongly supercooled transitions. Quoting $\alpha$ without its definition is therefore not sufficient for a precision comparison \cite{Athron2024}.

The inverse duration is conventionally
\begin{equation}
\frac{\beta}{H_*}\equiv \left. T\frac{d}{dT}\frac{S_3(T)}{T}\right|_{T_*}
\end{equation}
for radiation domination and monotonically varying $S_3/T$, up to sign conventions. It approximates the logarithmic time derivative of the nucleation rate. When nucleation is broad, strongly supercooled, or vacuum domination is important, a single $\beta$ can cease to describe the bubble-size distribution reliably. In such cases it is safer to compute a mean bubble separation directly from the nucleation history.

For an approximately exponential nucleation history, a commonly used scaling is
\begin{equation}
R_*\sim \mathcal O(1)\frac{v_w}{\beta},
\end{equation}
where $v_w$ is the bubble-wall speed. The precise order-one coefficient depends on the definition of $R_*$ and the nucleation history. This scale is exceptionally important for cosmic magnetism because it typically sets both the stirring scale of the plasma and the initial magnetic correlation length.

\subsection{Wall velocity, fluid profiles and hydrodynamic regimes}

The velocity of an expanding bubble wall cannot be read off from the effective potential alone. It depends on the driving pressure, particle friction, and the fluid profile ahead of and behind the wall. 
The expanding bubble wall transfers energy to the surrounding plasma, but the resulting fluid motion depends on the wall velocity relative to the speed of sound, $c_s\simeq 1/\sqrt{3}$, and the hydrodynamic solution. Three regimes are commonly distinguished.

In a \emph{deflagration}, the wall moves subsonically and pushes the plasma ahead of it, producing a shock wave that heats and accelerates the fluid before the wall arrives. In a \emph{detonation}, the wall propagates sufficiently fast that the plasma ahead remains undisturbed, while the released energy drives fluid motion behind the wall, including a rarefaction wave. \emph{Hybrid solutions}, also called supersonic deflagrations, lie between these regimes and contain both a shock wave ahead of the wall and a rarefaction wave behind it.

These different fluid profiles determine how much of the released energy is converted into bulk kinetic energy rather than heat. This is commonly characterized by the efficiency factor
\begin{equation}
\kappa_v=\frac{\rho_{\rm kin}}{\Delta\rho},
\end{equation}
where $\Delta\rho$ is the released energy density under the chosen convention. Consequently, even for the same effective potential and phase-transition strength, different wall velocities and hydrodynamic regimes can produce different fluid kinetic energies, affecting the subsequent generation of turbulence, magnetic fields, and gravitational waves \cite{Espinosa2010,Athron2024}.

For magnetogenesis this distinction matters in three ways. First, the fluid velocity determines the available vorticity and turbulent kinetic energy. Second, the thickness and geometry of the sound shell set characteristic gradients that can seed magnetic fields or amplify pre-existing ones. Third, conventional diffusion-based electroweak baryogenesis is most efficient for particular subsonic or mildly supersonic wall regimes, whereas a magnetic-anomaly baryogenesis mechanism can remain viable in regimes where standard charge transport ahead of the wall is inefficient \cite{VandeVis2025}.

One should therefore distinguish the wall Lorentz factor $\gamma_w$, the bulk-fluid velocity behind/ahead of the wall and the rms turbulent velocity. They are related dynamically but are not the same input.

\subsection{From released energy to magnetic energy}

The energy released by a transition does not all become magnetic energy. Part is stored in bulk fluid motion, part in thermal excitations, and part may remain in the scalar sector. It is useful to introduce a magnetic conversion efficiency:
\begin{equation}
\epsilon_{B,*}\equiv\frac{\rho_{B,*}}{\rho_{\rm rad,*}}.
\end{equation}
For a transition-generated field this energy must ultimately come from the same free-energy reservoir that drives scalar gradients and bulk plasma motion. A useful schematic budget is
\begin{equation}
\epsilon_{\phi,*}+\epsilon_{K,*}+\epsilon_{B,*}+\epsilon_{\rm heat,*}
\lesssim \frac{\alpha}{1+\alpha},
\label{eq:energybudgetfopt}
\end{equation}
where the decomposition is model and time dependent. Equation \eqref{eq:energybudgetfopt} is not a universal identity with fixed efficiencies; it is a bookkeeping warning. One cannot maximize the standard sound-wave GW efficiency and independently assume an equipartition magnetic field without checking that both draw on the same energy.

The magnetic initial condition requires more information than $\epsilon_B$. At minimum one needs
\begin{equation}
\left\{E_B(k),\ H_B(k),\ \lambda_*,\ f_{h,*}\right\}.
\end{equation}
Bubble-collision calculations can generate a field even before fully developed turbulence, while later turbulent dynamo action may reshape both the amplitude and spectrum. Three-dimensional simulations by Di et al. demonstrated the simultaneous production of magnetic fields and GWs from an electroweak-scale FOPT and found a broken-power-law magnetic spectrum rather than a monochromatic seed \cite{DiEtAl2021}. Yang and Bian subsequently studied two- and three-bubble magnetic generation with bubble deformation \cite{YangBian2022}. The correct interface to cosmological MHD is therefore a spectrum, not a single field value.

\subsection{Why the same transition produces gravitational waves}

Once the plasma is set in motion, it can also generate gravitational waves. Scalar-field gradients, acoustic waves, and turbulence contribute on different time scales. For phenomenological purposes the total spectrum is commonly decomposed as
\begin{equation}
\Omega_{\rm GW}=\Omega_\phi+\Omega_{\rm sw}+\Omega_{\rm turb}+\Omega_{B}+\cdots,
\end{equation}
where the last term emphasizes the magnetic anisotropic stress. In many thermal transitions the acoustic contribution can dominate the standard FOPT signal, although its duration may be shorter than a Hubble time when shocks form. Numerical studies have greatly refined the sound-wave picture \cite{Hindmarsh2014,Hindmarsh2015}, while direct MHD simulations show that turbulent and magnetic stresses have their own time coherence and spectral shape \cite{RoperPol2020}.

The characteristic observed frequency associated with a source scale $R_*$ redshifts approximately as
\begin{equation}
f_0\sim \frac{a_*}{a_0}\frac{1}{R_*},
\end{equation}
so an electroweak-scale transition with $R_*H_*\ll1$ naturally maps into the milli-Hertz to higher-frequency range depending on the source and duration. Magnetic inverse transfer after the transition can generate stress on scales larger than the original stirring scale, so the magnetic GW spectrum need not be identical to the instantaneous bubble-collision spectrum.

\subsection{Supercooling changes the rules}

The usual phase-transition formulas assume that the Universe remains radiation dominated and that reheating is a small correction. Strong supercooling can invalidate both assumptions: vacuum energy may dominate before bubbles percolate, and rapid reheating can change the entropy and the physical scale of the magnetic field.
\begin{equation}
\rho_{\rm vac}\gtrsim \rho_{\rm rad},
\end{equation}
causing accelerated expansion, altered bubble-size statistics and substantial entropy production at reheating. The physically relevant redshift of a generated field can then be controlled by $T_{\rm reh}$ rather than the temperature at which the first bubbles appeared. Moreover, MHD cannot begin in its usual form until a sufficiently dense conducting plasma is restored.

This is particularly important for recent supercooled electroweak magnetogenesis models \cite{ArteagaGhoshalStrumia2025}. Their potentially large correlation lengths arise partly because the transition history differs qualitatively from the textbook weakly supercooled case. Applying the usual $R_*\simeq v_w/\beta$ and adiabatic-radiation redshift without verifying their regime of validity can misestimate both $\lambda_0$ and the GW peak.

\subsection{Perturbative-systematic issues}

One should also distinguish the true transition dynamics from the approximations used to calculate it. The finite-temperature effective potential is gauge dependent away from physical observables, and nucleation predictions may depend significantly on the perturbative order, infrared treatment, and renormalization scheme. Dimensional reduction and nonperturbative methods help control these uncertainties when applicable \cite{Athron2024}.

For a magnetic review the appropriate attitude is pragmatic: use $(T_*,\alpha,\beta/H_*,v_w,R_*)$ only as an interface after documenting how they were obtained and what their theory uncertainties are. A visually precise PMF or GW curve does not compensate for uncontrolled transition inputs.

\subsection{Minimum reporting standard for a magnetic FOPT model}

To compare models fairly, a paper that predicts both primordial fields and gravitational waves should report the following inputs and assumptions:
\begin{enumerate}
\item the transition temperatures $T_c,T_n,T_p$ and, if relevant, $T_{\rm reh}$;
\item the definition of $\alpha$ and the method used to obtain $\beta/H$ or the full bubble-size distribution;
\item the wall velocity and whether it is calculated or assumed;
\item the kinetic, magnetic and scalar-gradient energy fractions at a specified time;
\item the magnetic energy and helicity spectra, not only a peak amplitude;
\item the scale at which MHD evolution is initialized and the assumed plasma conductivity/viscosity;
\item the GW source decomposition and whether the acoustic, turbulent and magnetic pieces share a consistent energy budget.
\end{enumerate}

For electroweak magnetogenesis, a first-order phase transition cannot be fully characterized by $\alpha$ and $\beta/H_*$. The relevant initial conditions also include the spatial and temporal distribution of bubbles and plasma flows, which determine the magnetic energy spectrum and its helicity. Because the same stress history sources gravitational waves, the magnetic-field and gravitational-wave calculations must use compatible energy budgets and source lifetimes.

\section{Magnetogenesis at electroweak and nearby phase transitions}
\label{sec:magnetogenesis}

The electroweak scale provides several possible sources of primordial magnetic fields. The basic physics differs from one mechanism to another, but every calculation must determine how the field is generated, how much energy it carries, its typical wavelength, and whether it has a net helicity.

\subsection{General requirements}

To produce a primordial magnetic field, a model must generate currents or time-dependent gauge fields and must specify the scales on which they are correlated. At the electroweak scale, the proposed sources include Higgs gradients, colliding bubbles, charge separation, plasma turbulence, and CP-violating dynamics. The mechanisms differ not only in their amplitudes but also in the spectra and helicities they produce.

A meaningful initial magnetic condition should specify at least
\begin{equation}
\{\epsilon_{B,*},\lambda_*,n_B,f_{h,*}\},
\end{equation}
where $\epsilon_{B,*}$ is the initial magnetic-energy fraction, $\lambda_*$ is an integral scale, $n_B$ characterizes the infrared spectrum, and $f_{h,*}$ is the helicity fraction. Many phenomenological papers instead assume one or more of these quantities. That is acceptable for exploring consequences but must be distinguished from a first-principles magnetogenesis prediction.

\subsection{Higgs-gradient magnetogenesis}

One of the earliest electroweak magnetogenesis mechanisms was proposed by Vachaspati \cite{Vachaspati1991}. The basic idea is that, during electroweak symmetry breaking, the Higgs field may acquire different orientations in different spatial regions. The resulting gradients can generate electromagnetic fields through their coupling to the electroweak gauge bosons. To understand how this happens, consider the covariant derivative of the Higgs doublet,
\begin{equation}
D_\mu\Phi=
\left(\partial_\mu
-\frac{ig}{2}\sigma^a W_\mu^a
-\frac{ig'}{2}Y_\mu\right)\Phi,
\end{equation}
where $W_\mu^a$ and $Y_\mu$ are the $SU(2)_L$ and $U(1)_Y$ gauge fields, respectively. When the Higgs field varies in space, its covariant derivative contains information about both the Higgs gradients and the gauge fields.

To extract the electromagnetic component, we introduce the unit vector
\begin{equation}
\hat n^a=\frac{\Phi^\dagger\sigma^a\Phi}
{\Phi^\dagger\Phi},
\end{equation}
which specifies the local orientation of the Higgs field in weak-isospin space. A gauge-invariant definition of the electromagnetic field-strength tensor in the broken phase is then \cite{Tornkvist1998}
\begin{equation}
\begin{aligned}
\mathcal F_{\mu\nu}^{\rm em}
={}&-\sin\theta_W\,\hat n^a W_{\mu\nu}^a
+\cos\theta_W\,Y_{\mu\nu}\\
&+\frac{\sin\theta_W}{g}
\epsilon^{abc}\hat n^a
(D_\mu\hat n)^b(D_\nu\hat n)^c,
\end{aligned}
\end{equation}
where
\begin{equation}
(D_\mu\hat n)^a=
\partial_\mu\hat n^a
+g\epsilon^{abc}W_\mu^b\hat n^c.
\end{equation}
The last term shows explicitly how spatial variations of the Higgs orientation can contribute to the electromagnetic field.

For illustration, consider a configuration with vanishing gauge potentials, $W_\mu^a=Y_\mu=0$, and a nonzero Higgs magnitude. The magnetic field then becomes
\begin{equation}
B_i=\frac{1}{2}\epsilon_{ijk}
\mathcal F_{jk}^{\rm em}
=\frac{\sin\theta_W}{2g}
\epsilon_{ijk}\epsilon^{abc}
\hat n^a\partial_j\hat n^b
\partial_k\hat n^c.
\end{equation}
Thus, a magnetic field can arise from spatially varying Higgs orientations even when the gauge potentials initially vanish in this particular gauge. More explicitly, parameterizing
\begin{equation}
\hat{\bm n}=
(\sin\vartheta\cos\varphi,\,
\sin\vartheta\sin\varphi,\,
\cos\vartheta),
\end{equation}
gives
\begin{equation}
\mathcal F_{ij}^{\rm em}
=\frac{\sin\theta_W}{g}\sin\vartheta
\left(\partial_i\vartheta\,\partial_j\varphi
-\partial_j\vartheta\,\partial_i\varphi\right).
\end{equation}
This expression makes the mechanism transparent: a nonzero magnetic field requires spatial variations of the Higgs orientation in at least two independent directions. If the orientation varies only along one spatial direction, this contribution vanishes.

If the Higgs orientation changes over a characteristic length scale $L$, the corresponding field-strength estimate is
\begin{equation}
B\sim\frac{\sin\theta_W}{gL^2},
\end{equation}
up to geometric factors. This is only an estimate of the field associated with the Higgs configuration, not a prediction for the surviving primordial magnetic field. Indeed, the gauge fields can compensate the Higgs gradients, and the complete evolution of the gauge-Higgs system must be considered to determine the physical electromagnetic flux.

This distinction is important because the Higgs orientation itself is gauge dependent, even though $\mathcal F_{\mu\nu}^{\rm em}$ is gauge invariant. Moreover, the electromagnetic field is not uniquely defined where the Higgs field vanishes, as can occur near electroweak defects. T\"ornkvist discussed these subtleties and the conditions under which Higgs gradients correspond to physical electromagnetic fields \cite{Tornkvist1998}. More recently, Vachaspati and Brandenburg studied magnetic spectra generated from spatially varying Higgs configurations using the Kibble mechanism. They found that the infrared magnetic-energy spectrum behaves as $E_B(k)\propto k^4$, with nontrivial helicity fluctuations \cite{VachaspatiBrandenburg2025}.

For the Standard Model crossover, the picture of sharply separated Higgs domains is not directly applicable in the same way as for a strongly first-order phase transition. Nevertheless, the mechanism illustrates how electroweak symmetry breaking can seed magnetic structure without requiring a conventional astrophysical battery. A quantitative prediction must still determine the subsequent gauge-field dynamics, magnetic spectrum, and plasma evolution.

\subsection{First-order transitions and bubble collisions}

If the electroweak transition is first order, it proceeds through nucleation and growth of bubbles of the broken phase. Collisions and phase gradients can excite gauge fields, while the moving walls stir the plasma. The nucleation rate per unit volume is denoted $\Gamma_{\rm nuc}(T)$ and is often approximated near the transition as
\begin{equation}
\frac{\beta}{H_*}\equiv T_*\left.\frac{d}{dT}\frac{S_3(T)}{T}\right|_{T_*},
\end{equation}
where $S_3$ is the finite-temperature bounce action. The mean bubble separation is parametrically $R_*\sim v_w/\beta$, with convention-dependent numerical factors.

When bubbles carrying different field phases collide, gradients and gauge currents are generated. Baym, Bödeker and McLerran studied magnetic fields produced by phase-transition bubbles \cite{BaymBodekerMcLerran1996}. Ahonen and Enqvist demonstrated ring-like magnetic fields in Abelian-Higgs bubble collisions and discussed multiple collisions and subsequent turbulent enhancement \cite{AhonenEnqvist1998}. Copeland, Saffin and Törnkvist developed the gauge dynamics of bubble collisions and phase equilibration further \cite{CopelandSaffinTornkvist2000}.

The qualitative geometry is robust: the collision region naturally supports circulating currents and therefore ring-like magnetic flux. The quantitative extrapolation to the full electroweak theory is more model dependent because $SU(2)_L\times U(1)_Y$ dynamics, plasma conductivity, wall microphysics and bubble deformation all matter.

Modern simulations have revisited this problem. Yang and Bian studied two- and three-bubble collisions and showed that bubble deformation modifies the generated field amplitude and morphology \cite{YangBian2022}. Numerical work that evolves transition dynamics and magnetic/turbulent stresses in three dimensions has also demonstrated simultaneous production of magnetic spectra and gravitational waves \cite{Hindmarsh2014,RoperPol2020}. These developments move the subject away from simple dimensional estimates toward computable initial spectra.

\subsection{Turbulence-driven fields}

Bubble collisions are not the only source of magnetic fields. A transition can drive shocks, shear flows, and vorticity that amplify weak seed fields through turbulent plasma motion. The kinetic-energy fraction $\epsilon_K$ and its relation to $\kappa_v(\alpha,v_w)$ are therefore as important as the initial magnetic seed.

A schematic bound is
\begin{equation}
\epsilon_{B,*}\lesssim \epsilon_{\rm turb,*}\lesssim \kappa_v\frac{\alpha}{1+\alpha},
\end{equation}
where $\alpha$ is the vacuum-to-radiation energy ratio. The inequalities emphasize that an order-one magnetic fraction should not simply be assumed unless a dynamo or direct gauge-field source is shown to provide it.

The initial scale is usually set by $R_*$ or by a smaller eddy scale. Because a post-inflationary causal source cannot establish correlations beyond the horizon, the large-scale spectrum and coherence scale are constrained by causality and analyticity \cite{HindmarshEverett1998}. This is precisely the scale that also sets the peak of turbulence-sourced gravitational waves, creating an intrinsic PMF--GW correlation.

\subsection{Electroweak strings, monopoles and dumbbells}

Electroweak gauge theory also supports string-like and monopole-like field configurations. These are generally not protected by the topological stability familiar from simpler grand-unified monopoles. Nevertheless, monopole--antimonopole pairs connected by $Z$-flux strings, often called electroweak dumbbells, provide a concrete way to generate electromagnetic structure \cite{Nambu1977} \cite{AchucarroVachaspati2000}.

Grasso and Riotto emphasized that electroweak strings and $W$-condensate structures can source primordial magnetic fields \cite{GrassoRiotto1998}. The old defect literature is particularly relevant today because real-time electroweak evolution can temporarily populate such configurations even when they are not stable asymptotic objects.

A striking modern realization was proposed by Domcke, Ema and Patel \cite{DomckeEmaPatel2025}. With spontaneous CP violation supplied by a rolling axion-like field, electroweak dumbbell configurations acquire a twist. The resulting magnetic field is helical, with a helicity fraction tied parametrically to the dimensionless axion velocity. The same configuration produces a small baryon asymmetry. Conceptually, this is valuable because the helicity is not inserted by hand: its sign and approximate magnitude follow from the CP-odd dynamics.

\subsection{Parity violation and helicity production}

For baryogenesis, the handedness of the field can matter more than its energy density. Ordinary parity-symmetric turbulence may produce a large magnetic field with zero net helicity. A nonzero average $\langle\bm A\cdot\bm B\rangle$ instead requires a parity-asymmetric ingredient or initial condition.

Several routes exist:
\begin{itemize}
\item explicit CP-odd operators such as $|H|^2Y_{\mu\nu}\widetilde Y^{\mu\nu}/\Lambda^2$ or a time-dependent pseudoscalar coupling;
\item chiral charge already present in the plasma, which biases one magnetic helicity through the CME/CPI;
\item parity-odd defect structure or twisted gauge-Higgs configurations;
\item helical fluid forcing, which can transfer kinetic helicity into magnetic helicity.
\end{itemize}

The 2024--2026 work of Liu, Qin and Bian provides an explicit phase-transition example: a CP-violating effective operator creates a time-dependent effective hypercharge chemical potential during bubble evolution, causing the bubble-generated magnetic configuration to acquire helicity \cite{LiuQinBian2026}. Their analysis further shows that the CME can amplify the resulting baryon asymmetry.

\subsection{Strongly supercooled transitions}

Strong supercooling can make electroweak-scale magnetogenesis more efficient by allowing larger bubbles and more energetic collisions. But it also changes the thermal history, so estimates based on radiation domination or instantaneous reheating must be treated with care.

Arteaga Tupia, Ghoshal and Strumia studied dynamical electroweak symmetry breaking with strong supercooling and found parameter regions in which the transition can generate large primordial magnetic fields while also producing potentially observable gravitational waves and, in some regimes, primordial black holes \cite{ArteagaGhoshalStrumia2025}. A key conclusion is that sufficiently low reheating temperatures, below a few TeV in their constructions, facilitate large present-day magnetic relics.

The lesson is broader than the specific model. A phase-transition magnetogenesis paper should report both $T_n$ and the temperature relevant for post-transition plasma evolution, because the redshift and MHD history are controlled by the latter.

\subsection{ALP-assisted transitions and correlated observables}

An axion-like particle (ALP) can play more than one role in a cosmological transition. Its potential can help produce a first-order transition, while a parity-odd coupling to gauge fields can bias magnetic helicity. The same dynamics may then predict a magnetic field and a gravitational-wave signal \cite{BorahDevGhoshal2026}.

The main methodological caution is equally important: maximal helicity is a strong assumption unless the microscopic ALP dynamics is shown to generate it efficiently. This illustrates a recurring hierarchy in the literature:
\begin{equation}
\text{FOPT parameters}\quad\text{are often computed from a potential},
\end{equation}
while
\begin{equation}
\text{magnetic efficiency and helicity}\quad\text{are often phenomenological inputs}.
\end{equation}
A mature PMF--GW prediction should reduce this mismatch.

\subsection{What a complete phase-transition magnetogenesis calculation should output}

A complete calculation should give more than a magnetic amplitude at the transition temperature. The minimum useful output is the initial energy fraction, correlation scale, spectral shape, and helicity fraction, together with enough information to evolve these quantities forward.
\begin{align}
&T_*,\quad \alpha,\quad \beta/H_*,\quad v_w,\quad R_*H_*,\\
&\epsilon_{B,*},\quad \epsilon_{K,*},\quad \lambda_*H_*,\quad n_B,\quad f_{h,*},\\
&E_B(k),\quad H_B(k),\quad E_v(k).
\end{align}
The full spectra are preferable because BAU, MHD evolution and GW production weight wavenumbers differently. A single $B_*$ and $\lambda_*$ can hide physically important UV or IR structure.

\begin{table}[t!]
\centering
\small
\caption{Representative electroweak-scale magnetogenesis channels. ``Helicity'' indicates whether handedness follows from the microphysics or is usually imposed phenomenologically.}
\begin{tabularx}{\textwidth}{@{}p{0.23\textwidth}p{0.20\textwidth}p{0.19\textwidth}X@{}}
\toprule
Mechanism & Transition/source scale & Helicity & Main caveat \\
\midrule
Higgs gradients & domain/Hubble scale; FOPT not required in principle & not generic & gauge-invariant field definition and crossover dynamics \\
Bubble-collision currents & bubble separation; FOPT required & not generic & extrapolation from simplified gauge systems \\
Turbulent dynamo & stirring scale; usually FOPT-driven & not generic & magnetic efficiency and saturation \\
EW defects/dumbbells & defect separation & model dependent & abundance and real-time decay \\
CP-twisted dumbbells & EW defect scale & predicted & dependence on the CP-odd sector \\
CPV SMEFT bubble magnetogenesis & bubble/collision scale; FOPT & predicted & EFT validity and coupled plasma treatment \\
Supercooled EW transition & large bubble scale; FOPT & often model dependent & reheating and nonlinear MHD \\
ALP-sector FOPT & bubble scale & often assumed in PMF projection & derivation of the initial helicity fraction \\
\bottomrule
\end{tabularx}
\end{table}

The principal challenge for electroweak magnetogenesis is no longer simply to demonstrate that magnetic fields can be generated. A predictive model should determine the initial field spectrum and helicity from its microscopic dynamics and then evolve these quantities consistently through baryogenesis, nonlinear MHD, and gravitational-wave production.

\subsection{From a first-order transition to the magnetic power spectrum}
\label{sec:computed_magnetic_spectrum}

The preceding mechanisms explain why a phase transition can generate magnetic fields, but a calculation must also determine how the magnetic energy is distributed among wavelengths. This distinction matters because the energy density controls the magnetic stress, the infrared spectrum determines how the field is constrained by the CMB and gamma-ray observations, and the helicity spectrum enters anomalous baryogenesis. We first explain how a spectrum is extracted from a field configuration and then construct a concrete phenomenological example, following Ref.~\cite{ArteagaGhoshalStrumia2025}. These are two different levels of calculation and should not be confused.

\paragraph{A spectrum from the gauge-field dynamics.}
A numerical calculation begins with the scalar--gauge equations following from the Lagrangian, or their conducting-plasma/MHD limit, together with the bubble nucleation history. For a bosonic electroweak simulation one evolves the Higgs and gauge fields in space and time. Once the physical electromagnetic or hypermagnetic field has been defined, its Fourier components are
\begin{equation}
 B_i(\bm k,t)=\int_V d^3x\,e^{-i\bm k\cdot\bm x}B_i(\bm x,t),\qquad \bm k\cdot\bm B(\bm k,t)=0.
 \label{eq:new_fft_B}
\end{equation}
The spectrum follows by averaging the squared Fourier amplitudes over directions of $\bm k$ and over realizations of the bubble network. For the energy-spectrum convention used in Sec.~2, one chooses $E_B(k,t)$ such that
\begin{equation}
 \rho_B(t)=\frac{\langle B^2(t)\rangle}{2}=\int_0^\infty dk\,E_B(k,t),\qquad
 \frac{d\rho_B}{d\ln k}=k E_B(k,t).
 \label{eq:new_energy_spectrum_norm}
\end{equation}
For the finite-volume Fourier convention in Eq.~\eqref{eq:new_fft_B}, Parseval's theorem gives a direct shell estimator. If the volume is $V$ and Fourier modes are grouped into a bin of width $\Delta k$,
\begin{equation}
 E_B(k_a)\simeq\frac{1}{2V^2\Delta k}
 \sum_{\bm k\,\in\,a}|\bm B(\bm k)|^2,\qquad
 H_B(k_a)\simeq\frac{1}{V^2\Delta k}
 \sum_{\bm k\,\in\,a}\mathrm{Re}[\bm A(\bm k)\cdot\bm B^*(\bm k)],
 \label{eq:new_shell_estimators}
\end{equation}
where in a periodic box the vector potential for nonzero modes may be reconstructed in Coulomb gauge as $\bm A(\bm k)=i\bm k\times\bm B(\bm k)/k^2$. The $1/V^2$ factors refer to the unnormalized forward transform in Eq.~\eqref{eq:new_fft_B}; different FFT libraries require corresponding normalization adjustments. These formulae show precisely how the field configurations generated by a bubble simulation become measurable spectral data. Equivalently, splitting the transverse field into circular polarizations, $\bm B(\bm k)=B_+(\bm k)\bm e_++B_-(\bm k)\bm e_-$, isolates the magnetic-energy and helicity spectra:
\begin{equation}
 E_B(k)\propto k^2\left\langle |B_+|^2+|B_-|^2\right\rangle_k,\qquad
 H_B(k)\propto k\left\langle |B_+|^2-|B_-|^2\right\rangle_k,
 \label{eq:new_two_helicity_spectra}
\end{equation}
where the same Fourier conventions must be used for both expressions. In particular,
\begin{equation}
 |H_B(k)|\leq\frac{2E_B(k)}{k},\qquad
 f_h(k)\equiv\frac{k H_B(k)}{2E_B(k)},\quad |f_h(k)|\leq1.
 \label{eq:new_helicity_realisability}
\end{equation}
Earlier numerical work on magnetic-field generation during electroweak-scale preheating was presented by D\'iaz-Gil and collaborators \cite{DiazGil2008}; the helicity--baryogenesis connection was anticipated by Cornwall \cite{Cornwall1997}. These setups illustrate different generation channels and should not be identified with thermal first-order transitions. A nonhelical ensemble has $\langle |B_+|^2\rangle_k=\langle |B_-|^2\rangle_k$, whereas a maximally helical field populates one polarization only. Consequently, a magnetic-energy spectrum alone does not tell us whether a model generates the net helicity required for helicity-driven baryogenesis.

The lattice simulations of Zhang, Vachaspati and Ferrer \cite{ZhangVachaspatiFerrer2019} offer an example of the first approach. They evolve the classical bosonic electroweak equations with randomly nucleated broken-phase bubbles and measure the magnetic field as the transition proceeds. Their setup converts roughly ten percent of the latent heat into magnetic energy, with the spectrum developing a peak near the final bubble-percolation scale. This is a result of their particular simulation, not a universal magnetogenesis efficiency. Vachaspati and Brandenburg \cite{VachaspatiBrandenburg2025} separately examined the field spectrum generated in a Kibble-type electroweak construction and found the causal infrared behavior $E_B(k)\propto k^4$, rather than the $k^3$ behavior previously used in that context. Their study does not amount to a fully dynamical thermal bubble simulation. A model using either result should state which initial conditions and dynamical effects are actually included.

\paragraph{A worked broken-power-law model.}
A practical example is the supercooled electroweak scenario of Arteaga Tupia, Ghoshal and Strumia \cite{ArteagaGhoshalStrumia2025}. Their magnetic energy at generation is estimated, rather than directly simulated for each microscopic model, as
\begin{equation}
 \rho_{B,*}\simeq\epsilon\,\kappa\,\frac{\alpha}{1+\alpha}\,\Delta V_H,
 \label{eq:new_energy_partition_ags}
\end{equation}
where $\epsilon\sim0.1$ is the assumed magnetic-generation efficiency, $\kappa$ parametrizes the relevant dynamical channel and $\Delta V_H$ is the energy released in the \emph{Higgs} part of the scalar potential. In the very supercooled collision-dominated limit the paper adopts $\kappa\simeq1$. The replacement of $\Delta V_H$ by the total vacuum-energy release would be a different, more optimistic assumption, especially when the transition is driven mainly by a hidden singlet sector. Equation~\eqref{eq:new_energy_partition_ags} must therefore be identified as the energy-budget prescription of this model, rather than a consequence of the bounce calculation alone.

For an approximately exponential bubble-nucleation history the characteristic physical scale at generation is taken as
\begin{equation}
 \lambda_*\simeq\frac{(8\pi)^{1/3}v_w}{\beta},\qquad
 \lambda_* H_*\simeq(8\pi)^{1/3}\frac{v_w}{\beta/H_*}.
 \label{eq:new_bubble_peak_scale}
\end{equation}
The overall coefficient is conventional and depends on the bubble-separation prescription. For $v_w\simeq1$ and $\beta/H_*=30$, one finds $\lambda_*H_*\simeq0.10$. The magnetic energy per logarithmic wavelength interval is then modeled as
\begin{equation}
 \frac{d\rho_{B,*}}{d\ln\lambda}
 =\frac{10}{17}\rho_{B,*}\begin{cases}
 (\lambda/\lambda_*)^{2/3}, & \lambda<\lambda_*,\\
 (\lambda/\lambda_*)^{-5}, & \lambda>\lambda_*.
 \end{cases}
 \label{eq:new_ags_broken_law}
\end{equation}
The coefficient $10/17$ is fixed by requiring that integration over $d\ln\lambda$ reproduces $\rho_{B,*}$: the two branches contribute $3/2$ and $1/5$, respectively, before multiplication by $10/17$. In wavenumber space the large-scale branch gives $d\rho_B/d\ln k\propto k^5$ and hence $E_B(k)\propto k^4$, as expected for a causal, divergence-free field. The short-scale branch corresponds to $d\rho_B/d\ln k\propto k^{-2/3}$, or $E_B(k)\propto k^{-5/3}$. The latter is a phenomenological turbulent slope; neither it nor the position of the turnover follows uniquely from causality. In particular, this spectral prescription is \emph{not} a solution of the field equations for the three particle models considered in Ref.~\cite{ArteagaGhoshalStrumia2025}.

\begin{figure}[t]
\centering
\includegraphics[width=0.82\linewidth]{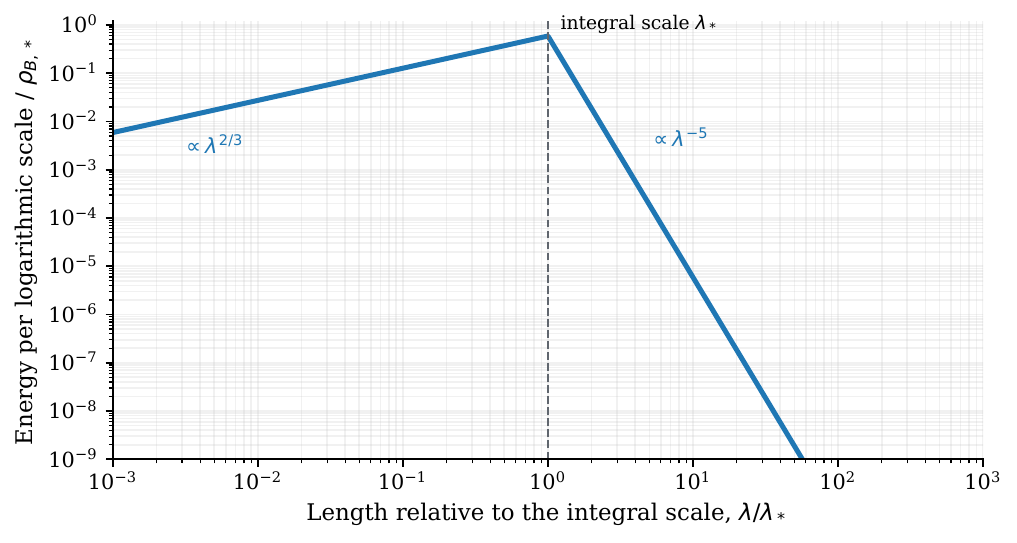}
\caption{A normalized example of the magnetic energy per logarithmic wavelength interval at the time of generation, using the broken-power-law ansatz of Ref.~\cite{ArteagaGhoshalStrumia2025}. The long-wavelength slope corresponds to the causal $E_B(k)\propto k^4$ limit, while the short-wavelength slope is a turbulent modeling assumption. This is an analytical reconstruction of the cited parametrization, not digitized numerical simulation data.}
\label{fig:new_generation_spectrum}
\end{figure}

To turn this into a complete model prediction, one must determine the temperature, transition strength, bubble statistics, and energy available to visible gauge fields from the thermal Lagrangian; evolve the gauge and plasma dynamics to obtain $E_B(k)$ and $H_B(k)$, or explicitly declare an assumed spectral ansatz; and finally propagate that spectrum to the present epoch. Most phenomenological studies control the first step much better than the second. Helicity, in particular, cannot consistently be assigned the value $f_h=1$ merely because the phase transition is strong.

\subsection{How helicity is generated or remains absent}
\label{sec:new_how_helicity_forms}

The energy and helicity of a magnetic field are independent initial data. Bubble collisions, Higgs gradients and a parity-symmetric turbulent dynamo can generate magnetic energy without producing a preferred \emph{sign} of helicity in an ensemble. A helical field instead requires a parity-odd ingredient, such as a rolling pseudoscalar coupled to $Y_{\mu\nu}\widetilde Y^{\mu\nu}$, a chiral chemical potential driving the chiral plasma instability, or a parity-odd gauge-Higgs configuration. An example is the electroweak dumbbell mechanism studied by Domcke, Ema and Patel \cite{DomckeEmaPatel2025}, while helical electroweak fields linked to topological evolution were studied by Zhang, Ferrer and Vachaspati \cite{ZhangFerrerVachaspati2017} and by Copi and collaborators \cite{CopiFerrerVachaspati2008}.

A convenient initial-condition language is
\begin{equation}
 H_{B,*}(k)=\frac{2f_{h,*}(k)}{k}E_{B,*}(k).
 \label{eq:new_fractional_helicity}
\end{equation}
Taking $f_{h,*}(k)=0$ means that the \emph{ensemble-averaged net} helicity vanishes. It does not imply that the field has no local patches of positive and negative helicity, nor does it guarantee the absence of nonhelical inverse transfer. Taking $f_{h,*}(k)=\pm1$ represents the fully helical limiting case. An intermediate value is physically meaningful only if a microscopic CP-odd source or a controlled initial-condition prescription determines it. These distinctions are essential when one wants to use the same magnetic field to predict a baryon asymmetry and a gravitational-wave polarization signal.

\section{Chiral transport and anomalous magnetohydrodynamics}
\label{sec:chiralMHD}

An ordinary magnetized plasma couples electric currents to magnetic fields. In a chiral plasma, the quantum anomaly adds new currents that can amplify one magnetic helicity and change fermion asymmetries. This two-way exchange is the basis of anomalous magnetohydrodynamics.

\subsection{From microscopic chirality to macroscopic current}

In a chiral plasma, right- and left-handed fermions may have different chemical potentials. We define their axial chemical potential by
\begin{equation}
 \mu_5\equiv\frac{\mu_R-\mu_L}{2}.
\end{equation}
The chiral magnetic effect generates a current parallel to a magnetic field,
\begin{equation}
 \bm J_{\rm CME}=\xi_{\rm CME}\bm B,
\qquad
 \xi_{\rm CME}\sim \frac{q^2}{2\pi^2}\mu_5,
\label{eq:CMEcurrent}
\end{equation}
for a simple relativistic fermion species, with the Standard Model generalization involving hypercharges and the full set of chemical potentials. The equilibrium interpretation of the CME requires care, but as an anomalous transport term in a dynamical chiral plasma it provides the standard route to the chiral plasma instability \cite{Vilenkin1980,JoyceShaposhnikov1997,Boyarsky2012,KamadaYamamotoYang2023}.

Anomaly-induced hydrodynamics places both magnetic and vortical transport in a common framework \cite{SonSurowka2009}. Fluid vorticity similarly sources the chiral vortical current,
\begin{equation}
 \bm J_{\rm CVE}=\xi_\omega\bm\omega,
\qquad
 \bm\omega\equiv\nabla\times\bm v,
\end{equation}
with $\xi_\omega$ containing chemical-potential and, depending on the current, temperature-dependent contributions associated with anomalies and equilibrium transport. The CME and CVE couple charge evolution to MHD rather than treating the magnetic field as an externally prescribed background.

\subsection{The induction equation with a chiral term}

The chiral magnetic current modifies the usual magnetic induction equation. To see how, begin with Ohm's law in the plasma rest frame:
\begin{equation}
\bm J=\sigma(\bm E+\bm v\times\bm B)+\xi_{\rm CME}\bm B+\xi_\omega\bm\omega.
\end{equation}
Neglecting the displacement current on MHD scales, $\bm J\simeq\nabla\times\bm B$. Solving for $\bm E$ and using Faraday's law yields
\begin{equation}
\partial_t\bm B
=\nabla\times(\bm v\times\bm B)
+\eta\nabla^2\bm B
+\eta\nabla\times(\xi_{\rm CME}\bm B)
+\eta\nabla\times(\xi_\omega\bm\omega),
\label{eq:chiralinduction}
\end{equation}
where $\eta=1/\sigma$.

To identify the instability, take $\xi_{\rm CME}$ constant, neglect fluid motion, and expand the magnetic field in circularly polarized Fourier modes,
\begin{equation}
 i\bm k\times\bm B_\pm=\pm k\bm B_\pm.
\end{equation}
Then
\begin{equation}
\partial_t B_\pm
=-\eta k^2B_\pm\pm\eta\xi_{\rm CME} kB_\pm.
\label{eq:CPI}
\end{equation}
One helicity grows whenever
\begin{equation}
 k<|\xi_{\rm CME}|,
\end{equation}
with the fastest growth near $k\sim|\xi_{\rm CME}|/2$ in this simplified treatment. This is the chiral plasma instability.

\subsection{The Joyce--Shaposhnikov mechanism}

Joyce and Shaposhnikov showed that a chiral asymmetry in the early-Universe plasma can destabilize one circularly polarized magnetic mode \cite{JoyceShaposhnikov1997}. This provides a direct mechanism for converting fermionic chirality into helical hypermagnetic fields \cite{TashiroVachaspatiVilenkin2012,SemikozSmirnovSokoloff2016}.

The basic conserved combination, in the absence of chirality flips and boundary fluxes, is a sum of fermionic chirality and magnetic helicity. Schematically,
\begin{equation}
 Q_5 + C\,\helicity_Y\simeq \mathrm{const}.
\label{eq:totalchirality}
\end{equation}
The coefficient $C$ depends on charges and conventions. The important point is physical: chirality is not simply destroyed by the anomaly; it can be transferred into gauge-field topology.

This insight underlies much later work, including the possibility that an asymmetry produced at a very high scale is temporarily stored as hypermagnetic helicity and partially recovered as baryon number near the electroweak epoch.

\subsection{Nonlinear chiral MHD}

The linear instability explains how the field initially grows, but not its late-time evolution. As magnetic helicity is generated, the anomaly changes the chiral chemical potentials that drove the instability. The field and the fermion asymmetries must therefore be evolved together \cite{RogachevskiiEtAl2017,Brandenburg2017,Schober2018,BrandenburgKamadaSchober2023,BrandenburgKamadaMukaida2023}.

Several qualitative lessons emerge:
\begin{enumerate}
\item CPI growth tends to populate the magnetic helicity favored by the sign of $\mu_5$.
\item As magnetic helicity grows, the fermionic chiral asymmetry decreases through the anomaly, limiting exponential growth.
\item Nonlinear turbulence can transfer energy and helicity across scales beyond the monochromatic instability band.
\item Chirality-flipping interactions eventually destroy the approximate conservation of total chirality, making the temperature at which the instability operates essential.
\end{enumerate}

The commonly used ``one-mode'' treatment is useful for intuition but can miss spectral effects. Continuous-spectrum studies are increasingly important when comparing to cosmological observables, because BBN and CMB constraints probe different ranges of $k$.

\subsection{Chiral vortical seeding}

The chiral magnetic effect requires a seed field. By contrast, a chiral vortical current can be present in a rotating or vortical fluid even before a sizable magnetic field is established. This makes the chiral vortical effect a possible seed mechanism when the plasma contains chiral asymmetries and vorticity.

A line of work by Abbaslu and collaborators has developed this possibility with Standard Model reaction networks. A 2025 study includes all three generations, perturbative chirality flips, Abelian and non-Abelian anomalies and weak sphalerons, and demonstrates simultaneous generation of a hypermagnetic field and matter asymmetries from transient temperature/vorticity fluctuations \cite{Abbaslu2025}. Such calculations are useful because they address a logical issue often left implicit: a chiral instability can amplify a seed, but a complete cosmology must say where the seed comes from.

\subsection{Hypermagnetic conductivity and diffusion}

Electrical conductivity competes with chiral amplification. Short-wavelength fields dissipate rapidly, whereas sufficiently long-lived modes can grow if the anomalous current overcomes resistive diffusion. Quantitative predictions therefore require temperature-dependent transport coefficients.
\begin{equation}
 k_{\rm diff}^2\sim \sigma_Y H,
\end{equation}
when the Ohmic diffusion time equals a Hubble time. Small-scale hypermagnetic fields above $k_{\rm diff}$ are erased, while large-scale fields survive expansion. This introduces a lower bound on the coherence scale of any helicity that must persist until the electroweak epoch.

For CPI-generated fields, the fastest instability scale must also grow quickly enough before chirality flips equilibrate the relevant fermion species. This produces a nontrivial window in $(\mu_5,T)$ rather than a universal instability condition.

\subsection{Anomaly equation and helicity evolution}

The reason chirality and magnetic helicity are exchanged is contained in two equations: the anomaly equation for a fermionic current and the helicity evolution equation. Their source terms both involve $\bm E\cdot\bm B$, making the connection explicit.
\begin{equation}
\frac{d h_Y}{dt}
=-2\left\langle\bm E_Y\cdot\bm B_Y\right\rangle.
\label{eq:helicitydot}
\end{equation}
Using the chiral Ohm law gives schematically
\begin{equation}
\frac{dh_Y}{dt}
\simeq -\frac{2}{\sigma_Y}
\left[
\left\langle \bm B_Y\cdot\nabla\times\bm B_Y\right\rangle
-\xi_Y\left\langle B_Y^2\right\rangle
\right].
\label{eq:helicityCME}
\end{equation}
The first term is resistive helicity loss, while the second can produce helicity when a chiral chemical potential is present. Through the anomaly, the same expectation value appears in the fermion-number equations,
\begin{equation}
\dot n_f+3Hn_f=-C_f\frac{g'^2}{4\pi^2}
\left\langle\bm E_Y\cdot\bm B_Y\right\rangle
+\text{reactions}.
\label{eq:fermionanomaly}
\end{equation}
Equations \eqref{eq:helicityCME} and \eqref{eq:fermionanomaly} make the feedback explicit.

\subsection{When is anomalous MHD required?}

Ordinary MHD is adequate when chiral chemical potentials are negligible or relax faster than the magnetic field changes. Anomalous MHD is needed when fermionic chirality remains dynamically important on the growth or decay time scale of the field.
\begin{equation}
\tau_{\rm CPI}^{-1}\sim \eta\xi_Y^2
\end{equation}
is competitive with Hubble and reaction times, anomalous terms must be included.

This criterion matters when a paper takes a helical field from inflation or a phase transition and evolves it with conventional MHD while simultaneously computing anomaly-driven baryogenesis. The generated fermion asymmetries can themselves modify the field through the CME. Neglecting that backreaction is justified only in a demonstrable regime.

The anomaly, the chiral magnetic effect, and the chiral plasma instability describe distinct but coupled aspects of the same system. The anomaly transfers charge between fermions and gauge-field helicity, the chiral magnetic effect modifies the constitutive relation for the electric current, and the chiral plasma instability amplifies a helicity mode when the relevant conditions hold. Retaining only one of these effects is justified only when the omitted rates or length scales are demonstrably negligible.

\section{Baryogenesis from hypermagnetic helicity}
\label{sec:baryogenesis}

Magnetic baryogenesis uses a special property of the electroweak theory: changing hypermagnetic helicity can generate fermionic asymmetry through the anomaly. The final baryon abundance is then determined by the coupled charge network and the timing of sphaleron freeze-out.

\subsection{The basic mechanism}

The Abelian anomaly allows an evolving hypermagnetic field to change fermion number. In particular, the decay of a helical hypermagnetic configuration can act as a source of baryon asymmetry. This possibility was developed in the early work of Giovannini and Shaposhnikov \cite{GiovanniniShaposhnikov1998,GiovanniniShaposhnikov1998BBN} \cite{Giovannini2000Knots}.

For a homogeneous treatment one may write
\begin{equation}
\frac{dY_B}{dt}=S_B[h_Y,E_B,\mu_i,T]-\Gamma_B(T)Y_B+\cdots,
\label{eq:simpleYB}
\end{equation}
where $S_B$ is proportional to the anomalous expectation value $\langle\bm E_Y\cdot\bm B_Y\rangle$ and $\Gamma_B$ is related to the weak-sphaleron rate after spectator effects are included. The ellipsis denotes couplings to other charge densities. A realistic calculation solves the full chemical-potential network rather than \cref{eq:simpleYB} alone.

\subsection{Why a helical field is special}

Not every primordial magnetic field produces a net baryon asymmetry. If the two magnetic helicities occur with equal weight, their anomalous contributions cancel on average. A net helical field avoids this cancellation, so its subsequent evolution can source fermion number.

For a near-maximally helical configuration peaked at a characteristic scale $\lambda_Y$, dimensional analysis gives
\begin{equation}
 h_Y\sim \lambda_Y B_Y^2,
\end{equation}
up to order-one factors. The combination $B^2\lambda$ therefore appears repeatedly in phenomenological estimates of the BAU.

\subsection{From qualitative proposal to quantitative kinetic equations}

A quantitative prediction must follow the coupled evolution of the magnetic field, fermion chemical potentials, and electroweak sphalerons. The final asymmetry depends on when hypermagnetic helicity changes relative to sphaleron freeze-out, not merely on the initial magnetic energy \cite{FujitaKamada2016,KamadaLong2016a,KamadaLong2016b}.

The physical picture can be separated into three regimes.

\paragraph{Above the electroweak crossover.}
The field is predominantly hypermagnetic. If its helicity changes, the $U(1)_Y$ anomaly sources fermionic charges. Weak sphalerons rapidly redistribute $B+L$, and spectator reactions enforce chemical equilibrium relations.

\paragraph{Through the crossover.}
The unconfined gauge field evolves from predominantly hypercharge to predominantly electromagnetic. The baryon source associated with hypercharge helicity therefore decreases. At the same time, the broken-phase sphaleron rate is falling rapidly.

\paragraph{After sphaleron freeze-out.}
Ordinary electromagnetic helicity can remain cosmologically long lived, but it no longer drives Standard Model baryon number in the same way. The final $Y_B$ is frozen, barring later baryon-violating new physics.

The relic asymmetry is consequently determined by a race:
\begin{equation}
\boxed{
\text{hypermagnetic source shuts off}
\quad\text{versus}\quad
\text{sphaleron washout shuts off}.}
\end{equation}

\subsection{A schematic kinetic derivation}

The anomaly supplies the source, but the plasma redistributes the resulting charge. A useful schematic equation for the baryon-to-entropy ratio is
\begin{equation}
 n_i-\bar n_i=\frac{g_iT^2}{6}\mu_i
\end{equation}
for relativistic fermions, with analogous factors for bosons. Hypercharge neutrality and fast interactions impose linear constraints among the $\mu_i$. The Abelian anomaly for species $i$ contributes
\begin{equation}
\left.\frac{dn_i}{dt}\right|_Y
=-C_i^Y\frac{g'^2}{4\pi^2}\langle\bm E_Y\cdot\bm B_Y\rangle,
\end{equation}
while weak sphalerons impose a rate proportional to a linear combination of left-handed quark and lepton chemical potentials. Eliminating the fast variables gives a reduced system of the form
\begin{equation}
\frac{d\bm Y}{dx}=\bm A(x)\bm Y+\bm b(x)\frac{d h_Y}{dx},
\label{eq:reducedcharges}
\end{equation}
where $x$ may be $T$, $\ln a$ or a normalized inverse temperature. The baryon asymmetry is a particular linear combination of $\bm Y$.

The importance of \cref{eq:reducedcharges} is methodological. A single formula $Y_B\propto B_0^2\lambda_0$ is not fundamental; it is the endpoint of a set of assumptions about the spectrum, plasma conductivity, electroweak mixing, spectator equilibrium and magnetic evolution.

\subsection{A phenomenological present-day scaling}

It is common to translate an early-Universe helicity source into a relation between the observed baryon asymmetry and today's magnetic-field parameters. Such relations are useful for estimates, provided one remembers that they incorporate assumptions about spectral shape, helicity conservation, and electroweak conversion.
\begin{equation}
Y_B\sim C_{\rm EW}\, B_0^2\lambda_0,
\label{eq:phenoscaling}
\end{equation}
where $C_{\rm EW}$ contains the electroweak conversion efficiency and unit factors. Such relations are useful for overlaying BAU bands on the present-day $B$--$\lambda$ plane, but $C_{\rm EW}$ is not universal. It depends on the helicity fraction, spectral shape, thermal mixing history and sphaleron evolution \cite{KamadaLong2016b,KamadaYamamotoYang2023,HamadaMukaidaUchida2026}.

This point is especially important when a model produces its magnetic field at a first-order transition. A BAU ``prediction'' obtained by inserting $B_0$ and $\lambda_0$ into a one-line estimator should be understood as a consistency estimate unless the charge network and crossover are solved for that model.

\subsection{The 2026 gauge-invariant reassessment}

The electroweak crossover is the most delicate part of this calculation. Hamada, Mukaida, and Uchida revisited the problem using a gauge-invariant description of magnetic flux in the hot Standard Model \cite{HamadaMukaidaUchida2026}. Their analysis identified three effects that can be missed by the conventional phenomenological conversion:
\begin{enumerate}
\item the magnetic mixing angle can be defined gauge invariantly rather than through a naive background-field projection;
\item approximate conservation of unconfined magnetic flux changes the relation between hypercharge and electromagnetic helicity through the crossover;
\item a nonperturbative process in the presence of magnetic flux, absent from earlier treatments, may affect the final baryon number.
\end{enumerate}
Their first two effects modify previous baryogenesis estimates by order-one factors; the third introduces a less-controlled uncertainty. The appropriate current status is therefore that helicity-decay baryogenesis is a well-motivated Standard Model mechanism whose \emph{precision efficiency} remains under active investigation.

The Kamada--Long treatment remains an important quantitative benchmark for baryogenesis from hypermagnetic helicity, but its conversion efficiency should not be regarded as exact. The gauge-invariant flux analysis of \cite{HamadaMukaidaUchida2026} changes some estimates and identifies additional uncertainty. Numerical predictions should therefore make their treatment of the electroweak crossover explicit and, where feasible, assess their sensitivity to alternative conversion prescriptions.

\subsection{Baryon isocurvature and BBN}

Reproducing the observed average $Y_B$ is not enough. A stochastic magnetic field also produces spatial fluctuations in the baryon asymmetry. If these fluctuations are too large when light elements form, they conflict with Big Bang nucleosynthesis (BBN) observations \cite{GiovanniniShaposhnikov1998BBN,KamadaUchidaYokoyama2021}.

The 2021 analysis of Kamada, Uchida, and Yokoyama found a strong tension between generating the observed mean baryon asymmetry and avoiding excessive baryon fluctuations in a class of pre-electroweak magnetic-field scenarios \cite{KamadaUchidaYokoyama2021}. The underlying physical concern is clear: a field powerful enough to explain an intergalactic signal may also produce large spatial variations in the anomaly source.

However, this result should not be treated as a model-independent exclusion. Hamada, Mukaida, and Uchida revisited both the mean baryon abundance and the isocurvature argument using their updated description of the electroweak crossover \cite{HamadaMukaidaUchida2026Constraints}. They identify parameter space in which maximally helical primordial fields can account for both the present intergalactic magnetic field and the baryon asymmetry. They also find a possible nonhelical-field window, but only with an extremely precise compensation during the crossover, at roughly the $10^{-9}$--$10^{-10}$ level. The latter possibility is therefore highly sensitive to the electroweak dynamics and should not be interpreted as a generic viable mechanism.

The status of the constraint is consequently conditional: the spatial fluctuation problem is real, but whether a particular model is excluded depends on its spectrum, helicity, electroweak matching, and subsequent magnetic evolution. A careful treatment should calculate both the mean $Y_B$ and the baryon-isocurvature spectrum rather than importing a universal exclusion curve.

\subsection{Can baryogenesis generate helicity instead?}

The anomaly works in both directions. A chiral fermion asymmetry can be transferred into magnetic helicity, so some magnetogenesis scenarios begin with particle asymmetry rather than with an established magnetic field. The final state depends on the competing chirality-flipping and magnetic growth rates \cite{Vachaspati2001}.

The quantitative late-time field strength depends strongly on the inverse transfer, but the conceptual statement survives: magnetic helicity can be a fossil of CP violation and baryon-number generation rather than their cause. This is one reason helicity observables---whether in gamma-ray morphology, the CMB or circularly polarized gravitational waves---are disproportionately valuable.

\subsection{Baryogenesis from an initially nonhelical field?}

A field with zero initial net helicity is not automatically irrelevant. Plasma evolution and chiral dynamics may generate helicity later, but the resulting baryon asymmetry depends on whether a net handedness develops and survives the electroweak crossover.

Thus ``nonhelical initial condition'' is not equivalent to ``no magnetic effect on baryogenesis''. It is a statement about one initial correlator, not a conserved property of the full anomalous plasma.

\subsection{Thermal leptogenesis in a helical hypermagnetic background}

Hypermagnetic fields can also affect mechanisms that generated lepton number at much higher temperatures. A preexisting helicity or chiral asymmetry changes the chemical-potential network and can either enhance or suppress the baryon number obtained after sphaleron processing \cite{Safari2024}.

This is a useful proof of principle for a broader research program: any baryogenesis mechanism operating above the electroweak crossover should be tested for sensitivity to a primordial hypermagnetic background if such a background is independently motivated.

\

\section{Magnetic feedback on electroweak dynamics and sphalerons}
\label{sec:feedback}

The interaction between magnetic fields and electroweak physics is not one-directional. A transition may create a field, but a preexisting field can modify the Higgs background, the transition dynamics, and the energy barrier for sphaleron processes.

\subsection{Why a background field can change the transition}

So far magnetic fields have mainly appeared as products of electroweak dynamics. The reverse effect is equally important: a sufficiently strong background field changes the free-energy balance between phases, the structure of the Higgs condensate, and potentially the sphaleron rate.

Elmfors, Enqvist and Kainulainen showed that a primordial hypermagnetic field can strengthen a first-order electroweak transition in parameter regions where such a transition exists \cite{ElmforsEnqvistKainulainen1998}. Lattice studies subsequently examined the effect nonperturbatively and found that although hypermagnetic fields can strengthen the transition, they do not rescue Standard Model electroweak baryogenesis at the physical Higgs mass \cite{KajantieMagnetic1998}.

The physics can be represented schematically by
\begin{equation}
\Delta F(B_Y,T)=\Delta V_{\rm eff}(T)+\Delta F_{\rm magnetic}(B_Y,T),
\end{equation}
where the second term depends on how the applied flux partitions between the unbroken and broken gauge sectors. In a BSM model this magnetic correction should be considered together with the finite-temperature scalar potential rather than added after the transition calculation.

\subsection{The sphaleron carries a magnetic dipole moment}

In the electroweak theory the sphaleron is not magnetically inert. It has a dipole interaction with an external field, which changes the energy barrier for anomalous $B+L$ violation. A simple way to see the leading effect is to write \cite{Comelli1999}
\begin{equation}
E_{\rm sph}(B)\simeq E_{\rm sph}(0)-\bm\mu_{\rm sph}\cdot\bm B,
\label{eq:sphdipole}
\end{equation}
so the field lowers the sphaleron barrier. Since the broken-phase sphaleron rate scales approximately as
\begin{equation}
\Gamma_{\rm sph}^{\rm br}\propto e^{-E_{\rm sph}(T,B)/T},
\end{equation}
a modest reduction in $E_{\rm sph}/T$ can significantly enhance washout.

The resulting competition is conceptually simple:
\begin{equation}
\boxed{
B_Y:\quad \text{stronger transition}
\qquad \text{but} \qquad
\text{lower sphaleron barrier}.}
\end{equation}
A calculation that includes only the first effect can therefore reach the wrong conclusion about BAU preservation.

At sufficiently large fields there is also a distinct nonperturbative gauge-field response. Ambj\o rn and Olesen showed that charged $W$ bosons can develop a condensate when the magnetic field approaches the scale $eB\sim m_W^2$; the resulting state is not captured by merely shifting the Higgs effective potential \cite{AmbjornOlesen1989,AmbjornOlesen1990}. This strong-field instability should be distinguished from small-field dipole corrections to the sphaleron barrier.

\subsection{Modern first-principles sphalerons in magnetic fields}

Perturbative arguments suggest that magnetic fields can change the sphaleron energy. Nonperturbative calculations are needed to quantify the effect in the thermal Standard Model. Modern lattice studies have revisited this problem with externally imposed magnetic backgrounds \cite{AnnalaRummukainen2023}.

For BSM first-order transitions, one should in principle recompute the sphaleron solution in the modified scalar/gauge sector and in the relevant magnetic background. Using the Standard Model dipole correction as a universal multiplicative factor is only an approximation.

\subsection{Real-time first-order transition dynamics in a magnetic background}

Magnetic backgrounds can also influence the transition while bubbles form and grow. Recent real-time simulations provide evidence that the field can slow a first-order transition and produce nontrivial Higgs configurations, demonstrating that the backreaction need not be a small correction \cite{DiBianCai2026}.

The appearance of Higgs vortices is especially interesting because it links the old electroweak-string literature to modern real-time transition simulations. A sufficiently strong background flux is not simply a perturbation of a homogeneous Higgs condensate; it can reorganize the spatial pattern of symmetry breaking.

\subsection{Magnetic fields in conventional electroweak baryogenesis}

The same feedback matters in conventional bubble-wall baryogenesis. A stronger sphaleron rate can help convert a chiral source into baryon number in the symmetric phase, but can also increase washout inside or behind bubbles. The net effect must be computed from the full space-time evolution \cite{MorrisseyRamseyMusolf2012,VandeVis2025}.

A primordial magnetic field can influence every stage:
\begin{itemize}
\item charged-particle dispersion relations and transport coefficients are modified;
\item Landau quantization becomes relevant for sufficiently strong fields;
\item the effective potential and nucleation history shift;
\item magnetically induced fluid structure can alter the wall environment;
\item the broken-phase sphaleron rate is enhanced through the sphaleron magnetic moment;
\item if the field is helical, an independent anomaly source contributes to the charge network.
\end{itemize}

These effects are rarely included simultaneously. Therefore, ``EWBG in a primordial magnetic field'' is a distinct problem from either conventional EWBG or helicity-decay baryogenesis alone.

\subsection{When is the magnetic field dynamically strong?}

A magnetic field can be observationally interesting without significantly affecting the electroweak plasma. To decide when its feedback is important, compare its energy density, pressure, and coupling to electroweak configurations with the corresponding thermal scales.
\begin{equation}
\epsilon_B=\frac{B^2/2}{\rho_{\rm rad}}.
\end{equation}
Even $\epsilon_B\ll1$ can be important for sphaleron physics because the barrier is exponentially sensitive to the field. Conversely, appreciable modification of bulk hydrodynamics generally requires magnetic pressure or tension to compete with the fluid kinetic or thermal pressure.

For a coherent background field, the stress is anisotropic,
\begin{equation}
T^{ij}_B=\frac{B^2}{2}\delta^{ij}-B^iB^j,
\end{equation}
so a strong field can make bubble growth direction dependent. For a stochastic field with coherence length well below the bubble size, an isotropically averaged magnetic pressure may be a better approximation. This scale ordering should be made explicit in simulations and effective descriptions.

\subsection{Feedback loops and potential nonlinearities}

The possible feedback is two-way: the transition generates magnetic structure, while the magnetic field changes the transition and the sphaleron rate. This coupling makes fully self-consistent simulations especially important in models with large magnetic energy fractions.
\begin{equation}
 B_Y\rightarrow\Gamma_{\rm sph}\rightarrow\mu_i
 \rightarrow\xi_Y\rightarrow B_Y.
\end{equation}
At the same time, the field changes the phase transition, which changes bubble flows and therefore the turbulence that evolves the field. This nested feedback is one reason separate-module calculations can be qualitatively informative yet quantitatively fragile.

A future real-time treatment would ideally evolve the Higgs/gauge sector and anomalous transport together. That is computationally demanding because the electroweak gauge dynamics is non-Abelian and quantum/thermal on microscopic scales, while the MHD cascade extends to much larger scales. Multiscale matching is therefore likely to be more realistic than a single brute-force simulation.

A primordial magnetic field need not be a passive background during electroweak symmetry breaking. It can modify the transition temperature or strength, alter the sphaleron barrier, produce spatially varying Higgs configurations, and contribute an independent anomaly source. A claim that magnetic fields assist electroweak baryogenesis must therefore account for their competing tendency to enhance sphaleron-induced washout.

\section{Pre-electroweak hypermagnetogenesis feeding electroweak baryogenesis}
\label{sec:preEW}

Helical fields relevant for electroweak baryogenesis need not originate near $100\,\mathrm{GeV}$. Fields produced during inflation or at higher temperatures may survive and influence the electroweak crossover. We focus here on the high-scale mechanisms whose later evolution feeds directly into electroweak physics.

\subsection{Why inflationary magnetogenesis is relevant here}

Not all fields that influence electroweak baryogenesis are generated at the electroweak scale. Inflation or earlier high-temperature processes can produce hypermagnetic fields whose helicity survives until the electroweak crossover. The baryon asymmetry can then be generated long after the magnetic field itself was created.
\begin{equation}
-\frac14 I^2(\phi)Y_{\mu\nu}Y^{\mu\nu}
\qquad\text{or}\qquad
-\frac{\phi}{4f}Y_{\mu\nu}\widetilde Y^{\mu\nu}.
\end{equation}
The first class can generate parity-even fields; the second preferentially amplifies one gauge-field helicity and is therefore particularly relevant to baryogenesis \cite{TurnerWidrow1988,Ratra1992,AnberSorbo2010,DurrerNeronov2013}.

If the inflaton or an axion-like field couples to hypercharge rather than to an unrelated hidden $U(1)$, the generated field participates directly in Standard Model anomalies. The same interaction can also produce chiral fermions through the anomaly and through Schwinger processes. Thus inflationary magnetogenesis and electroweak baryogenesis are not separable modules.

\subsection{Pseudoscalar inflation and baryon overproduction}

Pseudoscalar couplings to gauge fields can preferentially amplify one helicity during inflation. This is a natural route to helical primordial fields, but it can also produce too much baryon number after the electroweak crossover. The amount of amplification must therefore be confronted with the baryon abundance \cite{AnberSabancilar2015}.

This illustrates a general constraint on helical hypermagnetogenesis. Once the field is connected to $U(1)_Y$, the BAU becomes an unavoidable observable. A successful magnetogenesis model must satisfy
\begin{equation}
Y_B^{\rm pred}\simeq Y_B^{\rm obs},
\end{equation}
not merely a magnetic-field bound.

\subsection{Gauge-field and fermion production during axion inflation}

Axion-driven gauge production cannot always be treated independently of charged fermions. Schwinger pair creation and chiral currents can feed back on the growing gauge fields and modify their amplitude and helicity \cite{KobayashiAfshordi2014,DomckeMukaida2018,Sobol2018,Sobol2019,EckardsteinSchmitzSobol2025}.

A model-independent treatment of axion hypermagnetogenesis by Domcke, Ema and Mukaida and related work maps the competing requirements from backreaction, Schwinger production, CPI evolution and baryogenesis \cite{Domcke2022Hypermagnetogenesis}. The lesson is that a simple tachyonic mode equation during inflation is only the first step; the post-inflationary plasma can qualitatively reshape the result.

\subsection{Baryogenesis after pseudoscalar inflation}

Once an inflationary field has survived reheating, its subsequent evolution can generate baryon number through the hypercharge anomaly. The result depends on the reheating temperature, conductivity, plasma reactions, and the helicity that reaches the electroweak epoch \cite{DomckeHarlingMorganteMukaida2019}.

The final BAU is produced mainly around electroweak symmetry breaking even though the origin of the magnetic helicity is inflationary. This provides a useful distinction:
\begin{equation}
\text{magnetogenesis epoch}\neq\text{baryogenesis conversion epoch}.
\end{equation}
A review restricted only to fields physically generated at $T\sim100\,\mathrm{GeV}$ would therefore miss an important class of electroweak magnetic baryogenesis mechanisms.

\subsection{High-scale baryogenesis stored in helicity}

It is possible for a high-scale process to create an asymmetry whose information is subsequently carried by magnetic helicity. Later changes in the field then release part of that asymmetry into the fermion sector. This offers a different time ordering from conventional electroweak baryogenesis \cite{Kamada2018}.

The schematic chain is
\begin{equation}
\text{high-scale chiral asymmetry}
\rightarrow \text{CPI}
\rightarrow h_Y
\rightarrow \text{EW crossover}
\rightarrow Y_B.
\end{equation}
This can resurrect classes of GUT-baryogenesis models that would otherwise be dismissed because they generate $B-L=0$. It is a vivid example of gauge-field topology acting as a memory of high-scale particle physics.

\subsection{Leptogenesis and primordial magnetic fields}

Primordial magnetism can coexist with thermal leptogenesis, but their contributions are not generally independent. Both affect the charge network, and the anomaly can move asymmetry between helical fields and chiral fermions \cite{LongSabancilarVachaspati2014,Safari2024}.

These studies motivate three logically distinct categories:
\begin{enumerate}
\item \textbf{field $\to$ BAU}: a pre-existing helical field sources baryon/lepton charge;
\item \textbf{BAU/chirality $\to$ field}: a particle asymmetry seeds or amplifies helicity;
\item \textbf{co-generation}: anomalous transport and plasma dynamics generate both simultaneously.
\end{enumerate}
The third category is the most self-contained but also the hardest computationally.

\subsection{Early matter domination and nonstandard thermal histories}

Nonstandard expansion before the electroweak epoch changes more than the Hubble rate. It alters the relation between temperature and time, the dilution of asymmetries, and the redshifting and damping of primordial magnetic fields.

Magnetogenesis from baryon or chiral asymmetries has been considered in early matter-dominated cosmologies \cite{ElahiMehrabpour2021}. More generally, any relation inferred assuming radiation domination,
\begin{equation}
H(T)=1.66\sqrt{g_*}\frac{T^2}{M_{\rm Pl}},
\end{equation}
should be revisited if the model itself contains a long-lived scalar, modulus or vacuum-dominated supercooling stage.

A particularly instructive combined calculation is the pseudoscalar-inflation scenario of Jim\'enez, Kamada, Schmitz and Xu \cite{JimenezKamadaSchmitzXu2017}. The produced hypermagnetic field is helical and can later source baryon number, while gauge-field production also sources gravitational waves. Their model illustrates the importance of carrying the same field source through two observables: successful baryogenesis and a detectable gravitational-wave background need not occupy the same parameter region or frequency band. This is an inflationary counterexample to the idea that a shared magnetic source automatically implies an observable electroweak-scale GW signal.

\subsection{What survives from a high-scale field?}

The relevant question is not the magnetic field immediately after inflation, but the energy spectrum and helicity that remain when the electroweak plasma forms. This requires following reheating, conductivity, chiral dynamics, and subsequent MHD evolution.
\begin{enumerate}
\item Ohmic diffusion erases sufficiently short wavelengths;
\item CPI can transfer helicity between fermions and gauge fields;
\item chirality-flipping rates change as Yukawa interactions enter equilibrium;
\item turbulence can move the integral scale and reduce magnetic energy;
\item entropy production can dilute both field energy fractions and charge asymmetries;
\item the electroweak crossover converts the relevant unconfined flux into ordinary electromagnetism.
\end{enumerate}
A present-day field cannot therefore be propagated backward by $B\propto a^{-2}$ alone unless the field remained dynamically passive throughout its history.

Electroweak magnetic baryogenesis is characterized by the epoch at which the magnetic field transfers helicity into fermion number, not necessarily by the epoch at which that field was generated. Inflationary, grand-unified, and other high-scale mechanisms can supply hypermagnetic helicity to the electroweak plasma, provided that the field survives its intervening thermal history and respects backreaction and plasma constraints.

\section{From the electroweak epoch to today: MHD evolution}
\label{sec:mhd}

Connecting an electroweak magnetic field with present-day observations requires a long period of cosmological evolution. Expansion redshifts the field, while turbulence, viscosity, and reconnection redistribute or dissipate its energy. The evolution depends strongly on magnetic helicity.

\subsection{Why the evolution problem is central}

A field produced at $T\sim100\,\mathrm{GeV}$ is not observed directly. Its strength and coherence length change as the Universe expands and as the plasma moves, reconnects, and dissipates magnetic structure. The connection between an electroweak prediction and present-day observations therefore depends critically on MHD evolution.
\begin{equation}
\{B_*,\lambda_*,E_B(k),H_B(k)\}\rightarrow\{B_0,\lambda_0,E_{B,0}(k),H_{B,0}(k)\}
\end{equation}
is controlled by expansion, turbulence, viscosity, drag, reconnection and helicity conservation. There is no single universal power law valid from $T\sim100\,\mathrm{GeV}$ to the present.

Early work already showed that decaying MHD turbulence can grow the magnetic correlation length faster than pure expansion \cite{Son1999MHD}. The foundational cosmological treatments by Banerjee and Jedamzik and the later numerical literature provide the standard starting point \cite{BrandenburgEnqvistOlesen1996,BanerjeeJedamzik2004,Kahniashvili2013,Brandenburg2015,Subramanian2016}. Electroweak-scale hydromagnetic turbulence has also been evolved directly in expanding-universe simulations \cite{BrandenburgEtAl2017}. Recent reconnection-based work suggests that some widely used late-time scalings should be reconsidered \cite{HoskingSchekochihin2023}.

\subsection{Ideal MHD and helicity conservation}

In a highly conducting plasma, magnetic flux is approximately frozen into the fluid. Helicity is even more resistant to dissipation under suitable boundary conditions. This approximate conservation is what allows helical fields to transfer magnetic energy toward larger scales.
\begin{equation}
\bm E\simeq-\bm v\times\bm B,
\end{equation}
so $\bm E\cdot\bm B\simeq0$. Using
\begin{equation}
\frac{d\helicity_B}{dt}=-2\int d^3x\,\bm E\cdot\bm B,
\end{equation}
magnetic helicity is approximately conserved. With finite diffusivity,
\begin{equation}
\frac{d\helicity_B}{dt}=-2\eta\int d^3x\,\bm B\cdot(\nabla\times\bm B),
\end{equation}
which is suppressed relative to the magnetic-energy dissipation rate for large-scale helical structures.

This selective conservation is the reason helicity drives inverse transfer. If
\begin{equation}
h_B\sim2f_h\rho_B\xi_B\simeq\mathrm{const},
\end{equation}
then a decrease in magnetic energy requires a compensating increase in the integral scale.

\subsection{Self-similar decay and inverse cascade}

A convenient description of turbulent magnetic evolution assumes a self-similar spectrum. Magnetic energy then decreases while the characteristic correlation scale grows. For a maximally helical field, approximate helicity conservation fixes a simple relation between these quantities.
\begin{equation}
\rho_{B,c}\propto \tau^{-p},\qquad \xi_{B,c}\propto\tau^q.
\end{equation}
For fully helical turbulence, helicity conservation gives approximately $p=q$ because $\rho_B\xi_B$ is fixed. Depending on assumptions about the decay time and turbulence phenomenology, commonly used radiation-era scalings give $q$ near $2/3$ for helical decay in conformal variables. The precise exponent should not be elevated to a universal law; analytic and numerical treatments differ in detail \cite{Campanelli2007,BanerjeeJedamzik2004,Kahniashvili2013,Subramanian2016}.

A useful intuitive estimate follows from an eddy-turnover time
\begin{equation}
\tau_{\rm eddy}\sim \frac{\xi_B}{v_A},
\qquad
v_A\sim\frac{B}{\sqrt{\rho+p}}.
\end{equation}
If the field evolves on $\tau_{\rm eddy}\sim t$ while conserving helicity, one obtains a growing coherence length and decaying field amplitude. This is the classical inverse-cascade picture used in many cosmological projections.

\subsection{Partial helicity and the transition to a helical regime}

Many mechanisms produce a field that is only partly helical. In that case the early evolution can resemble nonhelical turbulence, but selective decay of magnetic energy may gradually increase the helicity fraction until the field becomes effectively maximally helical.
\begin{equation}
\text{nonhelical-like decay}\quad\rightarrow\quad f_h\to1\quad\rightarrow\quad\text{helical inverse transfer}.
\end{equation}
The time at which the second stage begins depends on the initial helicity fraction and spectrum. This is important for phase-transition models that predict a small but nonzero $f_{h,*}$: setting $f_h=0$ or $1$ from the start can bracket the outcome but does not reproduce the intermediate trajectory.

\subsection{Nonhelical inverse transfer}

Even a field with negligible total helicity can transfer power toward larger scales under some conditions. Numerical studies have found such behavior, but the scaling depends on the initial spectrum and plasma regime, so it should not be assumed to be universal \cite{Brandenburg2015}.

Later studies examined how the effect depends on Reynolds number, magnetic Prandtl number, and initial conditions \cite{ReppinBanerjee2017}. The conservative statement is that nonhelical inverse transfer is a real phenomenon in important MHD regimes, but it should not be treated as an automatic substitute for the robust helicity-driven inverse cascade in every cosmological plasma.

\subsection{Viscous and drag-dominated epochs}

The early-Universe plasma does not remain in one turbulence regime. Changes in particle mean free paths, viscosity, and free-streaming drag can suppress fluid motions and alter magnetic decay. A simple extrapolation of one turbulent scaling law to the present is therefore unreliable \cite{BanerjeeJedamzik2004}.

This matters because the coherence scale can effectively freeze during a viscous interval while magnetic energy decays differently, then resume nonlinear growth when turbulent motions return. A single power law extended over twenty orders of magnitude in scale factor is therefore best viewed as an effective fit, not a literal description of the entire history.

\subsection{Reconnection-controlled magnetic decay}

An important recent development is the realization that reconnection can determine the magnetic energy-decay rate in regimes where older estimates assumed eddy-turnover control. This changes the predicted relation between field strength and coherence length and is especially relevant when extrapolating to late cosmological times \cite{HoskingSchekochihin2023}.

The essential conceptual change is the timescale. Magnetic topology cannot relax faster than reconnection permits, so
\begin{equation}
\tau_{\rm decay}\sim \tau_{\rm rec}
\end{equation}
can be longer than the naive $\xi_B/v_A$. This slows decay and changes the trajectory in the $B$--$\lambda$ plane.

Uchida, Kamada and Tashiro adopted reconnection-driven decay when revisiting CMB spectral-distortion constraints \cite{UchidaKamadaTashiro2025}. Their results demonstrate that the observational limits themselves depend on the nonlinear decay law. A model should therefore propagate MHD-theory uncertainty into observational parameter inference.

\subsection{A useful $B$--$\lambda$ representation}

Much of the observational comparison can be shown in a plane of present-day magnetic strength and coherence length. Such plots are intuitive, but different MHD prescriptions can move the same initial electroweak field to substantially different locations in this plane.
\begin{equation}
B_c^2\lambda_c\simeq\mathrm{const},
\end{equation}
so a trajectory in log--log space has approximately
\begin{equation}
B_c\propto\lambda_c^{-1/2}.
\end{equation}
This is why many PMF plots contain diagonal ``helicity conservation'' lines. Other decay laws generate different slopes. When translating an electroweak field to today, the slope and endpoint both depend on which invariant and timescale control the decay.

It is often useful to normalize the generation scale to the Hubble radius,
\begin{equation}
\ell_*\equiv H_*\lambda_*,
\end{equation}
and the energy to radiation,
\begin{equation}
\epsilon_{B,*}=\frac{B_*^2}{2\rho_{\rm rad,*}}.
\end{equation}
These dimensionless quantities can be evolved and compared across transition temperatures without repeatedly converting huge field strengths in Gauss.

\subsection{Spectral memory and causal tails}

The coherence scale does not contain all the information about a primordial field. Its spectral slope and infrared tail matter for smoothing to megaparsec scales and for the anisotropic stresses that source CMB fluctuations and gravitational waves.

Therefore two models with the same $(B_0,\lambda_0)$ can still produce different observational signatures if their spectra differ. This is particularly relevant when comparing a narrow-band transition-generated field with a broad inflationary spectrum.

\subsection{Magnetic helicity through recombination and structure formation}

After recombination, the plasma changes from a tightly coupled radiation--baryon fluid to an ionized or partly neutral medium with emerging nonlinear structure. The assumptions behind early-Universe MHD must therefore be reconsidered when following a primordial field to galaxies or cosmic voids \cite{Subramanian2016,PlanckPMF2015}.

For the present review, the main point is that the electroweak-to-present evolution is not complete at recombination: the same field can continue to alter the thermal history and matter power spectrum. Recent simulations of PMF-induced small-scale structure find nonlinear behavior that differs from earlier analytic extrapolations, reinforcing the need for updated late-time transfer functions when connecting electroweak magnetogenesis to future 21-cm probes.

\subsection{A worked evolution from the transition to the present}
\label{sec:new_worked_MHD_evolution}

We now apply the spectrum in Eq.~\eqref{eq:new_ags_broken_law} to a specific, deliberately simplified evolution model. It is useful to work with the comoving field $B_c=a^2B$ and comoving integral scale $\lambda_c=\lambda/a$. If the magnetic spectrum decays self-similarly, its characteristic amplitude and coherence scale may be parametrized during the radiation era as
\begin{equation}
 B_c\propto a^{-p_B/2},\qquad \lambda_c\propto a^{p_\lambda},
 \label{eq:new_comoving_decay_laws}
\end{equation}
where we have used $a\propto\tau$ in radiation domination and $\tau$ is conformal time. The corresponding physical quantities obey $B\propto a^{-2-p_B/2}$ and $\lambda\propto a^{1+p_\lambda}$. The classical phenomenological choices adopted in Ref.~\cite{ArteagaGhoshalStrumia2025}, following the earlier cosmological MHD literature \cite{BanerjeeJedamzik2004,Kahniashvili2013,BrandenburgEtAl2017}, are
\begin{equation}
 \begin{array}{c|cc}
 &p_B&p_\lambda\\\hline
 \text{maximally helical}&2/3&2/3\\
 \text{nonhelical (conventional)}&1&1/2
 \end{array}.
 \label{eq:new_decay_exponent_table}
\end{equation}
The helical choice satisfies $B_c^2\lambda_c\simeq\mathrm{constant}$, as required by approximate conservation of comoving magnetic helicity. The nonhelical choice is an approximate decay law, not a helicity-conservation theorem. A partially helical field may initially follow a nonhelical-like evolution before its net fractional helicity becomes dynamically important; the crossover time depends on the initial fraction, spectrum and intervening viscosity.

With the simplifying assumption that nonlinear magnetic processing effectively ceases around recombination, subsequent evolution is taken to follow $B\propto a^{-2}$ and $\lambda\propto a$. These assumptions give the following useful transfer formulas between the generation epoch $*$ and the present epoch $0$:
\begin{equation}
 B_{0,\mathrm{char}}=B_{*,\mathrm{char}}\left(\frac{a_*}{a_0}\right)^2
 \left(\frac{a_*}{a_{\rm rec}}\right)^{p_B/2},\qquad
 \lambda_0=\lambda_*\left(\frac{a_0}{a_*}\right)
 \left(\frac{a_{\rm rec}}{a_*}\right)^{p_\lambda}.
 \label{eq:new_MHD_transfer}
\end{equation}
Here $a_{\rm rec}$ denotes the scale factor around recombination; in more realistic calculations, the onset and end of viscous, drag-dominated and turbulent regimes have to be tracked separately. Adiabatic radiation-era redshifting gives $a_*/a_{\rm rec}\simeq (T_{\rm rec}/T_*)[g_{*s}(T_{\rm rec})/g_{*s}(T_*)]^{1/3}$ and $a_0/a_{\rm rec}\simeq1100$. For a fixed broken-power-law shape, the present-day field per logarithmic wavelength interval is proportional to
\begin{equation}
 B_0(\lambda)=B_{0,\rm peak}\begin{cases}
 (\lambda/\lambda_0)^{1/3},&\lambda<\lambda_0,\\
 (\lambda/\lambda_0)^{-5/2},&\lambda>\lambda_0.
 \end{cases}
 \label{eq:new_present_B_spectral_shape}
\end{equation}
The field $B_{0,\rm peak}$ is a characteristic spectral amplitude; it must not be confused with the field smoothed over $1\,\mathrm{Mpc}$ or with an rms field integrated over all scales. The amplitude convention can absorb the $\sqrt{10/17}$ factor from Eq.~\eqref{eq:new_ags_broken_law}, so source normalizations must be compared consistently.

For its specific magnetic-efficiency and reheating assumptions, Ref.~\cite{ArteagaGhoshalStrumia2025} gives the useful estimates
\begin{align}
 \lambda_{0,\rm hel}&\simeq\lambda_*H_*\,0.06\,\mathrm{Mpc}
 \left(\frac{100\,\mathrm{GeV}}{T_*}\right)^{1/3},&
 B_{0,\rm hel}^{\rm peak}&\simeq8\times10^{-12}\,\mathrm{G}
 \left(\frac{100\,\mathrm{GeV}}{T_*}\right)^{7/3},\label{eq:new_AGS_helical_endpoints}\\
 \lambda_{0,\rm nonhel}&\simeq\lambda_*H_*\,0.6\,\mathrm{kpc}
 \left(\frac{100\,\mathrm{GeV}}{T_*}\right)^{1/2},&
 B_{0,\rm nonhel}^{\rm peak}&\simeq8\times10^{-14}\,\mathrm{G}
 \left(\frac{100\,\mathrm{GeV}}{T_*}\right)^{5/2}.
 \label{eq:new_AGS_nonhelical_endpoints}
\end{align}
These estimates were obtained under the conventions and fiducial efficiencies of the cited paper, including $\epsilon\simeq0.1$ and an approximately collision-dominated supercooled transition. They are not universal fits to all first-order phase transitions and should not be applied to an arbitrary value of $\epsilon_B$ without restoring its normalization dependence.

For $T_*=1\,\mathrm{TeV}$ and $\lambda_*H_*=0.1$, they give $\lambda_{0,\rm hel}\simeq2.8\,\mathrm{kpc}$ and $B_{0,\rm hel}^{\rm peak}\simeq3.7\times10^{-14}\,\mathrm{G}$, compared with $\lambda_{0,\rm nonhel}\simeq19\,\mathrm{pc}$ and $B_{0,\rm nonhel}^{\rm peak}\simeq2.5\times10^{-16}\,\mathrm{G}$. The orders-of-magnitude difference reflects the assumed evolution law, not a difference in the underlying scalar potential. In particular, a field that is maximally helical from the beginning is a favorable limiting case; whether a microscopic model reaches it must be established independently.

\paragraph{A dynamical consistency check of the conventional extrapolation.}
The power laws above implicitly assume that an actively processed magnetic spectrum follows the same turbulent regime from reheating to recombination. This assumption is not automatic when the initial magnetic-energy fraction is very small. For a relativistic plasma, the Alfv\'en speed is approximately
\begin{equation}
 v_{A,*}^2\simeq\frac{2\rho_{B,*}}{\rho_*+p_*}\simeq\frac32\epsilon_{B,*},\qquad
 \epsilon_{B,*}\equiv\frac{\rho_{B,*}}{\rho_{\rm rad,*}},\qquad
 H_*\tau_{A,*}\sim\frac{H_*\lambda_*}{v_{A,*}}.
 \label{eq:new_alfven_test}
\end{equation}
For the specific prescription in Eq.~\eqref{eq:new_energy_partition_ags}, using $\Delta V_H=m_h^4/(16\lambda_H)$ with $m_h=125\,\mathrm{GeV}$, $\lambda_H=0.126$, $g_*=106.75$, $\epsilon=0.1$ and $T_*=1\,\mathrm{TeV}$, we obtain
\begin{equation}
 \epsilon_{B,*}\simeq 3.45\times10^{-7},\qquad v_{A,*}\simeq7.2\times10^{-4},\qquad
 H_*\tau_{A,*}\simeq 1.4\times10^2\quad (\lambda_* H_*=0.1).
 \label{eq:new_alfven_benchmark}
\end{equation}
Thus the magnetic Alfv\'en crossing time is much longer than a Hubble time \emph{at generation}. These numbers do not prove that nonlinear evolution never occurs: kinetic turbulence, later magnetic amplification, viscosity changes, and reheating dynamics may alter the state. They do show that continuously imposing an Alfv\'enic decay law from $a_*$ to $a_{\rm rec}$ is not derived by the benchmark's magnetic initial conditions alone.

A useful diagnostic of actively processed relic fields is the schematic ``largest processed eddy'' relation $B_{0,\rm eddy}\simeq C_{\rm eddy}(\lambda_0/\mathrm{Mpc})$ with $C_{\rm eddy}$ of order $10^{-8}\,\mathrm{G}$ in one commonly adopted convention; its normalization depends on the epoch and plasma history \cite{BanerjeeJedamzik2004,HoskingSchekochihin2021,HoskingSchekochihin2023}. The continuously extrapolated $1\,\mathrm{TeV}$ helical endpoint above lies a factor of roughly $750$ below the $C_{\rm eddy}=10^{-8}\,\mathrm{G}$ diagnostic. \emph{Being below this line is not an observational exclusion}; it instead means that the field would not have been actively Alfv\'enically processed up to that scale under the assumptions behind the line.

For illustration only, one may follow a conserved \emph{comoving} helicity trajectory $B_c^2\lambda_c=K$ until it intersects that diagnostic line. Starting from the published helical extrapolation, $K=(3.7\times10^{-14}\,\mathrm{G})^2(2.8\times10^{-3}\,\mathrm{Mpc})$, the algebraic intersection gives
\begin{equation}
 \lambda_{0,\rm int}=\left(\frac{K}{C_{\rm eddy}^2}\right)^{1/3}\!\simeq 3.4\times10^{-5}\,\mathrm{Mpc},\qquad
 B_{0,\rm int}=C_{\rm eddy}\lambda_{0,\rm int}\simeq3.4\times10^{-13}\,\mathrm{G}.
 \label{eq:new_intersection}
\end{equation}
Here $C_{\rm eddy}$ is understood in units of G/Mpc. This is a \emph{conditional helicity-preserving estimate}, not an independently solved MHD trajectory. In particular, nothing guarantees that the system becomes turbulent at precisely this intersection, or that the same turnover law applies during all viscous epochs. The distinct flux-freezing limit, with no turbulent processing, should likewise be retained as a reference scenario.

The reconnection-controlled theory of Hosking and Schekochihin \cite{HoskingSchekochihin2021,HoskingSchekochihin2023} introduces still another dynamical timescale and can change both helical and nonhelical decay. We therefore keep Eqs.~\eqref{eq:new_AGS_helical_endpoints}--\eqref{eq:new_AGS_nonhelical_endpoints} as a \emph{reproduction of a published phenomenological prescription}, but do not regard their endpoints as validated predictions for the specific weak initial fields. Figures~\ref{fig:new_observational_comparison}--\ref{fig:new_three_model_mass_map} distinguish those extrapolations from the conditional helicity-intersection estimate. A quantitatively reliable replacement requires evolving the velocity and magnetic spectra with the actual forcing, viscosity, drag and reconnection history.

\subsection{Recommended treatment of MHD uncertainty}

For phenomenological studies, the safest approach is to show how the prediction changes under distinct physically motivated MHD evolution prescriptions. A single redshift scaling or inverse-cascade law is not an adequate error estimate when the plasma may pass through different regimes.
\begin{enumerate}
\item a conventional turbulence-based evolution calibrated to the standard Banerjee--Jedamzik / numerical-MHD literature;
\item a reconnection-controlled evolution representative of the Hosking--Schekochihin theory \cite{HoskingSchekochihin2021,HoskingSchekochihin2023}, with explicit identification of the assumed reconnection regime.
\end{enumerate}
The spread should be treated as a theory systematic, not as a choice of plotting style. If the phenomenological conclusion survives both, it is considerably more robust.

For fields generated near the electroweak scale, uncertainty in the present-day coherence length may be dominated by nonlinear MHD evolution rather than by the generation mechanism. Helicity can facilitate the survival of power on large scales. Reconnection-controlled decay laws may also change previous conclusions for more general initial spectra; their applicability must be checked in the relevant plasma regime.

\section{Gravitational waves from phase transitions and primordial magnetism}
\label{sec:gw}

Phase transitions, turbulent flows, and magnetic stresses can all produce a stochastic gravitational-wave background. If these processes share a common electroweak origin, their signals may be correlated. A reliable comparison must nevertheless keep the different gravitational-wave sources distinct.

\subsection{Tensor perturbations sourced by anisotropic stress}

Magnetic fields and fluid motions gravitate. Their anisotropic stresses can excite tensor perturbations, so an electroweak process that generates turbulence or magnetic structure can also generate gravitational waves. In Fourier space, the tensor perturbation equation reads
\begin{equation}
 h_{ij}''+2\mathcal H h_{ij}'+k^2h_{ij}
 =16\pi G a^2\Pi^{TT}_{ij},
\label{eq:gweq}
\end{equation}
where primes denote conformal-time derivatives and $\Pi^{TT}_{ij}$ is the transverse-traceless anisotropic stress. For a magnetic field,
\begin{equation}
T_{ij}^{(B)}=B_iB_j-\frac12\delta_{ij}B^2
\end{equation}
up to sign/convention choices for the spatial stress, so stochastic PMFs provide a direct source of GWs. The source is quadratic in the field and therefore sensitive to unequal-time four-point correlators, usually reduced under Gaussian or turbulence closure assumptions.

Early quantitative studies established turbulence and magnetically generated fields as important cosmological GW sources \cite{KosowskyMackKahniashvili2002}. Caprini and Durrer developed influential treatments of gravitational radiation from causal magnetic fields and turbulence, including the strong integrated constraint that GW production can place on energetic primordial fields \cite{CapriniDurrer2002,CapriniDurrer2006}; the long-lived MHD source and its unequal-time correlations were subsequently treated in detail by Caprini, Durrer and Servant \cite{CapriniDurrerServant2009}.

\subsection{Standard first-order phase-transition sources}

A first-order phase transition can generate gravitational waves (GWs) through three main processes: collisions of expanding bubble walls, acoustic waves in the surrounding plasma, and magnetohydrodynamic (MHD) turbulence. These sources have different physical origins, characteristic time scales, and spectral shapes. Their relative contributions depend on the transition strength $\alpha$, inverse duration $\beta/H_*$, bubble-wall velocity $v_w$, and the fraction of released energy transferred to each source. The resulting stochastic GW background is commonly approximated by
\begin{equation}
\Omega_{\rm GW}(f)
\simeq
\Omega_{\phi}(f)
+\Omega_{\rm sw}(f)
+\Omega_{\rm turb}(f),
\label{eq:gw_three_sources}
\end{equation}
where the three terms denote the contributions from scalar-field gradients, sound waves, and turbulence, respectively. This decomposition is approximate, since acoustic and turbulent stresses may be correlated.

The GW energy-density spectrum observed today is defined as
\begin{equation}
\Omega_{\rm GW}(f)
\equiv
\frac{1}{\rho_{c,0}}
\frac{d\rho_{\rm GW,0}}{d\ln f},
\label{eq:gw_spectrum_definition}
\end{equation}
where $\rho_{c,0}$ is the present critical energy density. Assuming radiation domination and approximately adiabatic expansion after the transition, the characteristic frequency at production is redshifted to the present. A useful reference frequency is
\begin{equation}
h_*=
1.65\times10^{-5}\,{\rm Hz}
\left(\frac{T_*}{100\,{\rm GeV}}\right)
\left(\frac{g_*}{100}\right)^{1/6},
\label{eq:gw_redshift_frequency}
\end{equation}
where $T_*$ is the temperature relevant for GW production and $g_*$ is the number of relativistic degrees of freedom. The precise choice of $T_*$ is important for strongly supercooled transitions, where the reheating temperature can differ significantly from the nucleation and percolation temperatures.

\paragraph{Gravitational waves from bubble collisions.}

During a first-order transition, bubbles of the stable phase nucleate and expand into the surrounding false vacuum. The scalar field changes rapidly across the bubble walls, producing anisotropic stresses that source gravitational waves. When a substantial fraction of the released energy remains in the accelerating walls, their collisions can provide an important GW contribution.

In the commonly used envelope approximation, the overlapping regions of colliding bubbles are assumed to stop contributing to the source. The corresponding present-day GW spectrum is fitted by \cite{CapriniFigueroa2018,Caprini2020,Athron2024}
\begin{equation}
\begin{aligned}
h^2\Omega_{\phi}(f)
={}&1.67\times10^{-5}
\left(\frac{H_*}{\beta}\right)^2
\left(\frac{\kappa_{\phi}\alpha}{1+\alpha}\right)^2
\left(\frac{100}{g_*}\right)^{1/3}\\
&\times
\frac{0.11v_w^3}{0.42+v_w^2}
S_{\phi}(f),
\end{aligned}
\label{eq:gw_bubble_spectrum}
\end{equation}
where $\kappa_{\phi}$ is the fraction of released energy stored in the scalar-field gradient configuration. The spectral shape is
\begin{equation}
S_{\phi}(f)=
\frac{3.8(f/f_{\phi})^{2.8}}
{1+2.8(f/f_{\phi})^{3.8}}.
\label{eq:gw_bubble_shape}
\end{equation}
This function is normalized to unity at $f=f_{\phi}$. The peak frequency today is given by
\begin{equation}
\begin{aligned}
f_{\phi}={}&1.65\times10^{-5}\,{\rm Hz}
\left(\frac{\beta}{H_*}\right)
\left(\frac{T_*}{100\,{\rm GeV}}\right)
\left(\frac{g_*}{100}\right)^{1/6}\\
&\times
\frac{0.62}{1.8-0.1v_w+v_w^2}.
\end{aligned}
\label{eq:gw_bubble_peak}
\end{equation}
The spectrum rises approximately as $f^{2.8}$ below the peak and decreases as $f^{-1}$ above it. The infrared exponent is a fit to envelope-approximation simulations rather than an exact asymptotic prediction.

The amplitude depends quadratically on $H_*/\beta$, reflecting the relatively short duration of the collision source. In a thermal transition with significant plasma friction, most of the released energy is transferred to the plasma rather than retained in the scalar walls, so $\kappa_{\phi}$ can be small. Conversely, scalar-field collisions may dominate in approximately vacuum transitions or sufficiently strongly supercooled scenarios. The envelope approximation should therefore not be applied indiscriminately to every fast-moving bubble wall.

\paragraph{Gravitational waves from sound waves.}

The expansion of bubbles also transfers energy to the surrounding plasma, generating compression and rarefaction waves. These acoustic disturbances can persist after neighboring bubbles collide and continue to source gravitational waves. For many non-runaway thermal transitions, this is expected to be the dominant GW contribution \cite{Hindmarsh2014,Hindmarsh2015,Caprini2020}.

A widely used fit for the present-day acoustic spectrum is
\begin{equation}
\begin{aligned}
h^2\Omega_{\rm sw}(f)
={}&2.65\times10^{-6}
\left(\frac{H_*}{\beta}\right)
\left(\frac{\kappa_v\alpha}{1+\alpha}\right)^2\\
&\times
\left(\frac{100}{g_*}\right)^{1/3}
v_w\,\Upsilon_{\rm sw}\,
S_{\rm sw}(f),
\end{aligned}
\label{eq:gw_sound_spectrum}
\end{equation}
where $\kappa_v$ denotes the efficiency of converting the released energy into bulk plasma motion and $\Upsilon_{\rm sw}$ accounts approximately for the finite lifetime of the acoustic source. The spectral shape is
\begin{equation}
S_{\rm sw}(f)=
\left(\frac{f}{f_{\rm sw}}\right)^3
\left[
\frac{7}{4+3(f/f_{\rm sw})^2}
\right]^{7/2}.
\label{eq:gw_sound_shape}
\end{equation}
The peak is located at
\begin{equation}
f_{\rm sw}=
1.9\times10^{-5}\,{\rm Hz}
\frac{1}{v_w}
\left(\frac{\beta}{H_*}\right)
\left(\frac{T_*}{100\,{\rm GeV}}\right)
\left(\frac{g_*}{100}\right)^{1/6}.
\label{eq:gw_sound_peak}
\end{equation}
The spectral shape behaves as $f^3$ at frequencies well below the peak and approximately as $f^{-4}$ above it.

An important feature of the acoustic contribution is its dependence on the duration of the sound-wave period. The original long-lasting-source approximation effectively assumes that the fluid motion persists for a time comparable to or longer than the Hubble time. However, sufficiently strong fluid perturbations can form shocks and become turbulent much earlier. An estimate for the shock-formation time is
\begin{equation}
\tau_{\rm sh}\sim\frac{R_*}{\overline U_f},
\label{eq:gw_shock_time}
\end{equation}
where $R_*$ is the mean bubble separation and $\overline U_f$ is the root-mean-square fluid velocity. For a radiation-like plasma, the latter may be estimated as
\begin{equation}
\overline U_f^{\,2}
\simeq
\frac{3}{4}
\frac{\kappa_v\alpha}{1+\alpha}.
\label{eq:gw_rms_velocity}
\end{equation}
For approximately exponential bubble nucleation,
\begin{equation}
R_*\simeq
(8\pi)^{1/3}\frac{v_w}{\beta},
\label{eq:gw_bubble_separation}
\end{equation}
up to factors depending on the nucleation history.

A frequently used approximate suppression factor for an acoustic source of duration $\tau_{\rm sw}$ is
\begin{equation}
\Upsilon_{\rm sw}
=
1-\frac{1}{\sqrt{1+2H_*\tau_{\rm sw}}}.
\label{eq:gw_sound_lifetime}
\end{equation}
For $H_*\tau_{\rm sw}\ll1$, this reduces to
\begin{equation}
\Upsilon_{\rm sw}\simeq H_*\tau_{\rm sw},
\end{equation}
so short-lived acoustic motion produces a smaller GW amplitude. The estimate $\tau_{\rm sw}\sim\tau_{\rm sh}$ is commonly used when shocks terminate the approximately linear acoustic regime.

The simple spectral shape in Eq.~\eqref{eq:gw_sound_shape} does not capture every feature found in simulations. In particular, the sound-shell model identifies two characteristic scales: the mean bubble separation $R_*$ and the sound-shell thickness $\Delta R_*$. The latter can control the location of the spectral peak, while the former introduces another break in the spectrum \cite{HindmarshHuberRummukainenWeir2017,Hindmarsh2018SoundShell}. More recent calculations also show that cosmic expansion and the time evolution of the kinetic energy can modify the amplitude and spectral shape. Thus, Eqs.~\eqref{eq:gw_sound_spectrum}--\eqref{eq:gw_sound_lifetime} should be regarded as useful phenomenological approximations rather than universal precision formulas.

\paragraph{Gravitational waves from MHD turbulence.}

After bubble collisions, nonlinear fluid motions and shocks can generate vorticity and turbulence. In a conducting plasma, the turbulent velocity field is coupled to the magnetic field, producing additional anisotropic stresses. Both kinetic turbulence and magnetic stresses can therefore contribute to the GW background.

A commonly adopted phenomenological spectrum is \cite{CapriniDurrerServant2009,Caprini2020,RoperPol2020}
\begin{equation}
\begin{aligned}
h^2\Omega_{\rm turb}(f)
={}&3.35\times10^{-4}
\left(\frac{H_*}{\beta}\right)
\left(\frac{\kappa_{\rm turb}\alpha}{1+\alpha}\right)^{3/2}\\
&\times
\left(\frac{100}{g_*}\right)^{1/3}
v_w\,S_{\rm turb}(f),
\end{aligned}
\label{eq:gw_turb_spectrum}
\end{equation}
where $\kappa_{\rm turb}$ characterizes the fraction of released energy transferred into MHD turbulence. Its value is model dependent; phenomenological studies frequently adopt
\begin{equation}
\kappa_{\rm turb}
=\epsilon_{\rm turb}\kappa_v,
\qquad
\epsilon_{\rm turb}\sim0.05-0.1,
\label{eq:gw_turb_efficiency}
\end{equation}
although this is an assumption rather than a universal result.

The corresponding spectral function is
\begin{equation}
S_{\rm turb}(f)=
\frac{(f/f_{\rm turb})^3}
{\left(1+f/f_{\rm turb}\right)^{11/3}
\left(1+8\pi f/h_*\right)},
\label{eq:gw_turb_shape}
\end{equation}
with characteristic frequency
\begin{equation}
f_{\rm turb}=
2.7\times10^{-5}\,{\rm Hz}
\frac{1}{v_w}
\left(\frac{\beta}{H_*}\right)
\left(\frac{T_*}{100\,{\rm GeV}}\right)
\left(\frac{g_*}{100}\right)^{1/6}.
\label{eq:gw_turb_peak}
\end{equation}
The additional factor $1+8\pi f/h_*$ parametrizes the finite temporal coherence of the turbulent source in this template. The spectrum has a causal $f^3$ rise in the infrared and a high-frequency falloff determined by the assumed turbulent cascade and decorrelation model.

Strictly speaking, $f_{\rm turb}$ is a characteristic break frequency rather than the exact maximum of Eq.~\eqref{eq:gw_turb_shape}. The actual maximum depends on the ratio $f_{\rm turb}/h_*$. For $8\pi f_{\rm turb}/h_*\gg1$, it occurs near $f\simeq1.2f_{\rm turb}$.

Unlike the acoustic source, MHD turbulence is nonlinear and evolves through an energy cascade. Its GW spectrum depends on the initial kinetic and magnetic spectra, the stirring scale, helicity, and the unequal-time correlations of the anisotropic stresses. Numerical simulations have demonstrated that the GW production efficiency and spectral shape can differ appreciably from the simplified fit \cite{RoperPol2020,Kahniashvili2021}. The magnetic contribution is particularly important in the present review because a helical magnetic field can also generate circularly polarized GWs, as discussed below.

\paragraph{Comparison of the three sources.}

The three contributions have different dependences on the phase-transition parameters. In particular, the collision amplitude scales as $(H_*/\beta)^2$, while the usual long-lasting acoustic and turbulent templates scale approximately as $H_*/\beta$, before accounting for acoustic lifetime effects and changes in the turbulent source. The spectral peaks are controlled primarily by the characteristic bubble or fluid length scale, redshifted from the transition temperature to the present.

For illustration, taking
\begin{equation}
T_*=100\,{\rm GeV},\qquad
\frac{\beta}{H_*}=100,\qquad
v_w=0.8,\qquad g_*=100,
\end{equation}
the characteristic frequencies are
\begin{equation}
\begin{aligned}
f_{\phi}&\simeq4.3\times10^{-4}\,{\rm Hz},\\
f_{\rm sw}&\simeq2.4\times10^{-3}\,{\rm Hz},\\
f_{\rm turb}&\simeq3.4\times10^{-3}\,{\rm Hz}.
\end{aligned}
\label{eq:gw_example_frequencies}
\end{equation}
These scales lie in or near the millihertz frequency range targeted by LISA. Their amplitudes, however, cannot be determined from the transition temperature and duration alone; one must also calculate the relevant energy-conversion efficiencies and source lifetimes.

Finally, the expressions above should not be extrapolated without qualification to strongly supercooled or vacuum-dominated transitions. In such scenarios, the nucleation history may not be characterized by a single $\beta$, the bubble separation may differ from the usual estimate, and reheating can alter the subsequent redshift. Moreover, the scalar, acoustic, and turbulent efficiencies must obey a common energy budget. In particular, $\kappa_{\rm turb}$ describes energy within the fluid/MHD sector and should not be counted as an additional independent reservoir beyond $\kappa_v$. Recent studies have emphasized the need to model the evolving fluid stress rather than simply adding independently maximized template amplitudes \cite{Athron2024,Caprini2020}.

\subsection{Gravitational waves sourced directly by magnetic fields}
\label{sec:gw_magnetic_calculation}

A primordial magnetic field contributes to the gravitational-wave (GW) background through its anisotropic stress, even in the absence of a first-order phase transition. This contribution is distinct from the stresses carried by acoustic or vortical fluid motions. We denote the GW spectrum sourced by the magnetic stress by $\Omega_{\rm GW}^{(B)}$, reserving $\epsilon_B=\rho_B/\rho_{\rm rad}$ for the magnetic energy fraction. Unlike the familiar phenomenological fits for bubble collisions and sound waves, there is no universal template for $\Omega_{\rm GW}^{(B)}$: its shape and normalization depend on the magnetic power spectrum, its time evolution, and the duration and coherence of the source \cite{CapriniDurrer2002,CapriniDurrer2006,CapriniDurrerServant2009,RoperPol2022PMFGW}.

\paragraph{From the magnetic field to the tensor source.}

We use comoving magnetic fields $\mathbf B_c=a^2\mathbf B_{\rm phys}$ and Fourier conventions $B_i(\mathbf x)=\int d^3k\,B_i(\mathbf k)e^{i\mathbf k\cdot\mathbf x}/(2\pi)^3$. Neglecting electric-field stresses on MHD scales, the traceless spatial magnetic stress is

\begin{equation}
\Pi_{ij}^{(B)}(\mathbf x,\eta)
=\frac{1}{a^4(\eta)}
\left(\frac{\delta_{ij}}{3}B_c^2-B_{c,i}B_{c,j}\right).
\label{eq:mag_gw_stress}
\end{equation}

Only its transverse-traceless (TT) part sources GWs. Defining $P_{ij}(\hat{\mathbf k})=\delta_{ij}-\hat k_i\hat k_j$ and $\Lambda_{ij,lm}=P_{il}P_{jm}-\tfrac12P_{ij}P_{lm}$, we obtain

\begin{equation}
\Pi_{ij}^{(B),\rm TT}(\mathbf k,\eta)
=-\frac{\Lambda_{ij,lm}(\hat{\mathbf k})}{a^4(\eta)}
\int\frac{d^3p}{(2\pi)^3}
B_{c,l}(\mathbf p,\eta)B_{c,m}(\mathbf k-\mathbf p,\eta).
\label{eq:mag_gw_stress_fourier}
\end{equation}

Thus, the tensor source is a \emph{convolution} of two magnetic-field modes. A field whose energy is concentrated near $k_B$ can consequently source GWs over a range of wave numbers rather than only at $k_B$.

\paragraph{Magnetic power spectra and unequal-time correlations.}

For a statistically homogeneous and isotropic field, including possible helicity, write

\begin{equation}
\begin{aligned}
&\langle B_{c,i}(\mathbf k,\eta_1)B_{c,j}^*(\mathbf q,\eta_2)\rangle
=(2\pi)^3\delta^{(3)}(\mathbf k-\mathbf q)\,M_{ij}(\mathbf k;\eta_1,\eta_2),\\
&M_{ij}=\frac12 P_{ij}(\hat{\mathbf k})P_B(k;\eta_1,\eta_2)
+\frac{i}{2}\epsilon_{ijl}\hat k_l P_H(k;\eta_1,\eta_2).
\end{aligned}
\label{eq:mag_gw_two_point}
\end{equation}

This convention matches the magnetic correlator introduced in Sec.~2.3; in particular $E_B(k,\eta)=k^2P_B(k;\eta,\eta)/(4\pi^2)$ and $\rho_{B,c}=\int dk\,E_B(k,\eta)$. The parity-even function $P_B$ determines the magnetic energy, while $P_H$ describes the helical component. The corresponding TT-stress unequal-time correlator (UETC) is defined by

\begin{equation}
\left\langle\Pi_{c,ij}^{\rm TT}(\mathbf k,\eta_1)
\Pi_{c,ij}^{\rm TT*}(\mathbf q,\eta_2)\right\rangle
=(2\pi)^3\delta^{(3)}(\mathbf k-\mathbf q)
P_{\Pi_B}(k;\eta_1,\eta_2),
\label{eq:mag_gw_uetc}
\end{equation}

where $\Pi_{c,ij}^{\rm TT}\equiv a^4\Pi_{ij}^{(B),\rm TT}$. If the magnetic field is Gaussian, Wick's theorem expresses $P_{\Pi_B}$ entirely in terms of the two-point functions:

\begin{equation}
\begin{aligned}
P_{\Pi_B}(k;\eta_1,\eta_2)
={}&\Lambda_{ij,lm}\Lambda_{ij,rs}
\int\frac{d^3p}{(2\pi)^3}\Big[
M_{lr}(\mathbf p)M_{ms}(\mathbf k-\mathbf p)\\
&\hspace{41mm}+M_{ls}(\mathbf p)M_{mr}(\mathbf k-\mathbf p)\Big],
\end{aligned}
\label{eq:mag_gw_stress_convolution}
\end{equation}

where the common time arguments $(\eta_1,\eta_2)$ of the $M$ functions are suppressed. For a non-Gaussian field, the connected four-point function must also be included. In either case, knowledge of the equal-time spectrum $E_B(k)$ alone is insufficient: one must specify, or calculate, the time correlations of the magnetic stress.

\paragraph{From the tensor source to $\Omega_{\rm GW}^{(B)}$.}

The tensor equation introduced in Eq.~\eqref{eq:gweq} reads

\begin{equation}
h_{ij}''+2\mathcal H h_{ij}'+k^2h_{ij}
=16\pi G a^2\Pi_{ij}^{(B),\rm TT},
\qquad \mathcal H\equiv a'/a.
\label{eq:mag_gw_wave_eq}
\end{equation}

During radiation domination $a''=0$. For $\mu_{ij}=a h_{ij}$ and vanishing initial sourced perturbations, its Green-function solution is

\begin{equation}
\mu_{ij}(\mathbf k,\eta)
=\frac{16\pi G}{k}\int_{\eta_i}^{\min(\eta,\eta_f)}d\eta'
\sin[k(\eta-\eta')]\,
\frac{\Pi_{c,ij}^{\rm TT}(\mathbf k,\eta')}{a(\eta')},
\label{eq:mag_gw_green}
\end{equation}

where $\eta_i$ and $\eta_f$ mark the beginning and end of the active magnetic source. At times after the source has ended, and for subhorizon modes $k\gg\mathcal H$, the GW energy density is

\begin{equation}
\rho_{\rm GW}^{(B)}\simeq\frac{\langle h_{ij}'h_{ij}'\rangle}{32\pi G a^2},
\qquad
\Omega_{\rm GW}^{(B)}(k,\eta)
\equiv\frac{1}{\rho_{\rm tot}}
\frac{d\rho_{\rm GW}^{(B)}}{d\ln k}.
\label{eq:mag_gw_energy_def}
\end{equation}

Using Eqs.~\eqref{eq:mag_gw_uetc}--\eqref{eq:mag_gw_green} and averaging over the rapid GW oscillations gives the desired spectrum,

\begin{equation}
\begin{aligned}
\Omega_{\rm GW}^{(B)}(k,\eta)
={}&\frac{(16\pi G)^2}{24a^2\mathcal H^2}
\frac{k^3}{2\pi^2}
\int_{\eta_i}^{\eta_f}\frac{d\eta_1}{a(\eta_1)}
\int_{\eta_i}^{\eta_f}\frac{d\eta_2}{a(\eta_2)}\\
&\qquad\times\cos[k(\eta_1-\eta_2)]
P_{\Pi_B}(k;\eta_1,\eta_2).
\end{aligned}
\label{eq:mag_gw_master}
\end{equation}

Here $\rho_{\rm tot}=3H^2/(8\pi G)$ and the expression assumes a radiation-dominated background during production; the UETC and the resulting spectrum must be computed with a consistent Fourier normalization. The double time integral makes the dependence on the source lifetime and temporal coherence explicit. A common approximation is to write $P_{\Pi_B}(k;\eta_1,\eta_2)=[P_{\Pi_B}(k;\eta_1,\eta_1)P_{\Pi_B}(k;\eta_2,\eta_2)]^{1/2}\mathcal C_{\Pi}(k;\eta_1,\eta_2)$ and adopt a physically motivated decorrelation function $\mathcal C_{\Pi}$. Different choices can produce appreciably different GW amplitudes and spectral slopes \cite{CapriniDurrerServant2009,RoperPol2020,RoperPol2022PMFGW}.

For illustration, if the comoving stress is approximately constant and fully coherent during a short interval $\Delta\eta_B$, with $a\simeq a_*$ over that interval, Eq.~\eqref{eq:mag_gw_master} simplifies to

\begin{equation}
\Omega_{\rm GW}^{(B)}(k,\eta)
\simeq\frac{(16\pi G)^2}{24a^2\mathcal H^2a_*^2}
\frac{k^3 P_{\Pi_B,*}(k)}{2\pi^2}
\left[\frac{2\sin(k\Delta\eta_B/2)}{k}\right]^2.
\label{eq:mag_gw_coherent_example}
\end{equation}

This example shows that a prescribed instantaneous magnetic spectrum cannot determine the GW signal without an assumption about the duration of magnetic stresses. Numerical MHD calculations can instead supply the full time-dependent stress directly. For fixed spectral shape and source history, $\Pi_B\propto B^2$ implies approximately $\Omega_{\rm GW}^{(B)}\propto\epsilon_B^2$, but there is no model-independent numerical coefficient multiplying this scaling.

\paragraph{Peak frequency and present-day normalization.}

The characteristic wave number $k_{\rm GW}^{\rm peak}$ is obtained by maximizing Eq.~\eqref{eq:mag_gw_master}, not simply by locating the peak of $E_B(k)$. The frequency observed today is

\begin{equation}
f_{\rm GW,0}^{(B)}=\frac{k_{\rm GW}^{\rm peak}}{2\pi a_0}
=\frac{a_*}{a_0}\frac{k_{\rm GW,phys}^{\rm peak}}{2\pi}.
\label{eq:mag_gw_fpeak}
\end{equation}

For a field injected on a physical length $L_*$, one generally expects $k_{\rm GW,phys}^{\rm peak}L_*$ to be of order unity to a few tens, depending on the stress spectrum, the source duration, and the adopted wave-number convention. Taking $L_*\sim R_*\sim v_w/\beta$ therefore places electroweak-scale magnetic GW signals broadly near the mHz range, although the actual peak must be determined from the calculated spectrum.

If the source ends during radiation domination and entropy is conserved afterwards, its spectrum redshifts as free radiation:

\begin{equation}
h^2\Omega_{\rm GW,0}^{(B)}(f)
\simeq1.67\times10^{-5}
\left(\frac{100}{g_*}\right)^{1/3}
\Omega_{\rm GW,*}^{(B)}(k=2\pi a_0f).
\label{eq:mag_gw_today}
\end{equation}

The quantity on the right is the GW fraction at the end of production, \emph{not} the magnetic energy fraction. The numerical prefactor assumes standard relativistic degrees of freedom and an adiabatic thermal history. Nonstandard expansion or entropy production requires a different redshift factor.

Finally, the separation between magnetic GWs and turbulence-generated GWs requires care. The complete TT stress contains $\Pi_{ij}^{\rm TT}=\Pi_{ij}^{(v),\rm TT}+\Pi_{ij}^{(B),\rm TT}$; hence its UETC contains kinetic, magnetic, and kinetic--magnetic cross terms. If the phenomenological $\Omega_{\rm turb}$ template already includes MHD stresses, adding $\Omega_{\rm GW}^{(B)}$ would double count part of the same source. The safest procedure is to compute the total tensor source from the joint velocity and magnetic-field evolution and separate its components only when the cross correlations are also retained. Helical magnetic fields additionally produce a parity-odd TT correlator and may generate circularly polarized GWs, as discussed in the following subsections \cite{KahniashviliGogoberidzeRatra2005,Kahniashvili2021}.

\subsection{Magnetic and turbulent GW sources are not identical}

A turbulent plasma can source gravitational waves even without a dynamically important magnetic field. Conversely, a strong magnetic field contributes its own anisotropic stress. These two sources may be correlated, but treating them as identical can lead to double counting or an incorrect estimate of the spectrum.

For a strongly magnetized plasma one should in principle evolve
\begin{equation}
E_v(k,t),\qquad E_B(k,t),\qquad H_v(k,t),\qquad H_B(k,t)
\end{equation}
and compute the total anisotropic stress. Magnetic inverse transfer can shift source power to lower $k$, altering both the peak frequency and the low-frequency shape. Kahniashvili and collaborators explored inverse-cascade helical MHD as a GW source and found enhanced low-frequency power relative to nonhelical turbulence \cite{Kahniashvili2008}.

\subsection{Circular polarization from helical turbulence}

Helical magnetic turbulence distinguishes left- and right-handed circular polarizations. This parity asymmetry can appear in the gravitational-wave background, making circular polarization a particularly interesting observable for mechanisms that also generate baryon number.
\begin{equation}
\Pi_{\rm GW}(k)=\frac{\Omega_R(k)-\Omega_L(k)}{\Omega_R(k)+\Omega_L(k)}.
\label{eq:pol}
\end{equation}
Kahniashvili, Gogoberidze and Ratra showed that helical turbulence generated in a first-order phase transition can produce circularly polarized cosmological GWs \cite{KahniashviliGogoberidzeRatra2005}. Direct numerical simulations have since confirmed and refined the relationship between turbulent helicity and GW polarization \cite{Kahniashvili2021}.

This establishes a particularly attractive CP-sensitive chain,
\begin{equation}
\text{CP-odd microphysics}
\rightarrow f_h\neq0
\rightarrow \Pi_{\rm GW}\neq0.
\end{equation}
Unlike the unpolarized energy spectrum, $\Pi_{\rm GW}$ can retain information about the sign and amount of parity violation in the source.

\subsection{Relation to baryogenesis}

Anomalous baryogenesis and polarized gravitational waves can both depend on magnetic helicity, but that does not by itself imply a model-independent relation between them. Establishing a quantitative correlation requires following the helicity spectrum through the electroweak crossover and through the subsequent turbulent phase.
\begin{equation}
\mathrm{sign}(Y_B),\qquad \mathrm{sign}(h_B),\qquad \mathrm{sign}(\Pi_{\rm GW}).
\end{equation}
The first two signs are related in anomaly-driven scenarios, while the second and third are related by the parity structure of the turbulent stress. Establishing a robust three-way sign prediction requires consistent conventions and real-time evolution, and should not be assumed from qualitative handedness arguments alone.

\subsection{Magnetic GWs versus phase-transition GWs}

It is useful to separate gravitational waves produced directly during a transition from those driven later by magnetic stresses. Their source durations, peak frequencies, and polarization properties need not be the same.
\begin{equation}
\Omega_{\rm GW}
=\Omega_{\phi}+\Omega_{\rm sw}+\Omega_{\rm turb}+\Omega_{B}
+\text{cross correlations}.
\end{equation}
The decomposition is physically intuitive but not exact because magnetic and velocity turbulence are correlated, and because the source can transition continuously from acoustic to vortical/MHD dynamics. Adding independent template amplitudes can double count energy or neglect cross terms.

A robust strategy is to use phase-transition simulations to determine the kinetic and magnetic spectra and then compute the stress correlators consistently. This is beginning to be feasible in controlled toy models but remains challenging for realistic electroweak gauge-Higgs dynamics.

\subsection{Recent multi-messenger model studies}

Recent models have begun to compute gravitational waves and primordial magnetic fields from the same transition. These studies are important because a gravitational-wave measurement can constrain the transition parameters that also determine the field-generation mechanism \cite{ArteagaGhoshalStrumia2025,BorahDevGhoshal2026}.

Conversely, an IGMF inference provides information about $\epsilon_{B,*}$ and $f_h$ that conventional GW fits often leave unconstrained. A combined likelihood can therefore break degeneracies that neither messenger resolves alone.

\subsection{Detection prospects for circular polarization}
\label{sec:gw_polarization_detection}

The detection of circular polarization in a stochastic gravitational-wave background (SGWB) would provide an important probe of parity-violating processes in the early Universe. In the context of primordial magnetogenesis, a helical magnetic field can generate an asymmetry between the right- and left-handed GW modes through its parity-odd anisotropic stress. However, measuring this asymmetry is considerably more challenging than detecting the total GW energy density. The difficulty arises not only from the smaller polarized signal, but also from the response of gravitational-wave interferometers to the two helicities.

\paragraph{How circular polarization is measured.}

For a stochastic background, we introduce the two circularly polarized GW components, $h_R$ and $h_L$. The corresponding intensity and circular-polarization spectra can be written as
\begin{equation}
\Omega_{\rm GW}^{I}(f)
=\Omega_R(f)+\Omega_L(f),
\qquad
\Omega_{\rm GW}^{V}(f)
=\Omega_R(f)-\Omega_L(f).
\label{eq:gw_stokes_IV}
\end{equation}
The degree of circular polarization is then
\begin{equation}
\Pi_{\rm GW}(f)
=\frac{\Omega_{\rm GW}^{V}(f)}
{\Omega_{\rm GW}^{I}(f)},
\qquad
|\Pi_{\rm GW}(f)|\leq 1.
\label{eq:gw_detect_polarization_fraction}
\end{equation}
An unpolarized background has $\Pi_{\rm GW}=0$, whereas $|\Pi_{\rm GW}|=1$ corresponds to a fully circularly polarized signal.

A detector measures a combination of the two GW polarizations, weighted by its antenna response. For two detector channels $a$ and $b$, the expected cross-correlation of the stochastic signals can be expressed as
\begin{equation}
\begin{aligned}
\langle\widetilde s_a(f)\widetilde s_b^*(f')\rangle_{\rm GW}
={}&\frac12\delta(f-f')
\frac{3H_0^2}{10\pi^2f^3}\\
&\times\left[
\gamma_{ab}^{I}(f)\Omega_{\rm GW}^{I}(f)
+\gamma_{ab}^{V}(f)\Omega_{\rm GW}^{V}(f)
\right],
\end{aligned}
\label{eq:gw_polarized_correlation}
\end{equation}
where $H_0$ is the present Hubble parameter and $\gamma_{ab}^{I,V}$ are the overlap reduction functions, normalized according to the convention used in Eq.~\eqref{eq:gw_polarized_correlation}. These functions depend on the separation, orientation, and frequency-dependent antenna responses of the detectors. The intensity response $\gamma_{ab}^{I}$ measures the sum of the two GW helicities, while $\gamma_{ab}^{V}$ measures their difference \cite{Seto2006Circular,SmithCaldwell2017,Orlando2021}.

The sensitivity to polarization therefore depends crucially on the detector geometry. In particular, a single planar interferometer has no direct sensitivity to the circular-polarization monopole of an exactly isotropic SGWB. This follows from reflection symmetry: right-handed waves arriving from one side of the detector plane cannot be distinguished from left-handed waves arriving from the opposite side. This degeneracy can be broken by measuring anisotropies or by cross-correlating suitably oriented, spatially separated interferometers \cite{Seto2006Circular,Orlando2021}.

For two detectors with uncorrelated instrumental noise, and in the weak-signal approximation, a useful estimate of the polarization signal-to-noise ratio is
\begin{equation}
{\rm SNR}_V^2\simeq
2T_{\rm obs}
\left(\frac{3H_0^2}{10\pi^2}\right)^2
\int_0^\infty df\,
\frac{|\gamma_{ab}^{V}(f)|^2
|\Omega_{\rm GW}^{V}(f)|^2}
{f^6P_a(f)P_b(f)},
\label{eq:gw_polarization_snr}
\end{equation}
where $T_{\rm obs}$ is the observation time and $P_a$, $P_b$ are the detector noise power spectra. This expression assumes that the intensity contribution has been separated or otherwise controlled. In a realistic network, the intensity and polarization should be estimated jointly, including their covariance and the correlations between detector channels.

Equation~\eqref{eq:gw_polarization_snr} illustrates two important features. First, the polarization sensitivity depends on $\gamma_{ab}^{V}$ and can therefore be much smaller than the sensitivity to the total SGWB. Second, for fixed noise and spectral shape, the signal-to-noise ratio scales as ${\rm SNR}_V\propto|\Pi_{\rm GW}|\sqrt{T_{\rm obs}}$. Thus, even a strong SGWB may not yield a measurable polarization unless the helicity fraction and detector response are sufficiently large.

\paragraph{LISA.}

The Laser Interferometer Space Antenna (LISA), adopted by ESA with a planned launch around 2035, will consist of three spacecraft separated by approximately $2.5\times10^6\,{\rm km}$. Its principal sensitivity lies in the millihertz band, making it a natural target for GWs generated by electroweak-scale first-order phase transitions.

For a transition near $T_*\sim100\,{\rm GeV}$ and $\beta/H_*\sim10$--$100$, the characteristic frequencies from bubble collisions, acoustic waves, and turbulence can fall within LISA's sensitive range. However, its triangular configuration is approximately planar, and a direct measurement of the isotropic circular-polarization monopole is not possible using its internal correlations alone. One possibility is to exploit the dipolar anisotropy induced by the motion of the Solar System relative to the cosmological rest frame. This introduces a fractional anisotropy of order $v/c\sim10^{-3}$, whose modulation can make an otherwise inaccessible polarized component observable \cite{Seto2006Circular,Domcke2019Circular}.

The feasibility of this approach is model dependent. Ellis et al.~\cite{Ellis2020Polarization} studied circularly polarized GWs from helical turbulence following a first-order phase transition, including the direct and inverse cascades. They found that polarization detection is substantially more demanding than detecting the total background, although favorable models can produce potentially measurable signals. More detailed MHD simulations by Roper Pol et al.~\cite{RoperPol2022Polarization} indicate that the polarization generated by electroweak-scale helical turbulence is generally difficult to detect using LISA's kinematic-dipole response alone, even when the total GW spectrum is observable.

Therefore, a prediction that the GW spectrum lies above LISA's ordinary sensitivity curve does not establish that its circular polarization will be detected. The latter requires a separate calculation using the parity-odd detector response.

\paragraph{LISA--Taiji and TianQin networks.}

A particularly promising possibility is to combine LISA with another space-based interferometer. Taiji is a proposed Chinese mission with sensitivity in a frequency range similar to LISA. Because the two constellations would be spatially separated and could have different orientations, their joint response need not possess the reflection symmetry that suppresses the isotropic polarization signal in a single planar detector.

Orlando, Pieroni, and Ricciardone \cite{Orlando2021} studied the sensitivity of the LISA--Taiji network to the Stokes parameters $I$ and $V$. For a maximally chiral, approximately scale-independent SGWB, they found that a clear polarization measurement could be possible for a signal with $h^2\Omega_{\rm GW}^{I}\sim10^{-12}$, under their assumed detector configurations and observation time. They also found that the uncertainty on the polarization amplitude can be one to two orders of magnitude larger than that on the total intensity.

The relative orientation of the two constellations is particularly important. Chen et al.~\cite{Chen2025LISATaiji} compared different LISA--Taiji configurations and found that a network with oppositely inclined detector planes can improve the sensitivity to parity violation by approximately an order of magnitude at low frequencies relative to a configuration with similarly inclined planes. These results emphasize that the ability to measure polarization depends not only on instrumental noise, but also on the geometry of the network.

The TianQin mission provides another possible partner for LISA. A joint LISA--TianQin network could improve the reconstruction of stochastic backgrounds through complementary orbital geometries and independent noise streams. Studies of such networks have investigated the separation of different GW polarization components, although sensitivity to the isotropic circular-polarization signal must be evaluated using the appropriate parity-odd response rather than inferred from the ability to detect additional non-tensor polarization modes.

For the specific case of helical MHD turbulence, Roper Pol et al.~\cite{RoperPol2022Polarization} found that the polarization could be accessible to a LISA--Taiji network for turbulent energy fractions of order $5\%$ and substantial magnetic helicity, with fractional helicity between approximately $0.5$ and $1$, under their source and detector assumptions. Such numerical conditions are illustrative forecasts, not general detection thresholds.

More recently, Furusawa and Tashiro \cite{FurusawaTashiro2026} performed a polarization forecast for helical primordial magnetic fields using the LISA--Taiji network. Their analysis showed that combining intensity and circular-polarization measurements can help constrain the helical fraction of primordial fields. Their calculation concerns inflationary magnetic fields and passive tensor perturbations rather than turbulence generated by an electroweak transition, but provides a useful example of how magnetic helicity could be inferred from an observed GW background.

\paragraph{DECIGO and BBO.}

The proposed Deci-hertz Interferometer Gravitational-wave Observatory (DECIGO) and Big Bang Observer (BBO) are designed to explore frequencies around $0.1$--$10\,{\rm Hz}$, between the most sensitive bands of LISA and ground-based interferometers. Such frequencies can be relevant for phase transitions occurring at temperatures above the electroweak scale or for transitions with sufficiently small characteristic source lengths.

Their proposed multiple-constellation configurations offer the possibility of cross-correlating independent detectors with a geometry sensitive to circular polarization. Seto \cite{Seto2007Circular} showed that appropriately breaking the symmetry of a multiple-interferometer configuration can make the isotropic circular-polarization monopole measurable. This makes DECIGO/BBO-like concepts particularly interesting for parity-violating early-Universe sources.

Their improved sensitivity could also help probe magnetic-field configurations whose GW signals are too weak for LISA. However, their polarization reach depends on the final constellation design, detector geometry, and achievable noise levels. They should therefore be regarded as proposed experimental opportunities rather than guaranteed future sensitivities.

\paragraph{Ground-based observatories.}

Current ground-based interferometers, including LIGO, Virgo, and KAGRA, operate primarily at frequencies from several tens to thousands of hertz. These frequencies are generally above the expected peak of conventional electroweak-scale GW signals, but may be relevant for much higher-temperature transitions or for sources with exceptionally short characteristic time scales.

A network of detectors with different orientations and geographical locations can distinguish the intensity and polarization components of a SGWB through cross-correlation measurements. Future third-generation observatories, particularly the Einstein Telescope (ET) and Cosmic Explorer (CE), are expected to improve the sensitivity to stochastic backgrounds and their angular structure.

Smith and Caldwell \cite{SmithCaldwell2017} studied the sensitivity of different GW experiments to frequency-dependent circular polarization, including ground-based networks. Mentasti, Contaldi, and Peloso \cite{Mentasti2023Polarization} subsequently developed a framework for detecting anisotropies and chirality with present and future ground-based interferometers. Their forecasts show that third-generation networks could measure small polarization fractions for sufficiently strong stochastic backgrounds, although the sensitivity deteriorates rapidly for weaker signals.

A recent full-Stokes analysis by Boybeyi and Mandic \cite{BoybeyiMandic2026} has also demonstrated how ground-based detector data can be used to constrain polarized anisotropies of the SGWB. These are constraints on stochastic polarization components, not a detection of primordial helical gravitational waves.

\paragraph{Implications for electroweak magnetogenesis.}

The prospects for detecting circularly polarized GWs from electroweak magnetogenesis depend on three distinct ingredients: the production of magnetic helicity, the conversion of that helicity into parity-odd gravitational-wave stresses, and the polarization sensitivity of the detector network.

The GW polarization fraction is generally frequency dependent and should not be identified directly with the initial magnetic helicity fraction. Numerical simulations show that the polarization may be largest near the GW spectral peak and can decrease away from it, depending on whether turbulence is initially present or continuously driven \cite{RoperPol2022Polarization,Kahniashvili2021}.

Consequently, a complete prediction should provide both $\Omega_{\rm GW}^{I}(f)$ and $\Omega_{\rm GW}^{V}(f)$, including the evolution of the helical magnetic spectrum during the period when GWs are produced. These spectra should then be convolved with the detector response functions to calculate the expected ${\rm SNR}_I$ and ${\rm SNR}_V$ separately.

The magnetic inverse cascade can transfer energy toward larger scales while the source remains active, potentially altering the low-frequency GW spectrum. However, the late-time evolution of the magnetic field does not modify GWs that have already been emitted. Thus, the relevant input is the time-dependent magnetic stress during GW production, rather than the present-day magnetic spectrum alone.

The most promising observational strategy for electroweak-scale signals is therefore to combine a measurement of the GW intensity by LISA with polarization-sensitive cross-correlations involving additional space-based detectors. If a nonzero circular polarization were observed together with a spectral shape consistent with a cosmological phase transition, it would provide evidence for parity-violating source dynamics. Establishing a quantitative connection with the baryon asymmetry would require the additional anomalous-transport and sphaleron calculations discussed in Sec.~7.

\section{Observational and cosmological constraints}
\label{sec:observations}

Primordial magnetic fields can leave signatures long after their production. Their effects are sought in the CMB, light-element abundances, gamma-ray cascades, large-scale structure, and gravitational waves. Each observable probes a different range of magnetic scales and relies on different astrophysical assumptions.

The observational status of primordial magnetic fields is unusual: there are stringent upper limits on some scales and model-dependent lower limits or hints on others, but no universally accepted direct detection of a primordial component. The constraints probe different moments of the spectrum and depend differently on helicity and coherence length. A viable electroweak model must therefore be compared to a multi-dimensional constraint set, not to a single number called ``the magnetic-field bound.''

\subsection{CMB anisotropies and polarization}

Primordial magnetic fields affect the CMB through their stress-energy tensor, through modifications of the ionized plasma, and through Faraday rotation. The constraints depend on the magnetic spectrum, so a limit quoted at $1\,\mathrm{Mpc}$ cannot always be applied directly to a field whose power is concentrated on much smaller scales \cite{PlanckPMF2015,PaolettiFinelli2011}.

For nearly scale-invariant spectra the bounds can be particularly stringent, while causal blue spectra require careful interpretation because their energy resides near a much smaller integral scale. A value such as
\begin{equation}
B_{1\,\mathrm{Mpc}}<\mathrm{few\ nG}
\end{equation}
should therefore not be converted into a total magnetic-energy bound without specifying $P_B(k)$.

Helical fields can generate parity-odd CMB correlations, including $TB$ and $EB$, and parity-odd higher-point functions \cite{CapriniDurrerKahniashvili2004,KahniashviliRatra2005}. Present CMB data constrain helicity but do not provide a compelling detection. Future high-sensitivity polarization surveys can improve these tests, although foreground control and cosmic birefringence degeneracies are important.

\subsection{BBN and baryon isocurvature}

Big Bang nucleosynthesis constrains both additional energy density and spatial variations in the baryon abundance. In magnetic baryogenesis, the second effect can be especially restrictive because an inhomogeneous hypermagnetic field generates an inhomogeneous baryon-number source.

Kamada, Uchida and Yokoyama found that deuterium abundance strongly constrains pre-electroweak primordial fields in scenarios where the same field sources baryogenesis \cite{KamadaUchidaYokoyama2021}. Independently, inhomogeneous recombination induced by primordial magnetic structure can substantially tighten CMB constraints compared with homogeneous treatments \cite{JedamzikSaveliev2019}. Their result is stronger than a simple bound on the mean magnetic energy because it probes fluctuations of the baryon source.

This constraint deserves special attention in model building. A model that reproduces the observed \emph{mean} $Y_B$ can still fail because it predicts excessive $\delta Y_B(\bm x)$. However, the exclusion is sensitive to the electroweak conversion prescription: the 2026 reanalysis of Hamada, Mukaida, and Uchida finds viable maximally helical configurations under revised crossover assumptions \cite{HamadaMukaidaUchida2026Constraints}. The correct comparison should therefore evolve the mean abundance and its fluctuations consistently.

\subsection{CMB spectral distortions}

Magnetic dissipation injects energy into the photon plasma and can distort the CMB spectrum. The resulting $\mu$- and $y$-type signals probe different redshift ranges and depend on the distribution of magnetic power across scales.

Uchida, Kamada and Tashiro revisited the calculation using reconnection-driven turbulent decay \cite{UchidaKamadaTashiro2025}. The resulting constraints differ from older treatments based on linear damping or simpler nonlinear decay laws and depend on the helicity fraction. This illustrates a general rule: an observational PMF constraint is often conditional on a transfer function supplied by MHD theory.

Future spectral-distortion missions could therefore test phase-transition magnetogenesis in regions difficult to access through $B_{1\,\mathrm{Mpc}}$ alone.

\subsection{Faraday rotation}

A magnetic field rotates the polarization plane of radiation passing through ionized matter. Faraday rotation therefore probes the line-of-sight magnetic field weighted by the free-electron distribution, providing a complementary handle on primordial magnetism.
\begin{equation}
\Delta\chi\propto\lambda_{\rm obs}^2\int dl\,n_e B_\parallel.
\end{equation}
Faraday rotation provides direct magnetic sensitivity but is weighted by the free-electron density and is therefore most powerful in magnetized structures rather than empty voids. CMB Faraday rotation constrains primordial fields, while radio-source rotation measures can constrain large-scale intergalactic magnetism. The interpretation requires modeling of Galactic and host contributions, as well as the electron-density distribution; representative extragalactic rotation-measure limits were derived by Pshirkov, Tinyakov and Urban \cite{PshirkovTinyakovUrban2016}.

\subsection{Gamma-ray cascades from TeV sources}

One of the main motivations for primordial void fields comes from very-high-energy gamma-ray observations. TeV photons interacting with background light produce electron--positron pairs, which can upscatter CMB photons. Magnetic deflection may spread or delay the resulting GeV cascade, so a missing cascade can constrain the intergalactic field under specified assumptions.
\begin{equation}
\gamma_{\rm TeV}+\gamma_{\rm EBL}\rightarrow e^+e^-.
\end{equation}
The pairs inverse-Compton scatter CMB/EBL photons back to GeV--TeV energies. In a weak IGMF, the pairs are deflected before radiating, changing the angular profile, time delay and spectrum of the cascade.

The scaling with coherence length depends on whether the pairs traverse many coherent domains. For coherence lengths longer than the cooling length, the deflection is approximately coherent; for shorter coherence lengths it becomes a random walk. This leads to a characteristic degeneracy between $B$ and $\lambda_B$.

Neronov and Vovk reported early evidence for a nonzero extragalactic field from the absence of expected cascade emission \cite{NeronovVovk2010}. Many subsequent analyses have emphasized source variability, spectral assumptions and EBL uncertainties \cite{TaylorVovkNeronov2011,Dermer2011,Dolag2011}. A 2023 combined Fermi-LAT and H.E.S.S. analysis found
\begin{equation}
B\gtrsim7.1\times10^{-16}\,\mathrm{G}
\end{equation}
for a $1\,\mathrm{Mpc}$ coherence length even for a $10$-yr duty cycle, with stronger lower limits for much longer assumed source activity \cite{Aharonian2023}.

MAGIC and Fermi-LAT observations of 1ES~0229+200 provide a more conservative lower bound at the $10^{-17}\,\mathrm{G}$ scale under their adopted assumptions \cite{Acciari2023}. The same MAGIC analysis quotes the distinct bound $B>10^{-14}\,\mathrm{G}$ for a field of cosmological origin; these are conditional limits for different field and source assumptions \cite{Acciari2023}. The spread between such results is a reminder that the source-duty-cycle prior matters.

\
\subsection{GRB pair echoes}

Gamma-ray bursts (GRBs) provide a complementary probe of
intergalactic magnetic fields through the delayed emission
of secondary gamma rays, commonly referred to as pair
echoes. Very-high-energy photons emitted by a GRB interact
with the extragalactic background light and produce
electron--positron pairs. These pairs subsequently
upscatter cosmic microwave background photons through
inverse Compton scattering, generating secondary gamma
rays at GeV energies. The intergalactic magnetic field
deflects the pairs, producing a time delay and suppressing
the observable cascade flux within a given observation
window.

The characteristic magnetic deflection angle scales as
\begin{equation}
\theta_B \sim
\begin{cases}
D_{\rm IC}/r_L,
& \lambda_B \gtrsim D_{\rm IC},\\[2mm]
\sqrt{D_{\rm IC}\lambda_B}/r_L,
& \lambda_B \lesssim D_{\rm IC},
\end{cases}
\end{equation}
where $D_{\rm IC}$ is the inverse-Compton cooling distance,
$r_L$ is the Larmor radius of the electron or positron,
and $\lambda_B$ denotes the magnetic coherence length.
Consequently, both the strength and coherence length
of the magnetic field determine the expected pair-echo
signal.

The exceptionally bright GRB 221009A has provided an
important opportunity to test this mechanism. Using
Fermi-LAT observations together with the multi-TeV
emission detected by LHAASO, Vovk et al.\ obtained a
lower bound of approximately $B\gtrsim10^{-19}\,\mathrm{G}$
for coherence lengths exceeding $1\,\mathrm{Mpc}$
\cite{Vovk2024GRB}. A subsequent analysis by Burmeister
et al.\ obtained the stronger conditional constraint
$B\gtrsim2.5\times10^{-17}\,\mathrm{G}$ at $95\%$
confidence level for comparable coherence lengths
\cite{GRB221009AIGMF}.

Unlike steady blazar observations, GRBs have relatively
well-defined emission episodes, reducing uncertainties
associated with the long-term activity of the gamma-ray
source. Nevertheless, the inferred magnetic bounds
remain sensitive to the assumed intrinsic TeV spectrum,
high-energy cutoff, afterglow contribution, and cascade
modeling. The difference between the published limits
therefore reflects, in part, their different modeling
assumptions. Future observations of bright, TeV-detected
GRBs, combined with extended GeV monitoring, could
provide increasingly useful constraints on magnetic
fields in cosmic voids.

\subsection{A 2026 anisotropic pair-halo hint}

Some recent analyses search for an extended, anisotropic gamma-ray halo around energetic sources rather than only for missing cascade flux. Such claims are interesting but require careful treatment of foregrounds and instrumental response before being interpreted as evidence for an intergalactic magnetic field.
\begin{equation}
B_0\simeq2.8\times10^{-16}\,\mathrm{G},
\end{equation}
with a quoted $99\%$ interval spanning roughly $0.9$--$8.9\times10^{-16}\,\mathrm{G}$ \cite{ZhangChenErrando2026}. If confirmed, a directional pair-halo morphology would be qualitatively stronger evidence than a spectral deficit alone because it uses the spatial imprint expected from magnetic deflection.

At present this should be described as evidence within a cascade model, not as a settled detection of a primordial field. Independent analyses, instrumental systematics and the plasma-cooling issue remain relevant.

\subsection{The beam-plasma-instability caveat}

The absence of a GeV cascade does not uniquely imply magnetic deflection. Plasma instabilities may extract energy from the pair beam before inverse-Compton scattering. The importance of this effect remains disputed and directly affects the interpretation of gamma-ray lower bounds. The original plasma-instability concern and its potential impact on cascade cooling were emphasized by Broderick, Chang and Pfrommer \cite{BroderickChangPfrommer2012}.

The efficiency of the instability depends on kinetic effects, beam inhomogeneity, nonlinear saturation and the state of the IGM. The literature contains both arguments for strong plasma cooling and calculations finding it insufficient in realistic regimes. This is an active problem rather than a resolved nuisance parameter. Therefore gamma-ray ``lower bounds'' should be labeled with their assumption that inverse-Compton cascade physics is not replaced by dominant plasma energy loss.

\subsection{Structure formation and 21-cm signatures}

Primordial magnetic fields can also heat gas, modify small-scale density perturbations, and affect the thermal history probed by 21-cm observations. These constraints become model dependent once astrophysical heating and reionization are included \cite{CruzEtAl2024,BhaumikPaulPal2025}.

Dark-Ages global-signal forecasts can already be competitive for selected magnetic spectral indices \cite{MohapatraNatwariyaNayak2024}, while explicit 2026 forecasts show that HERA- and SKA-like measurements can constrain PMFs through their enhancement of early small-scale structure, especially when combined with CMB priors \cite{WorkuCruzKamionkowski2026}.

\subsection{Gravitational-wave constraints and searches}

Gravitational-wave observations test the transition and turbulent energy scales rather than measuring a magnetic field directly. Existing upper limits and future sensitivities must be compared with a source spectrum computed using consistent transition and MHD parameters \cite{RoperPol2022PMFGW}.

The absence of a detected mHz stochastic background currently constrains only the most optimistic electroweak models indirectly. LISA will provide a much more direct test of strongly first-order electroweak/BSM transitions.

\subsection{Comparing an actual magnetic spectrum with observations}
\label{sec:new_observation_comparison}

\begin{figure}[t]
\centering
\includegraphics[width=0.93\linewidth]{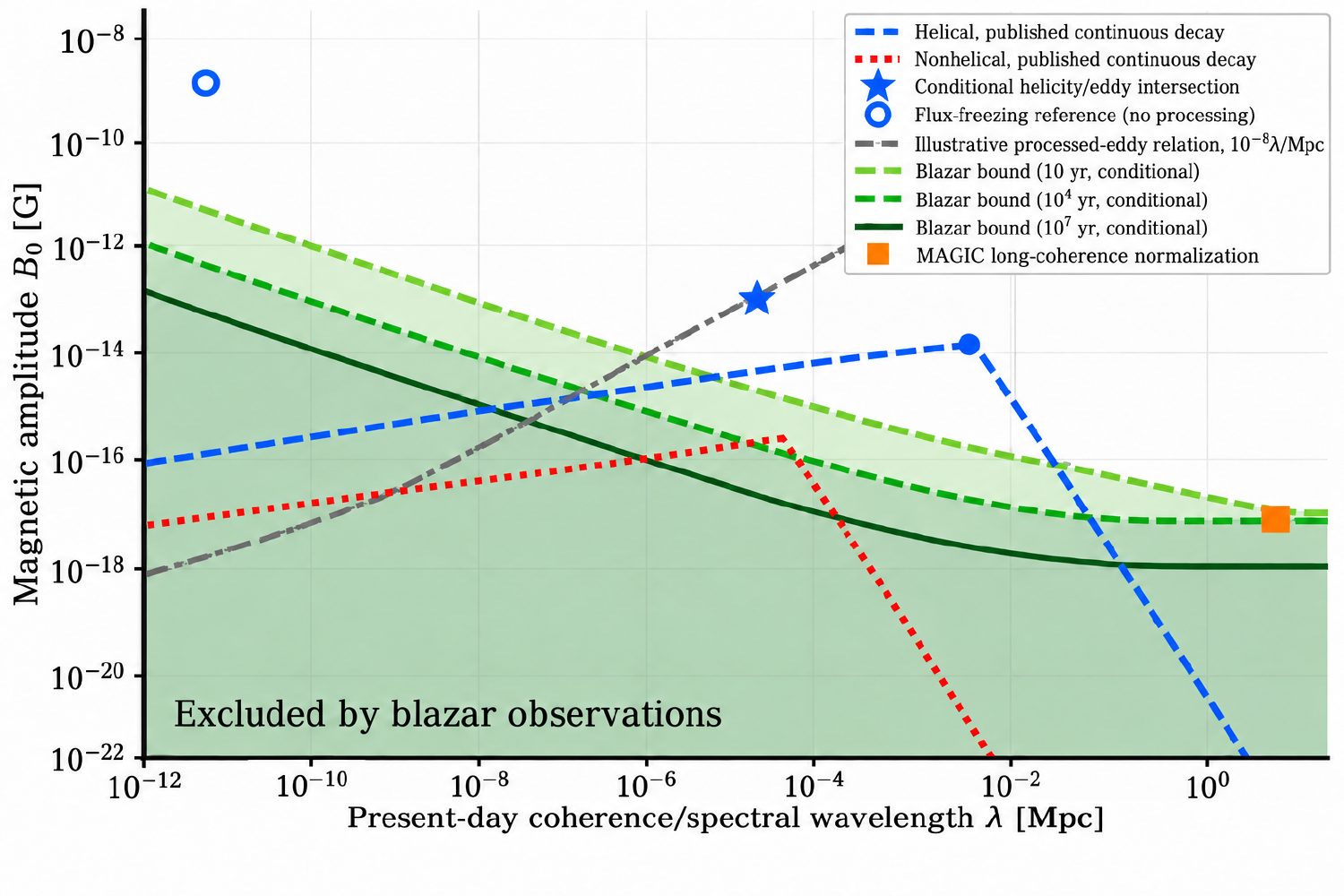}
\caption{Published conventional-decay extrapolations (dashed) and conditional helicity-processing diagnostic (marker) from Eqs.~\eqref{eq:new_present_B_spectral_shape}--\eqref{eq:new_AGS_nonhelical_endpoints}, for $T_*=1\,\mathrm{TeV}$, $\lambda_*H_*=0.1$ and the source paper's magnetic-efficiency prescription. The green dashed curve is the approximate blazar-guide relation in Eq.~\eqref{eq:new_blazar_guide}. The orange point indicates the MAGIC+Fermi-LAT long-coherence-length lower-limit normalization. The observations are not raw magnetic-field measurements, and lines connecting their reported limits at other correlation lengths have deliberately not been invented. The model curves and observational markers should be compared only within consistent spectrum, filling-factor and source assumptions.}
\label{fig:new_observational_comparison}
\end{figure}

Figure~\ref{fig:new_observational_comparison} makes the observational comparison explicit for the worked example of Sec.~\ref{sec:new_worked_MHD_evolution}. The helical and nonhelical dashed curves reproduce a published continuous-decay prescription for the \emph{same} transition inputs. An additional marker displays the conditional helicity-conserving intersection of Eq.~\eqref{eq:new_intersection}; it is a dynamical diagnostic, not a new MHD simulation. They are full illustrative spectral shapes in the $(B_0,\lambda)$ plane, not simply points specifying the integral scale. In particular, a present-day field at $\lambda_0$ cannot be compared directly to a $1\,\mathrm{Mpc}$-smoothed CMB bound unless the spectrum and smoothing window are specified.

The frequently reproduced blazar lower-bound curve is conditional on a model of the TeV source and electromagnetic cascade. A simple guide adopted in Ref.~\cite{ArteagaGhoshalStrumia2025} is
\begin{equation}
 B_{\rm blazar}^{\rm guide}(\lambda)\gtrsim2\times10^{-17}\,\mathrm{G}
 \sqrt{\max\left(1,\frac{0.2\,\mathrm{Mpc}}{\lambda}\right)}.
 \label{eq:new_blazar_guide}
\end{equation}
This is a \emph{phenomenological guide}, not a likelihood obtained by reanalyzing the gamma-ray events. It illustrates the random-walk scaling at coherence lengths shorter than the relevant pair-cooling length. It should not be applied without qualification to every source or to a field with arbitrary spatial filling factor and power spectrum. The more recent H.E.S.S.+Fermi-LAT analysis reports, for $\lambda=1\,\mathrm{Mpc}$, lower limits of $7.1\times10^{-16}\,\mathrm{G}$, $1.8\times10^{-14}\,\mathrm{G}$ and $3.9\times10^{-14}\,\mathrm{G}$ for assumed source-activity times of $10$, $10^4$ and $10^7$ years, respectively \cite{Aharonian2023}. MAGIC+Fermi-LAT instead obtains $B\gtrsim1.8\times10^{-17}\,\mathrm{G}$ in its long-coherence-length analysis of 1ES~0229+200, with assumptions about the source history and cascade spectrum that differ from the H.E.S.S. treatment \cite{Acciari2023}. For a magnetic field of cosmological origin, the same MAGIC analysis separately quotes $B\gtrsim10^{-14}\,\mathrm{G}$; this more restrictive, model-specific statement should not be replaced by the long-coherence value \cite{Acciari2023}. The differences among the limits reflect their source and cascade assumptions, not inconsistent direct measurements of a magnetic amplitude.

The CMB probes a different quantity. For example, Planck obtained $B_{1\,\mathrm{Mpc}}<4.4\,\mathrm{nG}$ at 95\% confidence for one nonhelical power-spectrum prescription, with other limits for different spectral indices, helicities and treatments of recombination \cite{PlanckPMF2015}. This is an \emph{upper} bound on a specified smoothed field, not an upper horizontal line in the plot of the field at its integral scale. A complementary Lyman-$\alpha$ forest analysis found $B_{1\,\mathrm{Mpc}}<0.30\,\mathrm{nG}$ (3$\sigma$) \emph{for the specified nearly scale-invariant index} $n_B=-2.9$ \cite{Pavicevic2025}. This is much tighter in that spectral class, but is not a substitute bound for the causal blue spectrum $E_B\propto k^4$ generated by a phase transition. The observations are therefore complementary rather than interchangeable. The same distinction applies to Faraday rotation, CMB spectral distortions and baryon-isocurvature constraints.

Parity-odd combinations of gamma-ray arrival directions have also been proposed as probes of the \emph{helicity} of intergalactic fields \cite{TashiroChenFerrerVachaspati2014,ChenChowdhuryFerrer2015}. Such observables depend on source populations, sky masks and cascade assumptions, and are not currently interchangeable with the magnetic-energy limits displayed in Fig.~\ref{fig:new_observational_comparison}.

The Fermi-LAT extended-halo search of Ackermann and collaborators provides an additional, morphologically motivated limit and reports no significant extension for its target sample \cite{FermiFHES2018}. Complementary constraints from radio rotation measures depend instead on the line-of-sight free-electron distribution; Pshirkov, Tinyakov and Urban obtain bounds at the nG level for their assumed coherence regimes \cite{PshirkovTinyakovUrban2016}. Future Cherenkov Telescope Array observations may strengthen joint TeV/GeV cascade studies, but forecast sensitivity is model-dependent. Finally, nonobservation of a GeV cascade need not uniquely establish magnetic deflection: competing beam--plasma energy losses and other source uncertainties remain under investigation \cite{BroderickChangPfrommer2012}. A well-defined model comparison should state the assumed blazar lifetime and cascade prescription, fold the predicted full magnetic spectrum into the appropriate observable, and quote the resulting conditional limit. It should not promote one convenient guide curve to a model-independent detection of primordial magnetism.

\subsection{No single ``allowed region'' is universal}

Observational constraints are often displayed as one allowed region in the magnetic-field amplitude--coherence-length plane. That picture can be misleading: each probe assumes a spectrum, spatial distribution, and source history, and different assumptions can shift the inferred bounds.

\begin{table}[H]
\centering
\caption{Representative observational probes and the physics that limits their interpretation. Numerical limits vary with spectrum and model assumptions.}
\begin{tabularx}{\textwidth}{@{}lXX@{}}
\toprule
Probe & Primary sensitivity & Main modeling dependence \\
\midrule
CMB anisotropy/polarization & nG-scale PMFs, spectral shape, helicity & magnetic spectrum, recombination/heating \\
BBN baryon isocurvature & small-scale helicity-induced $\delta Y_B$ & BAU conversion and magnetic statistics \\
CMB spectral distortions & small-scale magnetic dissipation & nonlinear MHD decay law \\
Lyman-$\alpha$ forest & near-scale-invariant PMF amplitude $B_{1\,\mathrm{Mpc}}<0.30\,\mathrm{nG}$ \cite{Pavicevic2025} & applies to $n_B=-2.9$; not a generic causal-spectrum limit \\
Faraday rotation & line-of-sight $n_eB_\parallel$ & foregrounds and baryon distribution \\
Blazar cascades & weak void IGMF, $B$--$\lambda$ & source lifetime/spectrum, EBL, plasma cooling \\
GRB pair echoes & transient IGMF deflection & intrinsic VHE emission and cascade model \\
21-cm / small-scale structure & blue PMF power and heating & nonlinear baryon/MHD evolution \\
Stochastic GWs & transition/turbulence/PMF stress & source lifetime and stress correlators \\
\bottomrule
\end{tabularx}
\end{table}

Current observations make intergalactic magnetic fields a compelling target, but gamma-ray indications of fields in cosmic voids do not yet amount to an unambiguous detection of primordial magnetism. Electroweak scenarios should instead be assessed jointly against the observed baryon abundance, nucleosynthesis and baryon-isocurvature limits, CMB and spectral-distortion constraints, gamma-ray data, and any associated gravitational-wave signal.

\section{The 2024--2026 frontier: what changed most recently}
\label{sec:modern}

Recent work has moved beyond qualitative links between magnetogenesis, baryogenesis, and gravitational waves. Several studies now model two or more of these effects together, while new nonperturbative results have questioned parts of the conventional electroweak matching calculation. Here we summarize what has genuinely changed and which results remain model dependent.

The recent literature is important enough to deserve a dedicated section because it changes the emphasis of the subject. Earlier work established that each arrow in the magnetism--BAU network can exist. The newest papers increasingly ask whether several arrows can be made quantitative in a common model.

\subsection{Spontaneous electroweak magnetogenesis from twisted dumbbells}

Domcke, Ema, and Patel proposed a mechanism in which CP-violating dynamics twist electroweak monopole--string configurations during symmetry breaking \cite{DomckeEmaPatel2025}. The twist generates a helical magnetic field, linking an electroweak gauge configuration to a possible baryon asymmetry.

The mechanism is notable because it addresses a common weakness of PMF phenomenology: $f_h$ is predicted parametrically rather than set to zero or one by hand. The same CP-odd dynamics generates a small baryon asymmetry, producing a direct co-generation scenario. Important questions for future work include the abundance and lifetime of the dumbbell network in a thermal electroweak environment, the detailed spectrum of the released magnetic field, and the matching to anomalous MHD.

\subsection{SMEFT bubble magnetogenesis and CME-amplified baryogenesis}

Liu, Qin, and Bian studied helical-field production during a first-order electroweak transition with a CP-violating SMEFT operator \cite{LiuQinBian2026}. Their analysis illustrates why direct anomaly conversion and subsequent chiral magnetic amplification should be treated as separate stages.

This work is useful as a prototype of a genuinely coupled calculation:
\begin{equation}
\text{CPV EFT operator}\to\text{bubble dynamics}\to h_Y
\to\text{CME}\to Y_B.
\end{equation}
The remaining theoretical systematics include the EFT validity range, the mapping from idealized collision geometries to a stochastic network, and the subsequent electroweak-crossover treatment.

\subsection{Supercooled dynamical electroweak magnetogenesis}

Strong supercooling has received attention because it can produce powerful bulk motion and large bubbles. Recent dynamical-electroweak studies show how this can affect both the primordial field and the gravitational-wave background, though the result depends on reheating and MHD assumptions \cite{ArteagaGhoshalStrumia2025,BalajiFairbairnOleaRomacho2024}.

The conceptual advance is the use of supercooling to enlarge the effective magnetogenesis scale and alter the energy budget. This deserves to be separated from ordinary $\beta/H\sim10^2$ thermal electroweak transitions: a vacuum-dominated stage changes both the bubble distribution and the redshifting history.

\subsection{ALP-assisted PMF--GW correlations}

Borah, Dev, and Ghoshal considered an ALP framework with a first-order transition and a coupling to the Higgs sector \cite{BorahDevGhoshal2026}. By predicting both a gravitational-wave background and a magnetic field, their work illustrates the appeal and the limitations of correlated multi-messenger studies.

For maximally helical initial conditions they find present-day fields reaching roughly $10^{-9}\,\mathrm{G}$ for coherence lengths around $10^{-3}$--$10^{-1}\,\mathrm{Mpc}$ in favorable regions, together with a potentially observable space-interferometer signal. The strongest caveat is precisely the one highlighted by the broader review: maximal initial helicity is phenomenologically powerful but must be derived from the microphysics if the model is to make a true BAU prediction.

\subsection{Gauge-invariant magnetic-flux baryogenesis}

The central theoretical advance of Hamada, Mukaida, and Uchida is not another magnetogenesis model. It is a revised description of the magnetic flux that actually survives the Standard Model electroweak crossover \cite{HamadaMukaidaUchida2026}. A gauge-invariant mixing angle and the approximate conservation of unconfined flux can change baryogenesis estimates.

For phenomenology the main implication is uncomfortable but healthy: an uncertainty that was often hidden inside an ``EW conversion efficiency'' is now exposed as an active nonperturbative problem. Any modern review or model paper should therefore separate
\begin{equation}
\text{uncertainty in generating }h_Y
\end{equation}
from
\begin{equation}
\text{uncertainty in converting }h_Y\text{ into }Y_B.
\end{equation}
These are conceptually independent systematics.

\subsection{Revisiting the baryon-isocurvature exclusion}

A second paper by Hamada, Mukaida, and Uchida reexamined an important phenomenological question: can a helical primordial field account for both the intergalactic magnetic field and the observed baryon asymmetry without violating baryon-isocurvature constraints? Their 2026 analysis finds that maximally helical configurations can be viable once the hot-electroweak treatment is updated \cite{HamadaMukaidaUchida2026Constraints}. It also identifies a possible nonhelical case, but the required cancellation during the crossover is so precise that it should be viewed as a highly tuned limit. This result does not eliminate the need for a full inhomogeneous transport calculation; rather, it shows why exclusions inferred from an older matching prescription must be revisited.

\subsection{Magnetic fields as active participants in a first-order EW transition}

Di, Bian, and Cai studied how a primordial magnetic background changes the dynamics of a first-order electroweak transition \cite{DiBianCai2026}. Their simulations found delayed transition dynamics and nontrivial Higgs structure, reminding us that a magnetic field need not be a passive byproduct.

Earlier phenomenological calculations often treated the magnetic field as an output of a predetermined transition:
\begin{equation}
\text{compute transition}\to\text{add magnetogenesis}\to\text{compute BAU}.
\end{equation}
The simulation results emphasize that for sufficiently strong fields the correct structure is instead iterative:
\begin{equation}
\text{transition}\leftrightarrow\text{magnetic field}\leftrightarrow\text{sphalerons}.
\end{equation}

\subsection{Chiral-vortical co-generation}

Vorticity provides another way to start the magnetic and baryon-asymmetry dynamics. Abbaslu and collaborators showed how vortical currents and weak sphalerons can generate hypermagnetic fields and matter asymmetries together \cite{Abbaslu2025}.

The broader significance is that ``seed-field origin'' can be addressed within anomalous hydrodynamics itself rather than outsourced to an unspecified battery. This type of co-generation calculation is a natural bridge between microscopic thermal fluctuations and macroscopic helicity.

\subsection{Thermal leptogenesis plus helical fields}

Leptogenesis and magnetic baryogenesis can interact through the plasma chemical-potential network. Recent work has shown that the sign and strength of hypermagnetic helicity may modify an asymmetry produced by thermal leptogenesis \cite{Safari2024}.

\subsection{Recent developments in the spectrum and chiral sources}
Seller and Sigl \cite{SellerSigl2026} have developed a complementary treatment of magnetic spectra from random Higgs configurations, sharpening the role of isotropy and causality in the infrared tail. The result is a follow-up to the 2025 Kibble-type analysis, not a replacement for fully dynamical finite-temperature gauge-Higgs simulations. Separately, Liu and Bian \cite{LiuBian2025} studied amplification of a bubble-wall CP source by the chiral magnetic effect. This is related to, but distinct from, the published Liu--Qin--Bian model of magnetogenesis with a CP-odd gauge operator \cite{LiuQinBian2026}; the two papers use different sources of CP violation. Neither result makes fully helical field generation an automatic consequence of a first-order phase transition.

\subsection{Reconnection changes the observational translation}

Changes in our understanding of MHD decay matter even when the particle-physics model is unchanged. In particular, reconnection-controlled decay can move a predicted field to a different location in the observational $B_0$--$\lambda_0$ plane \cite{HoskingSchekochihin2023,UchidaKamadaTashiro2025}.

\subsection{Higgs-coupling deformations and correlated field--GW signals}

A model-independent Higgs-coupling analysis by Abu-Ajamieh and Okada studies first-order electroweak transitions generated by deviations in Higgs interactions, together with their gravitational-wave spectra and primordial magnetic fields \cite{AbuAjamiehOkada2026}. This provides a complementary, bottom-up perspective on the magnetogenesis problem: collider limits on Higgs interactions restrict the transition dynamics and therefore the allowed magnetic and gravitational-wave signals. As in other phenomenological applications, a further baryogenesis prediction requires the helicity spectrum and anomalous charge evolution, not only the field amplitude.

\subsection{Applying the supercooled benchmark across three model classes}
\label{sec:new_worked_model_benchmarks}

The dynamical electroweak scenarios analyzed by Arteaga Tupia, Ghoshal and Strumia \cite{ArteagaGhoshalStrumia2025} illustrate how the same magnetic prescription can be embedded in distinct particle-physics models. Their examples contain a classically scale-invariant sector in which a singlet-like scalar develops a vacuum expectation value through radiative effects, followed by a strongly supercooled phase transition. They consider (i) a model with two extra scalar singlets, (ii) a model with a dark $SU(2)_X$ gauge interaction, and (iii) a gauged $U(1)_{B-L}$ model. For the illustrative parameter regions considered there, the reheating temperatures are approximately $T_*\simeq0.07m_{s'}$, $T_*\simeq0.12M_X$ and $T_*\simeq0.09M_{Z'}$, respectively. These are model-specific approximate relations, not universal temperature--mass conversions. Their comparison is especially useful because it makes clear that the transition can release substantial energy in a singlet sector without necessarily producing an equally large visible electromagnetic field.

\paragraph{An explicit benchmark calculation.}
Consider a strongly supercooled transition with $\beta/H_*=30$, $v_w\simeq1$ and the benchmark prescription $\epsilon=0.1$, $\kappa\simeq1$. Equation~\eqref{eq:new_bubble_peak_scale} gives $\lambda_*H_*\simeq0.1$. At $T_*=1\,\mathrm{TeV}$, the magnetic generation prescription of Eq.~\eqref{eq:new_energy_partition_ags} fixes its initial energy, Eq.~\eqref{eq:new_ags_broken_law} specifies the scale distribution, and Eqs.~\eqref{eq:new_AGS_helical_endpoints}--\eqref{eq:new_AGS_nonhelical_endpoints} propagate its characteristic amplitude and coherence length to today. This reproduces a specified phenomenological input prescription rather than a dynamical MHD calculation; Sec.~\ref{sec:new_worked_MHD_evolution} shows why the continuously turbulent extrapolation is not self-consistently established for these inputs.

\begin{figure}[t]
\centering
\includegraphics[width=0.78\linewidth]{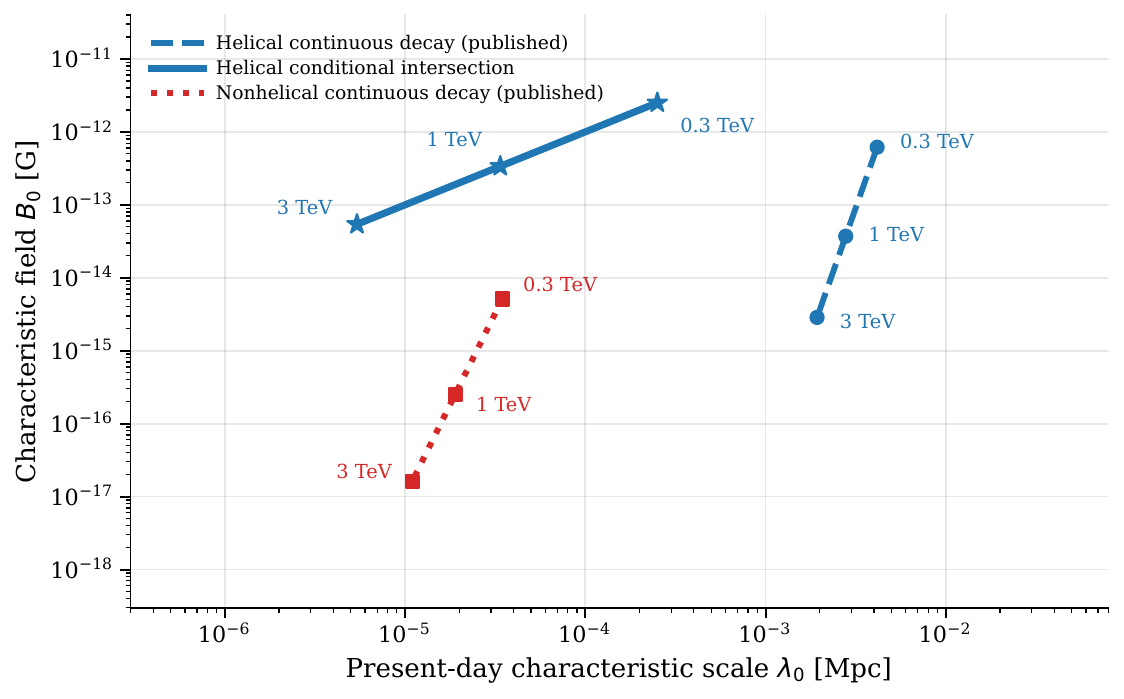}
\caption{Characteristic peak field strength and coherence length at three illustrative reheating temperatures, $T_*=0.3$, $1$ and $3\,\mathrm{TeV}$, for $\lambda_*H_*=0.1$ and the magnetic efficiency used in Ref.~\cite{ArteagaGhoshalStrumia2025}. Dashed branches use the source paper's continuous-helical/nonhelical prescriptions; the additional helical branch uses the conditional helicity-preserving processed-eddy intersection with $C_{\rm eddy}=10^{-8}\,\mathrm{G/Mpc}$. Neither is a solved MHD trajectory. This is not a parameter-space scan or new observational fit; in particular, alternative reconnection-controlled evolution could change the nonhelical branch.}
\label{fig:new_model_temperature_scan}
\end{figure}

For fixed $\lambda_*H_*=0.1$ and the same efficiency assumptions, the temperature dependence of this exercise is particularly transparent:
\begin{equation}
 B_{0,\rm hel}^{\rm peak}\propto T_*^{-7/3},\qquad
 B_{0,\rm nonhel}^{\rm peak}\propto T_*^{-5/2},\qquad
 \lambda_{0,\rm hel}\propto T_*^{-1/3},\qquad
 \lambda_{0,\rm nonhel}\propto T_*^{-1/2}.
 \label{eq:new_temperature_endpoints}
\end{equation}
We contrast the published continuous-decay endpoints with the conditional processed-eddy intersection (which preserves the same comoving helicity invariant) at three reheating temperatures in Fig.~\ref{fig:new_model_temperature_scan}; the numerical values follow directly from the cited approximate formulas and can be regenerated with the supplied plotting script. The curves show why increasing the transition scale generally makes it more difficult to obtain a sufficiently strong present-day visible magnetic field in this class of models. They do not constitute a new scan of the scalar masses or a demonstration that those model points satisfy collider, dark-matter and phase-transition-completion constraints.

It is instructive to compare this supercooled construction with other results rather than interpreting it as the unique successful mechanism. Ellis and collaborators \cite{EllisEtAl2019PMF} compared magnetic fields from bubble collisions and plasma turbulence in $|H|^6$ and $B-L$ extensions of the Standard Model and found that their relative viability depends on the microscopic model and on the chosen helicity assumption. Direct gauge-field simulations by Zhang, Vachaspati and Ferrer \cite{ZhangVachaspatiFerrer2019} provide a different route to the initial spectrum and efficiency. The recent Kibble-type study by Vachaspati and Brandenburg \cite{VachaspatiBrandenburg2025} instead emphasizes the causal infrared spectrum and the impact of helicity on the late-time result. Finally, the reconnection-controlled evolution of Hosking and Schekochihin \cite{HoskingSchekochihin2023} challenges the view that vanishing \emph{net} helicity automatically leads to a negligibly weak relic. These approaches address different parts of the calculation and should not be combined into one prediction without checking that their definitions and physical assumptions agree.

The joint $\YB$--PMF--GW program remains an open computation for these benchmark models: the inputs used above do not determine the sign or amplitude of the initial helicity spectrum, the anomaly-driven baryon source through the electroweak crossover, or the unequal-time gravitational-wave stress. It would therefore be incorrect to overlay a claimed successful baryon-asymmetry band on these figures without solving those additional equations.

\subsection{A synthesis of the frontier}

The recent literature highlights five questions that any comprehensive electroweak magnetism calculation should confront:
\begin{enumerate}
\item Can a particle model predict $f_{h,*}$ rather than assume it?
\item Can anomalous transport and sphalerons be evolved while the field itself changes dynamically?
\item Can the electroweak crossover be treated with a gauge-invariant, nonperturbatively controlled magnetic-flux prescription?
\item Can present-day PMF predictions be propagated with quantified MHD-theory uncertainty?
\item Can the same benchmark simultaneously predict $Y_B$, $B_0$, $\lambda_0$, $\Omega_{\rm GW}$ and $\Pi_{\rm GW}$?
\end{enumerate}

These questions define a more stringent standard than most existing PMF or FOPT studies. They also provide a concrete roadmap for turning the field into a precision multi-messenger subject.

\section{A unified BAU--PMF--GW framework}
\label{sec:unified}

The previous sections show that the same physics is often described using different parameters in the phase-transition, magnetic-field, baryogenesis, and gravitational-wave literatures. A common phenomenological framework is needed to compare predictions and track the approximations used at each stage.

\subsection{Why a common parameter language is needed}

Phase-transition studies quote $\alpha$, $\beta/H_*$, and wall velocity; magnetic-field studies quote field strength, coherence length, spectral shape, and helicity; baryogenesis studies quote reaction rates and chemical potentials. These quantities describe different stages of the same problem. To compare models, we need a common set of inputs and outputs.

A useful standardized input vector at generation is:
\begin{equation}
\mathcal P_*
=\left\{
T_*,g_*,\alpha,\frac{\beta}{H_*},v_w,R_*H_*,
\epsilon_{K,*},\epsilon_{B,*},\lambda_*H_*,n_B,f_{h,*}
\right\}.
\label{eq:Pstar}
\end{equation}
The transition quantities need not all exist for non-FOPT magnetogenesis, in which case the relevant subset is used. The magnetic spectra should be given whenever available.

The electroweak-anomaly stage adds
\begin{equation}
\mathcal P_{\rm EW}
=\left\{
\mu_i(T),\Gamma_{\rm sph}(T),\sigma_Y(T),\theta_{\rm flux}(T),
S_Y(T,k)
\right\},
\end{equation}
where $\theta_{\rm flux}$ stands for the gauge-invariant magnetic-flux interpolation rather than necessarily the vacuum weak mixing angle.

The late-time observables are
\begin{equation}
\mathcal O_0=
\left\{Y_B,\,P_{\delta B}(k),\,B_0,\lambda_0,f_{h,0},
\Omega_{\rm GW}(f),\Pi_{\rm GW}(f)\right\}.
\label{eq:observables}
\end{equation}
Here $P_{\delta B}$ denotes the baryon-isocurvature spectrum, which should be regarded as a primary prediction rather than an optional afterthought.

\subsection{The full pipeline}

We can organize a calculation into five steps. Each step connects a well-defined set of physical quantities to the next, so any missing microscopic calculation or uncertain transfer function becomes explicit:
\begin{enumerate}
\item \textbf{Microphysics $\to$ transition/plasma:} compute the finite-temperature potential or real-time symmetry-breaking dynamics, nucleation/percolation, wall behavior and CP-odd sources.
\item \textbf{Transition/plasma $\to$ magnetic initial conditions:} determine the magnetic energy, spectrum, coherence length and helicity fraction.
\item \textbf{Magnetic initial conditions $\to$ BAU:} solve anomalous transport and sphaleron kinetics through the electroweak conversion interval, including magnetic backreaction if relevant.
\item \textbf{Initial PMF $\to$ present PMF:} evolve the spectra through turbulent, viscous and reconnection regimes with quantified theory systematics.
\item \textbf{Stress history $\to$ GWs:} integrate the unequal-time tensor source, including acoustic, turbulent and magnetic contributions and polarization.
\end{enumerate}

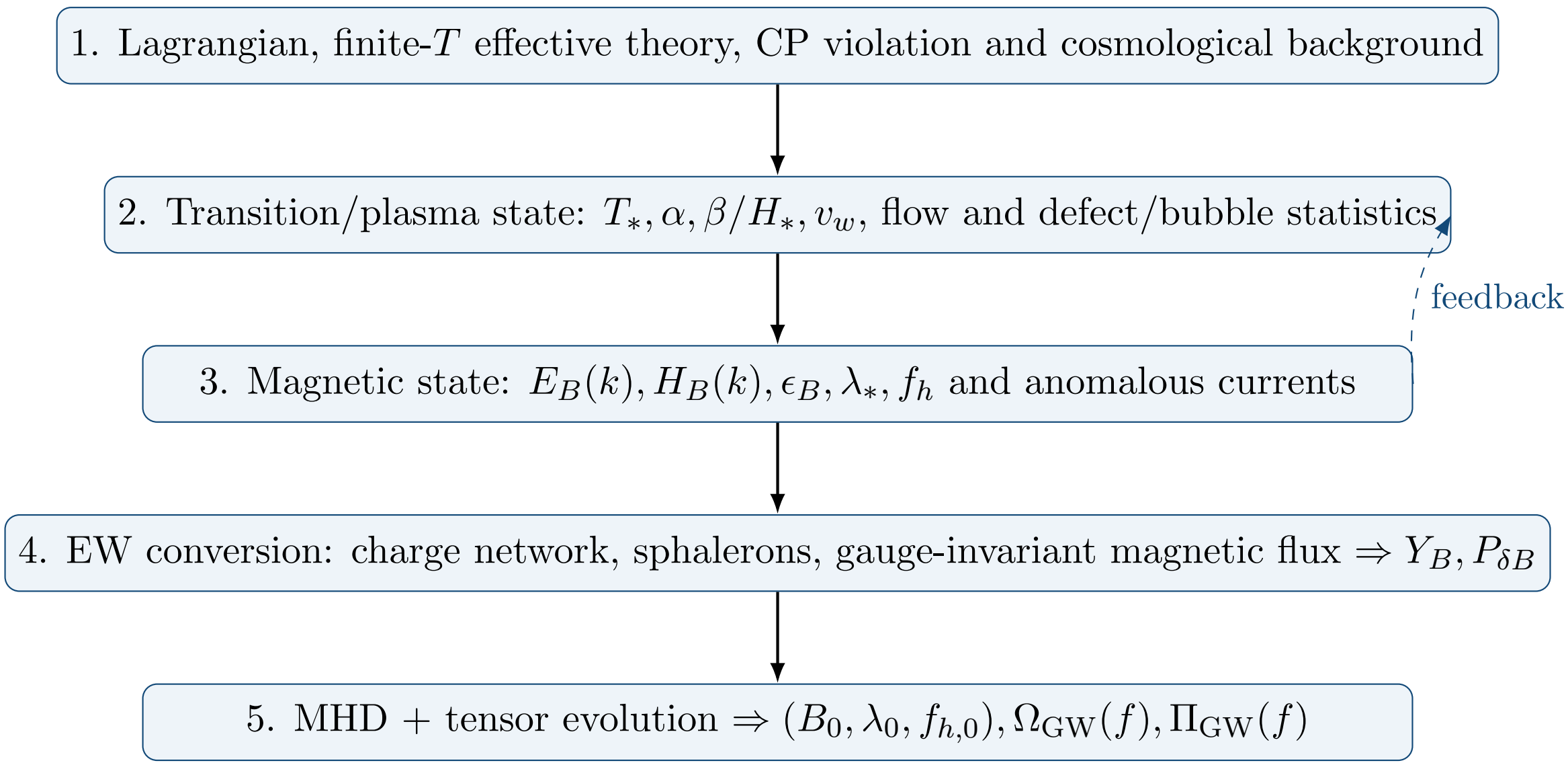
\begin{figure}[H]
\centering
\begin{tikzpicture}[node distance=0.9cm, every node/.style={align=center},box/.style={draw=accent,rounded corners,fill=soft,minimum width=12.5cm,minimum height=0.75cm},arr/.style={-{Latex[length=2mm]},thick}]
\node[box] (a) {1. Lagrangian, finite-$T$ effective theory, CP violation and cosmological background};
\node[box,below=of a] (b) {2. Transition/plasma state: $T_*,\alpha,\beta/H_*,v_w$, flow and defect/bubble statistics};
\node[box,below=of b] (c) {3. Magnetic state: $E_B(k),H_B(k),\epsilon_B,\lambda_*,f_h$ and anomalous currents};
\node[box,below=of c] (d) {4. EW conversion: charge network, sphalerons, gauge-invariant magnetic flux $\Rightarrow Y_B,P_{\delta B}$};
\node[box,below=of d] (e) {5. MHD + tensor evolution $\Rightarrow (B_0,\lambda_0,f_{h,0}),\Omega_{\rm GW}(f),\Pi_{\rm GW}(f)$};
\draw[arr] (a)--(b);\draw[arr] (b)--(c);\draw[arr] (c)--(d);\draw[arr] (d)--(e);
\draw[-{Latex[length=2mm]},dashed,accent,bend left=18] (c.east) to node[right,font=\small]{feedback} (b.east);
\end{tikzpicture}
\caption{A unified computational pipeline. The dashed arrow emphasizes that a sufficiently strong magnetic field can modify the transition and sphaleron dynamics, so the chain can require iteration.}
\label{fig:pipeline}
\end{figure}

The pipeline is depicted schematically in \cref{fig:pipeline}.

\subsection{A minimal benchmark table}

The following table is a practical template for comparing models. It separates parameters computed from the Lagrangian from those inserted phenomenologically, making missing pieces immediately visible.

\begin{longtable}{@{}p{0.26\textwidth}p{0.21\textwidth}p{0.45\textwidth}@{}}
\caption{Recommended benchmark data structure.}\label{tab:benchmark}\\
\toprule
Quantity & Symbol & Why it matters \\
\midrule
\endfirsthead
\toprule
Quantity & Symbol & Why it matters \\
\midrule
\endhead
Transition temperature & $T_*$ & redshift, horizon scale, GW frequency \\
Relativistic d.o.f. & $g_*,g_{*s}$ & expansion and entropy conversion \\
Transition strength & $\alpha$ & fluid/GW energy budget \\
Inverse duration & $\beta/H_*$ & bubble separation and GW peak \\
Wall velocity & $v_w$ & fluid shell, bubble scale, transport \\
Mean bubble scale & $R_*H_*$ & magnetic and GW correlation scale \\
Magnetic energy fraction & $\epsilon_{B,*}$ & initial PMF amplitude \\
Kinetic energy fraction & $\epsilon_{K,*}$ & turbulence and dynamo budget \\
Integral scale & $\lambda_*H_*$ & causal seed scale \\
IR magnetic index & $n_B$ & large-scale observational mapping \\
Helicity fraction & $f_{h,*}$ & BAU, inverse transfer, GW polarization \\
Baryon yield & $Y_B$ & mean BAU \\
Baryon isocurvature & $P_{\delta B}$ & BBN/CMB viability \\
Present field & $B_0$ & IGMF/CMB comparison \\
Present coherence & $\lambda_0$ & gamma-ray and Faraday observables \\
GW spectrum & $\Omega_{\rm GW}(f)$ & detector prediction \\
GW polarization & $\Pi_{\rm GW}(f)$ & parity/helicity observable \\
\bottomrule
\end{longtable}

\subsection{Consistency conditions}

Before interpreting a correlated baryon, magnetic-field, and gravitational-wave signal, one should check the underlying energy and time scales. Several consistency conditions are particularly useful.

\paragraph{Energy budget.}
At generation,
\begin{equation}
\epsilon_{B,*}+\epsilon_{K,*}+\epsilon_{\phi,*}\lesssim\frac{\alpha}{1+\alpha}
\end{equation}
for transition-sourced components, modulo the precise definition of $\alpha$ and any pre-existing field. One cannot independently maximize magnetic, sound-wave and scalar-collision efficiencies without checking the shared energy budget.

\paragraph{Causality.}
For post-inflationary production,
\begin{equation}
\lambda_*\lesssim H_*^{-1},
\end{equation}
and the infrared spectrum is constrained by analyticity/causality. A present Mpc coherence length requires nonlinear transfer, a sufficiently low generation temperature, supercooling, or inflationary origin.

\paragraph{Helicity.}
The realizability bound gives
\begin{equation}
|h_B|\le2\xi_B\rho_B.
\end{equation}
A model cannot choose an arbitrary $Y_B\propto h_B$ independently of its magnetic energy and coherence scale.

\paragraph{BAU and isocurvature.}
It is necessary to satisfy both
\begin{equation}
\langle Y_B\rangle=Y_B^{\rm obs}
\end{equation}
and the small-scale fluctuation limits from BBN/CMB. Increasing the PMF amplitude can worsen the second constraint even if it helps the first.

\paragraph{GW energy conservation.}
Acoustic, turbulent and magnetic GW templates must be sourced by a common stress-energy budget. Adding their standard peak amplitudes with independent maximal efficiency factors can overcount the available energy.

\subsection{A model taxonomy}

\begin{table}[H]
\centering
\small
\caption{A physically motivated taxonomy of electroweak cosmic-magnetism models. ``FOPT'' is stated in the source/mechanism column when it is essential.}
\begin{tabularx}{\textwidth}{@{}p{0.20\textwidth}p{0.24\textwidth}p{0.25\textwidth}X@{}}
\toprule
Class & Magnetic-field origin & BAU/helicity structure & Distinctive test \\
\midrule
Crossover helicity decay & pre-existing hypermagnetic field & helicity essential; BAU from $U(1)_Y$ anomaly & BAU--PMF relation and baryon isocurvature \\
Chiral instability & fermionic chiral asymmetry & helicity generated dynamically; anomaly and sphalerons redistribute charge & chiral-charge/field coevolution \\
EW bubble magnetogenesis & FOPT bubble currents and turbulence & helicity model dependent; BAU optional through EWBG/anomaly & mHz GW plus PMF \\
CP-helical FOPT & CP-odd bubble or defect dynamics & helicity predicted; BAU from anomaly and/or EWBG & BAU + polarized GW + PMF \\
Inflationary hypermagnetism & axion/gauge production during inflation & usually strongly helical; BAU from EW conversion & superhorizon spectrum plus BAU \\
Supercooled BSM transition & large FOPT bubbles and plasma flows & BAU model dependent; helicity often a separate assumption & PMF + GW + possible PBH \\
Leptogenesis + PMF & external or high-scale hypermagnetic field & RH-neutrino source modified by anomaly/spectators & altered leptogenesis parameter space \\
\bottomrule
\end{tabularx}
\end{table}


Instead of hiding all theoretical uncertainty in one normalization factor, it is more informative to vary four separate parts of the calculation:
\begin{equation}
\sigma_{\rm theory}=\{\sigma_{\rm mag},\sigma_{\rm EW},\sigma_{\rm MHD},\sigma_{\rm GW}\}.
\end{equation}
Here $\sigma_{\rm mag}$ represents uncertainty in the initial magnetic efficiency and helicity; $\sigma_{\rm EW}$ describes the hypermagnetic-to-BAU conversion; $\sigma_{\rm MHD}$ covers the evolution to today; and $\sigma_{\rm GW}$ covers the source model. This decomposition makes it obvious which new calculation would most improve a given prediction.

\subsection{A public-code opportunity}

The next practical step would be a public framework that propagates a common electroweak benchmark through alternative baryogenesis and magnetic-evolution prescriptions. It need not solve the microscopic dynamics from first principles; the immediate value would be transparent assumptions and consistent comparisons.
\begin{enumerate}
\item \texttt{transition}: imports $T_*,\alpha,\beta/H_*,v_w$ and spectral seeds;
\item \texttt{anomaly}: chemical-potential and sphaleron network;
\item \texttt{ew\_flux}: alternative crossover prescriptions including a Hamada--Mukaida--Uchida-inspired module;
\item \texttt{mhd}: helical/nonhelical conventional and reconnection-driven evolution;
\item \texttt{gw}: acoustic/turbulent/magnetic spectra and polarization;
\item \texttt{constraints}: CMB, BBN/isocurvature, spectral distortion and gamma-ray likelihood interfaces.
\end{enumerate}
Such a framework would make the review itself more useful and citable by turning literature conventions into reproducible calculations.

A useful comparison between different models should be based on their physical evolution through a common set of state variables, rather than on whichever final observable a particular paper emphasizes. Expressing the models in a common framework makes assumptions about initial helicity, magnetic efficiency, electroweak conversion, and subsequent MHD evolution explicit, so that their impact can be tested separately.

The calculation in Sec.~\ref{sec:new_worked_MHD_evolution} deliberately stops short of predicting the full vector $\mathcal O_0$: it fixes an assumed magnetic spectrum and compares alternative propagation prescriptions, but does not solve anomaly-driven baryogenesis or the gravitational-wave source correlators. A fully joint likelihood cannot therefore be inferred from the illustrative magnetic curves.

\section{Open problems and a research roadmap}
\label{sec:open}

The qualitative connections between electroweak dynamics, primordial fields, baryogenesis, and gravitational waves are well motivated. The main unresolved questions concern the quantitative matching between these stages. We organize the most useful open problems by the physics that is missing and by the calculations that would resolve it.

The literature now contains credible mechanisms for every qualitative connection in \cref{fig:network}. The high-value open problems are therefore mostly problems of \emph{matching, control and correlated prediction}. We rank them here by scientific leverage rather than by ease.

\subsection{Nonperturbative magnetic-flux conversion through the electroweak crossover}

The most urgent theoretical issue is to determine how hypermagnetic flux becomes electromagnetic flux through the electroweak crossover. The recent gauge-invariant analysis revealed effects missing from older interpolation models \cite{HamadaMukaidaUchida2026}. What remains is a quantitative nonperturbative calculation of the conversion and its impact on baryon number.

A decisive calculation would combine dimensionally reduced electroweak theory with controlled background magnetic flux and determine the relevant transport/relaxation coefficients across the crossover. The output should be a replacement for the phenomenological conversion factor used in older BAU estimators.

\subsection{Initial helicity from realistic first-order transitions}

It is common to assume either $f_h=0$ or $f_h=1$ at the end of a phase transition. These limits are useful, but they do not predict the helicity of a realistic transition. A calculation that follows the CP-violating source and the subsequent gauge-field dynamics is needed to determine the helicity spectrum \cite{DomckeEmaPatel2025,LiuQinBian2026}.

The target is a stochastic spectrum $H_B(k)$ generated by a realistic 3D transition, not merely an integrated helicity estimate from a two-bubble collision.

\subsection{Coupled gauge-Higgs--anomalous-MHD dynamics}

Existing simulations typically separate the microscopic electroweak fields from large-scale MHD. That separation can fail when magnetic backreaction modifies the Higgs configuration or sphalerons, and when chiral currents modify the field \cite{AnnalaRummukainen2023,DiBianCai2026}. The challenge is to match these scales without double counting.

This is technically difficult but potentially transformative because it would turn a chain of phenomenological interfaces into a controlled effective description.

\subsection{Revisit baryon isocurvature with modern crossover and MHD physics}

Magnetic baryogenesis must produce the right mean asymmetry without generating unacceptable spatial fluctuations. Since the original isocurvature analyses, both electroweak-flux matching and MHD decay modeling have changed \cite{KamadaUchidaYokoyama2021,HamadaMukaidaUchida2026,HoskingSchekochihin2023}. A new spectral analysis is therefore timely. The 2026 reanalysis of Hamada, Mukaida, and Uchida already finds that maximally helical fields may evade the earlier exclusion, while a nonhelical loophole requires exceptional fine tuning \cite{HamadaMukaidaUchida2026Constraints}.

This is a particularly attractive project because it can either strengthen a widely quoted exclusion or reopen motivated parameter space; either outcome is valuable.

\subsection{A common PMF--BAU--GW likelihood}

Comparing a model with several independent shaded regions is useful for orientation, but it is not a joint statistical test. A common likelihood would propagate correlated model parameters through baryon, magnetic, and gravitational-wave observables while varying the uncertain magnetic-evolution assumptions.

Even before a GW detection, upper limits from LISA-pathfinder-era projections or existing PTA/LIGO bands can be incorporated where relevant. Once LISA data arrive, the benefit of a common likelihood will become much larger.

\subsection{Circular polarization as a baryogenesis diagnostic}

If the same helical field helps generate baryon number and circularly polarized gravitational waves, it may be possible to correlate the two observables \cite{KahniashviliGogoberidzeRatra2005,Kahniashvili2021}. Such a relation is not automatic: the sign conventions, transport equations, and magnetic-to-tensor transfer function must all be established.

A robust sign correlation would be unusually powerful because the observed BAU has a known sign. It would turn GW polarization into a direct test of matter--antimatter origin rather than merely of a phase transition.

\subsection{The origin of gamma-ray-inferred void fields}

Even a firm detection of a weak intergalactic magnetic field would not by itself establish a primordial origin. Astrophysical processes can also magnetize low-density regions. A strong electroweak prediction should therefore include the field's spatial distribution and spectrum, not just its mean strength.

A primordial electroweak model should therefore predict more than a point in $(B_0,\lambda_0)$. It should predict spectral and parity information that can distinguish it from late-time astrophysical contamination.

\subsection{Plasma-instability control of cascade constraints}

Gamma-ray bounds rely on understanding how electron--positron beams lose energy in the intergalactic medium. If plasma instabilities compete efficiently with inverse-Compton cooling, a weak cascade may not require magnetic deflection. Better kinetic calculations are needed before treating all cascade lower bounds as universal.

\subsection{Supercooled transitions and reheating-aware MHD}

Supercooled transitions can create energetic sources, but their magnetic field cannot always be evolved with radiation-dominated formulas. Vacuum domination, rapid reheating, entropy production, and a temporarily dilute plasma affect both the initial field and the onset of MHD.

\subsection{Spectral rather than monochromatic baryogenesis}

A single field amplitude and correlation length may not capture the diffusion, instability, and anomaly source of a broad magnetic spectrum. A more complete calculation should evolve the magnetic energy and helicity spectra together with the chemical potentials.

%
%
%

\section{Conclusions}

Primordial magnetism provides a possible connection between three major problems in early-Universe physics: the origin of cosmic magnetic fields, the baryon asymmetry, and the production of gravitational waves. The connection is particularly interesting near the electroweak scale. Above electroweak symmetry breaking, the plasma contains hypermagnetic fields, while the anomaly allows changes in their helicity to act as sources of fermion number. At the same time, sphalerons violate $B+L$, and a first-order transition in a BSM theory can generate the fluid motion required for magnetogenesis and gravitational-wave production.

Several pieces of the physics are well understood. For the observed Higgs mass the Standard Model electroweak transition is a crossover. Chiral chemical potentials can amplify one magnetic helicity through the chiral plasma instability, and changing hypermagnetic helicity can source baryon number. The same fields may affect the Higgs background and the sphaleron energy barrier. During the subsequent cosmological evolution, magnetic helicity strongly influences the relation between field strength and correlation length. Finally, turbulent motion and magnetic anisotropic stresses can produce gravitational waves, with helical sources potentially generating circular polarization.

The main challenge is to turn these established connections into quantitative predictions. In particular, the conversion between hypermagnetic and electromagnetic flux through the crossover remains uncertain at a level relevant for baryogenesis. Recent gauge-invariant work shows that some ingredients of earlier phenomenological treatments need to be reconsidered. This also affects the interpretation of baryon-isocurvature exclusions: updated calculations now find viable maximally helical scenarios in regions previously considered disfavored, although nonhelical loopholes may require extreme tuning. The late-time evolution of primordial fields introduces a separate uncertainty because turbulent decay, reconnection, and viscous epochs can give different present-day amplitudes and coherence lengths.

The most informative output of a model is therefore not a single magnetic-field amplitude or gravitational-wave peak. It is a set of correlated observables,
\begin{equation}
\left\{Y_B,\;P_{\delta B}(k),\;B_0,\lambda_0,f_{h,0},\;\Omega_{\rm GW}(f),\Pi_{\rm GW}(f)\right\},
\end{equation}
where $P_{\delta B}$ describes spatial fluctuations in the baryon abundance and $\Pi_{\rm GW}$ denotes the gravitational-wave polarization fraction. A model that reproduces the mean baryon asymmetry may still be excluded by isocurvature. Likewise, a model that produces a strong magnetic field may predict too small a correlation length after cosmological evolution. These tests should be performed using the same initial spectrum and thermal history.

The natural theoretical program is to connect the microscopic model to the late-time observables in one calculation:
\begin{equation}
\begin{aligned}
\text{microphysics}&\longrightarrow\text{phase-transition and plasma dynamics}\\
&\longrightarrow\text{magnetic spectrum and helicity}\\
&\longrightarrow\text{anomalous charge evolution and sphalerons}\\
&\longrightarrow\text{cosmological MHD}\\
&\longrightarrow\text{baryon, magnetic-field, and GW observables}.
\end{aligned}
\end{equation}
Not every part of this chain can yet be calculated from first principles. Nevertheless, treating the uncertain stages explicitly, rather than folding them into a single efficiency factor, will make results easier to compare and improve as the theory develops. The opportunity is to use magnetic fields, baryon number, and gravitational waves as complementary measurements of the same early-Universe dynamics.

The revised worked example exposes a specific weakness of magnetic predictions that extend a self-similar turbulent-decay law from generation to recombination without checking the Alfv\'en crossing time. The illustrative helical and nonhelical endpoints should be regarded as conditional until the intervening plasma dynamics is solved.

\section*{Acknowledgments}
The author acknowledges the use of OpenAI's ChatGPT for assistance with literature organization, manuscript drafting, and language editing. The author assumes full responsibility for the scientific content, interpretations, and references presented in this review.

\appendix

\section{Notation, units and conversion conventions}
\label{app:notation}

\subsection{Cosmology}

We use $a_0=1$ for the present scale factor. Entropy conservation in the standard thermal history gives
\begin{equation}
aTg_{*s}^{1/3}=\text{const},
\end{equation}
so a physical length at generation redshifts to
\begin{equation}
\lambda_0=\lambda_*\frac{a_0}{a_*}
=\lambda_*\frac{T_*}{T_0}\left(\frac{g_{*s,*}}{g_{*s,0}}\right)^{1/3}
\end{equation}
only in the absence of nonlinear transfer. Likewise a passive magnetic field obeys $B_0=B_*(a_*/a_0)^2$.

During radiation domination,
\begin{equation}
H(T)=1.66\sqrt{g_*(T)}\frac{T^2}{M_{\rm Pl}},
\end{equation}
where $M_{\rm Pl}=1.22\times10^{19}\,\mathrm{GeV}$ is the unreduced Planck mass. Authors using the reduced Planck mass $\bar M_{\rm Pl}=2.435\times10^{18}\,\mathrm{GeV}$ write an equivalent formula with a different prefactor.

\subsection{Magnetic units}

A useful conversion is
\begin{equation}
1\,\mathrm{G}\simeq1.95\times10^{-20}\,\mathrm{GeV}^2
\end{equation}
in natural units. This number varies in appearance according to rationalized versus Gaussian electromagnetic conventions, so numerical calculations should keep one convention throughout.

The radiation-equivalent field at temperature $T$ is defined by $B_{\rm rad}^2/2=\rho_{\rm rad}$ in rationalized units,
\begin{equation}
B_{\rm rad}=\sqrt{\frac{\pi^2g_*}{15}}\,T^2.
\end{equation}
Then $B_*=\sqrt{\epsilon_{B,*}}B_{\rm rad}$.

\subsection{Baryon abundance}

We distinguish
\begin{equation}
Y_B=\frac{n_B-n_{\bar B}}{s}
\end{equation}
from
\begin{equation}
\eta_B=\frac{n_B-n_{\bar B}}{n_\gamma}.
\end{equation}
After electron--positron annihilation,
\begin{equation}
\eta_B\simeq7.04Y_B.
\end{equation}
Confusion between these two conventions produces factors of about seven in numerical comparisons.

\subsection{Phase-transition quantities}

The transition-strength parameter $\alpha$ has several definitions in the literature: ratio of released vacuum/trace-anomaly energy to radiation, latent heat normalized to radiation, or pseudotrace versions. Likewise $T_*$ can refer to nucleation, percolation, completion or reheating temperature. A quantitative PMF/GW comparison must state the definitions.

The mean bubble separation $R_*$ is often related to $\beta^{-1}$ by
\begin{equation}
R_*\sim \mathcal C\frac{v_w}{\beta},
\end{equation}
where $\mathcal C$ is convention dependent and can differ from unity by factors of a few.

\subsection{Acronyms}

\begin{longtable}{@{}ll@{}}
\toprule
Acronym & Meaning \\
\midrule
BAU & baryon asymmetry of the Universe \\
PMF & primordial magnetic field \\
IGMF & intergalactic magnetic field \\
EWPT & electroweak phase transition/crossover \\
FOPT & first-order phase transition \\
CME & chiral magnetic effect \\
CVE & chiral vortical effect \\
CPI & chiral plasma instability \\
MHD & magnetohydrodynamics \\
SGWB & stochastic gravitational-wave background \\
EBL & extragalactic background light \\
CMB & cosmic microwave background \\
BBN & Big-Bang nucleosynthesis \\
ALP & axion-like particle \\
SMEFT & Standard Model effective field theory \\
\bottomrule
\end{longtable}

\section{Useful derivations}
\label{app:derivations}

\subsection{Helicity evolution}

Let $\bm B=\nabla\times\bm A$ and choose a periodic volume. The helicity is
\begin{equation}
\helicity=\int d^3x\,\bm A\cdot\bm B.
\end{equation}
Using $\bm E=-\partial_t\bm A-\nabla\Phi$ and Faraday's law,
\begin{align}
\frac{d\helicity}{dt}
&=\int d^3x\left(\partial_t\bm A\cdot\bm B+\bm A\cdot\partial_t\bm B\right)\\
&=-\int d^3x\left[(\bm E+\nabla\Phi)\cdot\bm B+\bm A\cdot\nabla\times\bm E\right].
\end{align}
Integrating by parts and using $\nabla\cdot\bm B=0$, surface terms vanish and one obtains
\begin{equation}
\boxed{\frac{d\helicity}{dt}=-2\int d^3x\,\bm E\cdot\bm B.}
\end{equation}
This is the direct bridge to the Abelian anomaly because $F\widetilde F\propto\bm E\cdot\bm B$.

\subsection{Realizability bound}

Decompose a transverse magnetic field into helicity eigenmodes,
\begin{equation}
\bm B(\bm k)=B_+(\bm k)\bm e_+(\bm k)+B_-(\bm k)\bm e_-(\bm k),
\end{equation}
with $i\bm k\times\bm e_\pm=\pm k\bm e_\pm$. Magnetic energy at fixed $k$ is proportional to $|B_+|^2+|B_-|^2$, while helicity is proportional to $(|B_+|^2-|B_-|^2)/k$. Therefore
\begin{equation}
|H_B(k)|\le\frac{2E_B(k)}{k},
\end{equation}
with equality when only one polarization is populated.

\subsection{CPI growth rate}

For $\bm v=0$ and constant $\xi_{\rm CME}$, the anomalous induction equation is
\begin{equation}
\partial_t\bm B=\eta\nabla^2\bm B+\eta\xi_{\rm CME}\nabla\times\bm B.
\end{equation}
On helicity eigenmodes,
\begin{equation}
\partial_tB_\pm=-\eta(k^2\mp\xi_{\rm CME}k)B_\pm.
\end{equation}
The unstable mode has
\begin{equation}
\gamma(k)=\eta(k\xi_{\rm CME}-k^2),
\end{equation}
maximized at
\begin{equation}
k_{\rm max}=\frac{\xi_{\rm CME}}{2},\qquad
\gamma_{\rm max}=\frac{\eta\xi_{\rm CME}^2}{4}.
\end{equation}
This derivation neglects backreaction, velocity and expansion but makes the scale selection transparent.

\subsection{Why helical decay drives scale growth}

For a maximally helical field concentrated near $k_I\sim\xi_B^{-1}$,
\begin{equation}
h_B\sim2\rho_B\xi_B.
\end{equation}
If $h_B$ is conserved and dissipative processes reduce $\rho_B$, then
\begin{equation}
\xi_B\propto\rho_B^{-1}.
\end{equation}
Hence the characteristic scale must increase. Determining its time dependence requires an additional timescale relation (eddy turnover, reconnection, etc.); helicity conservation alone fixes only the trajectory in $(\rho_B,\xi_B)$ space.

\subsection{Baryon source from helicity change}

For a fermionic current with anomaly
\begin{equation}
\partial_\mu j_f^\mu=C_f\frac{g'^2}{16\pi^2}Y_{\mu\nu}\widetilde Y^{\mu\nu},
\end{equation}
and $Y_{\mu\nu}\widetilde Y^{\mu\nu}=-4\bm E_Y\cdot\bm B_Y$ in a standard convention,
\begin{equation}
\dot n_f+3Hn_f=-C_f\frac{g'^2}{4\pi^2}\langle\bm E_Y\cdot\bm B_Y\rangle.
\end{equation}
Using helicity evolution,
\begin{equation}
\dot h_Y=-2\langle\bm E_Y\cdot\bm B_Y\rangle,
\end{equation}
we obtain
\begin{equation}
\dot n_f+3Hn_f=C_f\frac{g'^2}{8\pi^2}\dot h_Y.
\end{equation}
The exact coefficient relating the sum of Standard Model species to $B$ depends on hypercharge assignments, multiplicities and spectator equilibrium. This is why the full chemical-potential network is required for a reliable numerical coefficient.

\section{Scaling relations and order-of-magnitude estimates}
\label{app:scalings}

\subsection{Electroweak horizon scale}

At $T_*=100\,\mathrm{GeV}$ with $g_*\simeq106.75$,
\begin{equation}
H_*\sim 10^{-14}\,\mathrm{GeV}\sim10^{10}\,\mathrm{s}^{-1},
\end{equation}
so the physical Hubble radius is of order centimeters. Redshift alone converts this into a present comoving scale vastly below a megaparsec. This immediately explains why inverse transfer is central to phase-transition magnetogenesis.

\subsection{Characteristic GW frequency}

A source with characteristic physical frequency $f_*$ redshifts to
\begin{equation}
f_0=\frac{a_*}{a_0}f_*.
\end{equation}
Using entropy conservation gives the familiar estimate
\begin{equation}
f_0\sim1.65\times10^{-5}\,\mathrm{Hz}
\left(\frac{f_*}{H_*}\right)
\left(\frac{T_*}{100\,\mathrm{GeV}}\right)
\left(\frac{g_*}{100}\right)^{1/6},
\end{equation}
up to source-dependent normalization. Bubble/flow scales with $f_*/H_*\sim\beta/H_*$ therefore naturally give mHz frequencies for electroweak transitions.

\subsection{Flux freezing}

If a field evolves passively,
\begin{equation}
B_0=B_*\left(\frac{a_*}{a_0}\right)^2,
\qquad
\lambda_0=\lambda_*\frac{a_0}{a_*}.
\end{equation}
The combination $B^2\lambda^4$ is then fixed by expansion alone, while helicity density redshifts with its appropriate power. Nonlinear MHD changes these relations in comoving variables.

\subsection{Helicity line}

For approximately conserved comoving helicity,
\begin{equation}
B_c^2\lambda_c\simeq\text{const},
\end{equation}
therefore
\begin{equation}
\log B_c=-\frac12\log\lambda_c+\text{const}.
\end{equation}
This is the origin of the slope $-1/2$ commonly shown for maximally helical evolution in $B$--$\lambda$ diagrams.

\subsection{Causal integral scale}

For a transition characterized by $\beta/H_*$ and wall velocity $v_w$,
\begin{equation}
\lambda_*H_*\sim\mathcal O(1)\frac{v_w}{\beta/H_*}.
\end{equation}
Thus $\beta/H_*=100$ naturally gives an initial integral scale at the percent-of-horizon level. Supercooling or unusually slow transitions can increase this substantially.

\subsection{Magnetic energy fraction}

Using $\rho_B=B^2/2$ and $\rho_{\rm rad}=\pi^2g_*T^4/30$,
\begin{equation}
B_*\simeq T_*^2\sqrt{\frac{\pi^2g_*}{15}\epsilon_{B,*}}.
\end{equation}
Hence enormous early-Universe field strengths in Gauss need not correspond to dynamically dominant energy fractions. Quoting $B_*\sim10^{22}$--$10^{24}\,\mathrm{G}$ without $\epsilon_B$ can be visually misleading.

\section{Chronological literature map}
\label{app:litmap}

This appendix provides a compact map of the papers that changed the conceptual or quantitative status of the subject. It is intentionally selective; the main bibliography contains additional model-specific and observational references.

\begin{longtable}{@{}p{0.11\textwidth}p{0.29\textwidth}p{0.52\textwidth}@{}}
\caption{Selected milestones in electroweak cosmic magnetism.}\label{tab:timelinepapers}\\
\toprule
Year & Reference & Main contribution \\
\midrule
\endfirsthead
\toprule
Year & Reference & Main contribution \\
\midrule
\endhead
1985 & Kuzmin, Rubakov, Shaposhnikov \cite{KuzminRubakovShaposhnikov1985} & Established the cosmological importance of anomalous electroweak baryon-number nonconservation. \\
1991 & Vachaspati \cite{Vachaspati1991} & Proposed magnetic-field generation from Higgs gradients during cosmological symmetry breaking. \\
1996--2000 & Baym--Bödeker--McLerran; Ahonen--Enqvist; Copeland--Saffin--Törnkvist \cite{BaymBodekerMcLerran1996,AhonenEnqvist1998,CopelandSaffinTornkvist2000} & Developed bubble-collision magnetogenesis and gauge-field dynamics. \\
1997 & Joyce--Shaposhnikov \cite{JoyceShaposhnikov1997} & Showed that a right-handed-electron asymmetry can trigger helical hypermagnetic-field growth through the Abelian anomaly. \\
1998 & Giovannini--Shaposhnikov \cite{GiovanniniShaposhnikov1998,GiovanniniShaposhnikov1998BBN} & Connected hypermagnetic helicity to baryogenesis and baryon inhomogeneity/BBN. \\
1998--1999 & Elmfors--Enqvist--Kainulainen; Comelli et al \cite{ElmforsEnqvistKainulainen1998,Comelli1999}. & Quantified magnetic effects on the EW transition and the sphaleron barrier. \\
2001 & Grasso--Rubinstein \cite{GrassoRubinstein2001} & Classic broad review of early-Universe magnetic fields including electroweak baryogenesis. \\
2001 & Vachaspati \cite{Vachaspati2001} & Proposed baryogenesis as a source of primordial magnetic helicity. \\
2004 & Banerjee--Jedamzik \cite{BanerjeeJedamzik2004} & Developed a standard cosmological MHD evolution framework including viscous stages. \\
2005--2008 & Kahniashvili et al \cite{KahniashviliGogoberidzeRatra2005,Kahniashvili2008}. & Established polarized GWs from helical turbulence and inverse-cascade MHD. \\
2010 & Neronov--Vovk \cite{NeronovVovk2010} & Opened the modern gamma-ray lower-bound program for void IGMFs. \\
2012--2013 & Boyarsky et al.; Durrer--Neronov \cite{Boyarsky2012,DurrerNeronov2013} & Developed chiral magnetic evolution and consolidated PMF generation/evolution/observation. \\
2014 & D'Onofrio--Rummukainen--Tranberg \cite{DOnofrio2014} & Lattice determination of the physical SM crossover and sphaleron freeze-out. \\
2015 & Brandenburg--Kahniashvili--Tevzadze \cite{Brandenburg2015} & Demonstrated nonhelical inverse transfer in decaying magnetically dominated turbulence. \\
2015--2016 & Anber--Sabancilar; Fujita--Kamada; Kamada--Long \cite{AnberSabancilar2015,FujitaKamada2016,KamadaLong2016a,KamadaLong2016b} & Modern quantitative era of hypermagnetic baryogenesis and electroweak-crossover evolution. \\
2018--2019 & Kamada; Domcke et al \cite{Kamada2018,DomckeHarlingMorganteMukaida2019}. & High-scale chirality/helicity storage and axion-inflation baryogenesis. \\
2021 & Kamada--Uchida--Yokoyama \cite{KamadaUchidaYokoyama2021} & Strong BBN baryon-isocurvature constraint on pre-EW PMF explanations of BAU+IGMF. \\
2023 & Kamada--Yamamoto--Yang \cite{KamadaYamamotoYang2023} & Comprehensive review of chiral effects, chiral MHD, PMFs, baryogenesis and chiral GWs. \\
2023 & Annala--Rummukainen \cite{AnnalaRummukainen2023} & Modern nonperturbative treatment of the electroweak sphaleron in magnetic fields. \\
2021--2023 & Hosking--Schekochihin \cite{HoskingSchekochihin2021,HoskingSchekochihin2023} & Reconnection-controlled decay framework (2021), with application to cosmic-void magnetism (2023). \\
2024--2026 & Liu--Qin--Bian \cite{LiuQinBian2026} & CP-helical FOPT magnetogenesis with CME-amplified baryogenesis. \\
2025 & Domcke--Ema--Patel \cite{DomckeEmaPatel2025} & Spontaneous CP-odd twisting of EW dumbbells generating helical fields and BAU. \\
2025 & Uchida--Kamada--Tashiro \cite{UchidaKamadaTashiro2025} & Recomputed CMB spectral-distortion limits using reconnection-driven PMF decay. \\
2026 & Hamada--Mukaida--Uchida \cite{HamadaMukaidaUchida2026} & Gauge-invariant hot-SM magnetic flux and reassessment of helicity-decay baryogenesis. \\
2026 & Di--Bian--Cai \cite{DiBianCai2026} & Real-time numerical study of magnetic effects on a first-order EW transition and sphalerons. \\
\bottomrule
\end{longtable}

\section{Chemical potentials, spectator constraints and anomaly bookkeeping}
\label{sec:chemicalnetwork}
\label{app:chemical}

This appendix gives a compact recipe for reconstructing the charge network used in magnetic baryogenesis calculations. Its purpose is not to impose one universal reduced system, but to make transparent where the numerical conversion between hypermagnetic helicity and baryon number comes from.

\subsection{Relativistic charge susceptibilities}

For a relativistic species with a small chemical potential $|\mu_i|\ll T$, the particle--antiparticle density difference is linear in $\mu_i$,
\begin{equation}
n_i-\bar n_i = \chi_i\mu_i+\mathcal O(\mu_i^3).
\end{equation}
For a massless Weyl fermion, $\chi_i=g_iT^2/6$, while for a complex bosonic degree of freedom the coefficient is twice as large. Interactions and finite-temperature masses correct these ideal-gas susceptibilities, but the linear system structure remains. It is often cleaner to formulate the network directly in terms of susceptibilities computed in the thermal theory than to identify number densities with chemical potentials by inspection.

For the Standard Model the useful hypercharge assignments are
\begin{center}
\begin{tabular}{@{}lcc@{}}
\toprule
field & representation & $Y$ with $Q=T_3+Y$ \\
\midrule
$Q_L$ & $(\mathbf 3,\mathbf 2)$ & $1/6$ \\
$u_R$ & $(\mathbf 3,\mathbf 1)$ & $2/3$ \\
$d_R$ & $(\mathbf 3,\mathbf 1)$ & $-1/3$ \\
$L_L$ & $(\mathbf 1,\mathbf 2)$ & $-1/2$ \\
$e_R$ & $(\mathbf 1,\mathbf 1)$ & $-1$ \\
$H$ & $(\mathbf 1,\mathbf 2)$ & $1/2$ \\
\bottomrule
\end{tabular}
\end{center}
A paper using the alternative convention $Q=T_3+Y/2$ doubles all tabulated hypercharges; anomaly coefficients must be transformed consistently.

\subsection{Fast interactions as linear constraints}

When a reaction is much faster than the Hubble rate it imposes a chemical-equilibrium relation. With the convention above, equilibrated Yukawa interactions imply schematically
\begin{align}
\mu_{u_i}&=\mu_{Q_i}+\mu_H,\\
\mu_{d_i}&=\mu_{Q_i}-\mu_H,\\
\mu_{e_i}&=\mu_{L_i}-\mu_H.
\end{align}
Weak sphalerons drive the combination
\begin{equation}
\sum_{i=1}^{3}\left(3\mu_{Q_i}+\mu_{L_i}\right)
\end{equation}
toward zero, while strong sphalerons impose
\begin{equation}
\sum_{i=1}^{3}\left(2\mu_{Q_i}-\mu_{u_i}-\mu_{d_i}\right)=0
\end{equation}
in equilibrium. Hypercharge neutrality requires
\begin{equation}
\sum_i Y_i(n_i-\bar n_i)=0,
\end{equation}
including the Higgs contribution and the proper multiplicities. These equations reduce a many-species system to a small number of approximately conserved charges.

Which relations may be imposed depends on temperature. Modern spectator-rate calculations place electron-Yukawa equilibration in a standard radiation-dominated cosmology at a characteristic temperature of order $10^5\,\mathrm{GeV}$, with order-one rate-definition uncertainties; older magnetic-baryogenesis papers often used a lower scale of order $10\,\mathrm{TeV}$ \cite{GarbrechtSchwaller2014,DomckeKamadaMukaida2023}. This distinction matters for the window in which a right-handed-electron asymmetry can trigger the chiral plasma instability. More generally, partially equilibrated spectators should be evolved with finite rates rather than switched on discontinuously \cite{GarbrechtSchwaller2014,Fong2022}.

\subsection{Abelian anomaly source}

For each chiral fermion current, the $U(1)_Y$ anomaly contributes a term proportional to its chirality sign, multiplicity and $Y_i^2$,
\begin{equation}
\partial_\mu j_i^\mu = C_i\frac{g'^2}{16\pi^2}Y_{\mu\nu}\widetilde Y^{\mu\nu}+\cdots,
\end{equation}
where $C_i$ should be computed once in the convention used for the current. Because
\begin{equation}
Y_{\mu\nu}\widetilde Y^{\mu\nu}=-4\bm E_Y\cdot\bm B_Y
\end{equation}
for a standard metric/sign convention, the corresponding yield equation contains
\begin{equation}
\frac{dY_i}{dt}\supset -C_i\frac{g'^2}{4\pi^2s}
\langle\bm E_Y\cdot\bm B_Y\rangle.
\end{equation}
Using
\begin{equation}
\dot h_Y=-2\langle\bm E_Y\cdot\bm B_Y\rangle,
\end{equation}
one may equivalently regard a change in helicity as an injection of anomalous charge. The sign is physical only after the conventions for $\epsilon^{0123}$, hypercharge and helicity have all been fixed. This is why sign claims connecting cosmic helicity to the sign of the BAU must state conventions explicitly.

\subsection{Matrix formulation}

Let $\bm q$ denote a vector of charge densities or yields and let $\bm\mu=\bm\chi^{-1}\bm q$ be the associated chemical potentials. A convenient general network is
\begin{equation}
\dot{\bm q}+3H\bm q
=\bm C_Y\,\frac{g'^2}{4\pi^2}\langle\bm E_Y\cdot\bm B_Y\rangle
-\sum_a \Gamma_a\,\bm v_a\left(\bm v_a^{\mathsf T}\bm\mu\right)
+\bm S_{\rm BSM},
\label{eq:generalchargenetwork}
\end{equation}
where each reaction $a$ relaxes a particular linear combination $\bm v_a^{\mathsf T}\bm\mu$, $\bm C_Y$ contains anomaly coefficients and $\bm S_{\rm BSM}$ denotes any new source such as heavy-neutrino decays or bubble-wall CP violation. Hypercharge neutrality can either be imposed as an algebraic constraint or included through an appropriate basis of independent charges.

Equation \eqref{eq:generalchargenetwork} is more robust than hard-coding a single set of coefficients because it permits rates to pass smoothly through partial equilibrium. It also makes clear how a pre-existing leptogenesis source and magnetic helicity source can interfere rather than simply add.

\subsection{From the charge network to baryon number}

The baryon density is a fixed linear combination of quark densities,
\begin{equation}
n_B=\frac13\sum_{i=1}^{3}\left[2n_{Q_i}+n_{u_i}+n_{d_i}\right]_{\rm asym},
\end{equation}
where the color and weak-isospin multiplicities are understood in the definition of each $n_i$. Once the fast constraints have been solved, one can write
\begin{equation}
Y_B=\bm c_B^{\mathsf T}\bm q_{\rm independent}.
\end{equation}
The familiar coefficient $B=(28/79)(B-L)$ follows only under a particular set of Standard Model equilibrium assumptions and is not the appropriate universal conversion formula during every temperature interval relevant for hypermagnetic baryogenesis.

In helicity-driven baryogenesis the weak sphaleron is simultaneously a conversion and washout process. A schematic reduced equation is
\begin{equation}
\dot Y_B=S_h(t)-\Gamma_B(T)Y_B+\sum_j a_j\Gamma_jY_j,
\end{equation}
where $S_h$ follows from the evolving hypermagnetic helicity and the final terms encode spectator charges. The Kamada--Long calculations implement such a network through the electroweak crossover \cite{KamadaLong2016a,KamadaLong2016b}; the 2026 generalized-symmetry analysis shows that the magnetic-flux conversion entering $S_h$ itself requires further nonperturbative control \cite{HamadaMukaidaUchida2026}.

\subsection{Chiral chemical potential relevant for the CPI}

The hypercharge chiral magnetic current is not controlled by a single lepton chemical potential. In the Standard Model it is a weighted combination of the chemical potentials of all chiral species,
\begin{equation}
\bm J_Y^{\rm CME}=\xi_Y\bm B_Y,
\qquad
\xi_Y\propto g'^2\sum_i c_iY_i^2\mu_i,
\end{equation}
with $c_i$ carrying chirality/multiplicity factors. Spectator processes can therefore move asymmetry into or out of the combination that drives the CPI even while conserving a total flavor or $B-L$ charge. This feature underlies the modern constraint on large primordial lepton-flavor asymmetries from CPI-triggered hypermagnetogenesis and subsequent electroweak baryogenesis \cite{DomckeKamadaMukaida2023}.

\subsection{Practical implementation checklist}

A reproducible numerical network should specify:
\begin{enumerate}
\item the hypercharge convention and anomaly-sign convention;
\item the susceptibility matrix, including whether thermal corrections are used;
\item which reactions are evolved with finite rates and which are imposed as equilibrium constraints;
\item the temperature dependence of weak/strong sphaleron and Yukawa rates;
\item the magnetic source $\langle\bm E_Y\cdot\bm B_Y\rangle$ or full spectral source;
\item the treatment of hypermagnetic-to-electromagnetic conversion through the crossover;
\item the initial conserved charges and any BSM source terms.
\end{enumerate}

The proportionality between magnetic helicity and the final baryon abundance is not fixed by the anomaly equation alone. Its numerical coefficient depends on the charge susceptibilities, spectator processes, and interaction rates active at each temperature. Precision calculations should therefore evolve the coupled charge network explicitly, rather than import a helicity-to-$Y_B$ conversion factor outside the regime where that coefficient was derived.




\end{document}